# RAY VELOCITY DERIVATIVES IN ANISOTROPIC ELASTIC MEDIA.
# PART I – GENERAL ANISOTROPY

*Zvi Koren and Igor Ravve (corresponding author), Emerson*

zvi.koren@emerson.com, igor.ravve@emerson.com

## ABSTRACT

We present an original, generic, and efficient approach for computing the first and second partial derivatives of ray (group) velocities along ray paths in general anisotropic elastic media. As the ray velocities deliver the ray element traveltimes, this set of partial derivatives constructs the so-called kinematic and dynamic sensitivity kernels (matrices) which are used in different key modeling and inversion methods, such as two-point ray bending methods and seismic tomography. The second derivatives are useful in the solution of the above-mentioned kinematic problems and they are essential for evaluating the dynamic properties along the rays (amplitudes and phases). The traveltime is delivered through an integral over a given Lagrangian defined at each point along the ray. In our approach, we use an arclength-related Lagrangian that depends on the ray velocity vectors. Although the Lagrangian (and hence the ray velocity magnitude) cannot be explicitly expressed in terms of the medium properties and the ray direction components, its derivatives can still be formulated analytically using the corresponding arclength-related Hamiltonian that can be explicitly expressed in terms of the medium properties and the slowness vector components. This requires first to invert for the slowness vector components, given the ray direction components. Computation of the slowness vector and the ray velocity derivatives is considerably simplified by using an auxiliary scaled-time-related Hamiltonian obtained directly from the Christoffel equation and connected to the arclength-related Hamiltonian by a simple scale factor.

This study consists of two parts. In Part I, we consider general anisotropic (triclinic) media, and provide the derivatives (gradients and Hessians) of the ray velocity, with respect to (1) the spatial/directional vectors and (2) the elastic model parameters. The derivatives are obtained for both quasi-compressional and quasi-shear waves, where other types of media, characterized with higher symmetries, can be considered particular cases. In Part II, we apply the theory of Part I explicitly to polar anisotropic media (transverse isotropy with tilted axis of symmetry, TTI), and obtain the explicit ray velocity derivatives for the coupled qP and qSV waves and for SH waves. The derivatives for polar anisotropy are obviously simplified, yielding more effective computations. The ray velocity derivatives are tested by checking consistency between the proposed analytical formulas and the corresponding numerical ones.

1. INTRODUCTION

The first and second partial derivatives of the traveltimes along ray trajectories in anisotropic elastic media construct the seismic-based kinematic and dynamic sensitivity kernels which are used in many seismic-driven modeling, imaging and tomographic methods. For example, assuming fixed model parameters, the partial traveltime derivatives with respect to (wrt) the spatial coordinates and the ray direction unit vectors along approximated (non-stationary) ray paths, are used for solving two-point ray bending problems (converging to stationary ray paths). Recently, Koren and Ravve (2021) and Ravve and Koren (2021) proposed an original two-point ray bending optimization method (referred to as the Eigenray method) for obtaining stationary ray paths and for computing the dynamic parameters along rays in general anisotropic media. A comprehensive list of references on ray bending methods, mainly existing for isotropic or transverse isotropic media, is included in these papers (e.g., Julian and Gubbins, 1977; Smith et al., 1979; Pereyra et al., 1980; Pereyra 1992, 1996; Moser et al. (1992), Snieder and Spencer, 1993; Grechka and McMechan, 1996; Bona and Slawinski, 2003; Zhou and Greenhalgh, 2005, 2006; Kumar et al., 2004; Casasanta et al., 2008; Wong, 2010; Sripanich and Fomel, 2014; Cao et al., 2017, among others). Also, assuming fixed ray trajectories in a background anisotropic velocity model, their time derivatives wrt the spatially varying elastic model parameters (the Fréchet derivatives) are used in the solution of seismic tomography for updating the model parameters. (Among the many publications on this topic, we mention the papers by Bishop et al., 1985; Farra and Madariaga, 1988; Williamson, 1990; Stork, 1992; Kosloff et al., 1996 and Koren et al., 2008; Zhang et al., 2013). The first derivatives, and optionally the second derivatives, are required for solving the above-mentioned kinematic problems, and the second derivatives are essential for computing the dynamic parameters, such as amplitudes and phases along the stationary rays, and their sensitivity to changes in the model parameters (e.g., Wang and Pratt, 1997 for isotropic media).

The primary objective of this study is to provide explicit analytic formulae for the derivatives of the ray (group) velocity of quasi-compressional and quasi-shear waves that construct the seismic-based kinematic and dynamic sensitivity kernels. It is assumed that the spatial distributions of the material elastic parameters, normally given on a fine 3D grid, are preconditioned (smoothed) to reliably provide their numerical first and second spatial derivatives. This study is motivated by the need to compute the spatial/directional gradients and Hessians of the ray velocity magnitudes which are the core computational components in the Eigenray method, presented recently by Koren and Ravve (2020) and Ravve and Koren (2020a). The kinematic Eigenray method is formulated as a solution of Fermat's principle for obtaining a stationary ray path between two fixed endpoints,

$$t = \int_S d\tau = \int_S L(\mathbf{x},\mathbf{r}) ds \rightarrow \text{stationary} \rightarrow \delta t = \int_S \delta L(\mathbf{x},\mathbf{r}) ds = 0 \quad , \quad (1.1)$$

where $t$ and $\delta t$ are the global traveltime and the traveltime variation (change) computed along the ray, respectively, and $d\tau$ is the integration running time. $S$ and $R$ are the fixed endpoints of the path: the source and receiver positions, $s$ is the arclength flow parameter, $\mathbf{x}(s)$ and $\mathbf{r}(s) \equiv \dot{\mathbf{x}} \equiv d\mathbf{x}/ds$ are the ray locations and ray directions (also the ray velocity directions), respectively, and $L(\mathbf{x},\mathbf{r})$ is the arclength-related Lagrangian, proposed in the Eigenray method to be,

$$L(\mathbf{x},\mathbf{r}) = \frac{\sqrt{\mathbf{r}\cdot\mathbf{r}}}{v_{\text{ray}}(\mathbf{x},\mathbf{r})} = \frac{\sqrt{\mathbf{r}\cdot\mathbf{r}}}{v_{\text{ray}}\left[\tilde{\mathbf{C}}(\mathbf{x}),\mathbf{r}\right]} \quad , \quad (1.2)$$

where $\tilde{\mathbf{C}}(\mathbf{x})$ is the spatially varying density-normalized fourth-order stiffness tensor. Parameter $v_{\text{ray}}$ is the magnitude of the ray velocity which explicitly depends on the ray direction, and implicitly on the ray coordinates (via the spatially varying anisotropic elastic parameters). Thus, traveltimes along the stationary ray paths are computed with the arclength-related Lagrangian (equation 1.2) that contains the reciprocal ray velocities along the ray elements, and the core computational part is reduced to evaluating the partial derivatives of the ray velocity magnitudes at discretized nodes along the rays. However, using the Lagrangian formulation, these derivatives can only be formulated implicitly (there is no explicit form for the ray velocity vs. its direction in general anisotropy), while, using the Hamiltonian formulation (with the dependency on the slowness vector $\mathbf{p}$ rather than the ray direction vector $\mathbf{r}$), the ray velocity derivatives can be obtained explicitly and analytically. The arclength-related Hamiltonian $H(\mathbf{x},\mathbf{p})$ can be defined through the scaled-time-related Hamiltonian $H^{\bar{\tau}}(\mathbf{x},\mathbf{p})$ which is obtained directly from the Christoffel equation,

$$H^{\bar{\tau}}(\mathbf{x},\mathbf{p}) = \det(\mathbf{\Gamma} - \mathbf{I}) \quad . \tag{1.3}$$

It will be considered in this study as the "reference Hamiltonian". The slowness vector $\mathbf{p}$ is related to the ray direction unit vector $\mathbf{r}$ through an inversion mechanism, implicitly formulated as $\mathbf{p} = \mathbf{p}\left[\tilde{\mathbf{C}}(\mathbf{x}),\mathbf{r}\right]$, where a constraint holds between the two vectors, $\mathbf{r}$ and $\mathbf{p}$, and the ray velocity magnitude, $\mathbf{r} \cdot \mathbf{p} = v_{\text{ray}}^{-1}$. Parameter $\mathbf{\Gamma}$ is the Christoffel matrix (tensor), $\mathbf{\Gamma} = \mathbf{p}\tilde{\mathbf{C}}\mathbf{p}$, $\mathbf{I}$ is the identity matrix, and superscript $\bar{\tau}$ indicates that the reference Hamiltonian's flow parameter is the scaled time (e.g., Koren and Ravve, 2021, to be explained later). This scaled-time-related (reference) Hamiltonian $H^{\bar{\tau}}(\mathbf{x},\mathbf{p})$ is connected to the arclength-related Hamiltonian $H(\mathbf{x},\mathbf{p})$ by,

$$H(\mathbf{x},\mathbf{p}) = \frac{H^{\bar{\tau}}(\mathbf{x},\mathbf{p})}{\sqrt{H_{\mathbf{p}}^{\bar{\tau}} H_{\mathbf{p}}^{\bar{\tau}}}} \quad ; \quad H_{\mathbf{p}}^{\bar{\tau}} \equiv \frac{\partial H^{\bar{\tau}}}{\partial \mathbf{p}} \quad , \tag{1.4}$$

where the latter is also related to the arclength-related Lagrangian via the Legendre transform,

$$L(\mathbf{x},\mathbf{r}) = \mathbf{r} \cdot \mathbf{p} - H(\mathbf{x},\mathbf{p}) \quad . \tag{1.5}$$

The reference Hamiltonian $H^{\bar{\tau}}(\mathbf{x},\mathbf{p})$ is first used in the slowness inversion to obtain (invert for) the slowness vectors $\mathbf{p}$, given the ray direction $\mathbf{r}$: A unique solution for quasi-compressional waves and multiple solutions for quasi-shear waves. Then, as indicated above, the ray velocity magnitude $v_{\text{ray}}$ can be explicitly expressed (and computed) as the reciprocal of the scalar product of the slowness vector and the ray direction vector,

$$v_{\text{ray}} = (\mathbf{p} \cdot \mathbf{r})^{-1} \quad , \quad \mathbf{p}(\mathbf{x},\mathbf{r}) = \mathbf{p}\big[\mathbf{m}(\mathbf{x}),\mathbf{r}\big] \quad , \qquad (1.6)$$

where, for general anisotropy, the spatially varying model parameters $\mathbf{m}(\mathbf{x})$ include the components of the elastic (stiffness) tensor $\tilde{\mathbf{C}}(\mathbf{x})$ or its equivalent stiffness matrix (Voight) representation $\mathbf{C}(\mathbf{x})$, while for higher symmetries, the model parameters may include, for example, the compressional and shear velocities in the local ("crystal") vertical direction, the normalized (unitless) anisotropic parameters (e.g., Thomsen or Tsvankin parameters), and the orientation angles of the crystal frame.

We explicitly obtain all the ray velocity derivative types for general anisotropy and for all wave types, as functions of the derivatives of the arclength-related Hamiltonian. The latter, in turn, are computed via the derivatives of the reference Hamiltonian; this operation considerably simplifies the computations. Note, however, that although the actual derivatives of the ray velocity magnitude are computed via the Hamiltonian derivatives, the ray velocity magnitude is still considered a "Lagrangian object", where the independent degrees of freedom (DoF) are the ray location and direction vectors, $\mathbf{x}$ and $\mathbf{r}$, respectively, while the Hamiltonian DoF are the location and the slowness vectors, $\mathbf{x}$ and $\mathbf{p}$. In particular, we emphasize that the spatial derivatives of the ray velocity assume a constant ray direction and a varying slowness, while the spatial derivatives of the Hamiltonian assume a constant slowness and a varying ray direction.

In this paper, we present a generic approach for efficiently computing the different types of the first and second derivatives of the ray velocities at specified ray nodes; they deliver the corresponding derivatives of traveltime ray elements in general anisotropic media: The spatial and directional gradient vectors, the spatial, directional and mixed Hessian matrices, and the first and second derivatives wrt the model parameters. These derivatives are computed at specific points $\mathbf{x}$ (nodes) along the ray trajectory (or its approximation, for example, in the case of ray bending problems) given the local physical (elastic) properties of the medium (or the background medium, for example, in the case of tomography), and the ray velocity direction $\mathbf{r}$. The spatial/directional derivatives constitute two gradient vectors of length 3 and three Hessian matrices of dimensions $3 \times 3$,

$$v_{\text{ray}}, \nabla_{\mathbf{x}} v_{\text{ray}}, \nabla_{\mathbf{r}} v_{\text{ray}}, \nabla_{\mathbf{x}} \nabla_{\mathbf{x}} v_{\text{ray}}, \nabla_{\mathbf{r}} \nabla_{\mathbf{r}} v_{\text{ray}}, \nabla_{\mathbf{r}} \nabla_{\mathbf{x}} v_{\text{ray}} = \left( \nabla_{\mathbf{x}} \nabla_{\mathbf{r}} v_{\text{ray}} \right)^T \quad , \tag{1.7}$$

where subscript $\mathbf{x}$ is related to the spatial derivatives, and $\mathbf{r}$ to the directional, which means derivatives wrt the Cartesian components of the ray direction $\mathbf{r}$.

The derivatives wrt the model parameters (the Fréchet derivatives) are,

$$\nabla_{\mathbf{m}} v_{\text{ray}}, \quad \nabla_{\mathbf{m}} \nabla_{\mathbf{m}} v_{\text{ray}} \quad . \tag{1.8}$$

These derivatives are, for example, the components of the sensitivity matrix used in traveltime tomography for updating the model parameters. In this case, the ray trajectory points $\mathbf{x}$ and the ray directions $\mathbf{r}$ are assumed fixed, where each partial derivative describes the ray velocity change wrt a change of a given medium's property $m_i$ at a given point, where all the other properties (at the given point and at all other points along the ray) are fixed. The non-diagonal second derivatives show the effect of a simultaneous change of two medium properties $m_i, m_j$ (at the same location) on the ray velocity magnitude in a fixed direction at that point. The medium properties $m_i$ (at the nodal points) may be the stiffness tensor components or their normalized equivalents (e.g., Thomsen, 1986, or Alkhalifah, 1998, for polar anisotropy; Tsvankin, 1997, 2012, or Alkhalifah, 2003, for orthorhombic media; Tsvankin and Grechka, 2011, for monoclinic media; Farra and Pšenčík, 2016, for general anisotropy up to triclinic). A prerequisite for the ray velocity derivatives is the knowledge of the slowness vector **p** at the considered point; otherwise, it should be computed (inverted).

The paper is structured as follows. We start with important background theoretical sections which are essential for establishing the desired derivatives. We begin with the variational formulations delivering the kinematic and dynamic differential equations of the Eigenray method and provide the explicit relations between the derivatives of the Lagrangian and those of the ray velocity magnitude. We then introduce the reference Hamiltonian, which will be used for the actual computations, and relate its derivatives to those of the arclength-related Hamiltonian. Next, we provide the relationships between the arclength-related Lagrangian and its derivatives, on one side, and the corresponding arclength-related Hamiltonian and its derivatives, on the other side. Additionally, since the actual computation of the set of the ray velocity derivatives requires the knowledge of the slowness vector components, we review the method for their inversion.

Finally, we check and confirm our relationships for the ray velocity derivatives using a given point and its small neighborhood in an inhomogeneous triclinic medium (with strong anisotropic effects), where the input includes (1) the model parameters: The 21 density-normalized elastic components of the stiffness matrix and their spatial gradients and Hessians, and (2) a given ray direction vector. In the example, we first invert for the slowness vectors: One solution for quasi-compression waves and (in this case) all eighteen solutions for quasi-shear waves (Grechka, 2017), and compute the corresponding ray velocity magnitudes, $v_{\text{ray}} = (\mathbf{p} \cdot \mathbf{r})^{-1}$. We then compute all the partial derivatives indicated in equations 1.6 and 1.7. We validate the correctness of the derivatives by comparing the analytic derivatives with the corresponding numerical derivatives approximated by finite differences. We then discuss the importance of the analytic derivatives and indicate the drawbacks in using numerical difference approximations (which, in this work, are only used for verification purposes).

Most of the mathematical proofs are moved to the appendices. In Appendix A we prove a key identity that relates the spatial gradient of the arclength-related Hamiltonian, $H_{\mathbf{x}}$, to its slowness gradient, $H_{\mathbf{p}} = \mathbf{r}$. In Appendices B, C and D we derive the directional gradient, directional Hessian, and the mixed Hessian of the ray velocity magnitude vs. the corresponding derivatives of the arclength-related Hamiltonian. In Appendix E we demonstrate that the sign of the reference Hamiltonian for qP waves is always plus, while for qS waves it can only be defined after performing the slowness inversion. In Appendix F we derive the relationships between the gradient and Hessian of the ray velocity magnitude wrt the ray direction vector, and the corresponding first and second derivatives of this magnitude wrt the dual ray direction polar angles: zenith and azimuth.

## 2. KINEMATIC AND DYNAMIC VARIATIONAL FORMULATIONS

Following the theory behind the Eigenray method, in order to find the stationary ray path corresponding to the vanishing first variation of the traveltime (kinematic problem) and to explore its second variation while solving the dynamic problem, we introduce the arclength-related Lagrangian (equation 1.2) to the Euler-Lagrange equation,

$$\frac{d}{ds} L_{\mathbf{r}} = L_{\mathbf{x}} \quad , \tag{2.1}$$

and obtain the governing kinematic relationship,

$$\frac{d}{ds}\left( \frac{\mathbf{r}}{v_{\text{ray}}} - \frac{\nabla_{\mathbf{r}} v_{\text{ray}}}{v_{\text{ray}}^2} \right) = -\frac{\nabla_{\mathbf{x}} v_{\text{ray}}}{v_{\text{ray}}^2} \quad , \tag{2.2}$$

where the expression in brackets represents the slowness vector $\mathbf{p}$, due to the momentum equation, $L_{\mathbf{r}} = \mathbf{p}$. Subscripts $\mathbf{x}$ and $\mathbf{r}$ mean the corresponding gradients, double subscripts mean the Hessians. Similarly, the second traveltime variation yields the dynamic equation in Lagrangian formulation, referred to as Jacobi equation (Bliss, 1916; Ravve and Koren, 2021),

$$\frac{d}{ds}\left( L_{\mathbf{rx}} \cdot \mathbf{u} + L_{\mathbf{rr}} \cdot \dot{\mathbf{u}} \right) = L_{\mathbf{xx}} \cdot \mathbf{u} + L_{\mathbf{xr}} \cdot \dot{\mathbf{u}} \quad , \tag{2.3}$$

where $s$ is the arclength of the central ray, vector $\mathbf{u}$ is a normal shift between the central ray and a paraxial ray, $\mathbf{u} \cdot \mathbf{r} = 0$. For a point source (where the initial conditions are based on the ray-normal eigenvectors of matrix $L_{\mathbf{rr}}$), two independent solutions for $\mathbf{u}$ ($\mathbf{u}_1$ and $\mathbf{u}_2$) are needed to obtain the Jacobian, $J = \mathbf{u}_1 \times \mathbf{u}_2 \cdot \mathbf{r}$ (the cross-section area of the ray tube), which, in turn, makes it possible to compute the (relative) geometric spreading (Červený, 2000; Ravve and Koren, 2021a).

Thus, the spatial and directional gradients of the Lagrangian along the ray path, $L_\mathbf{x}$ and $L_\mathbf{r}$, are needed for the kinematic ray bending, and the spatial, directional and mixed Hessians of the Lagrangian along the stationary path, $L_{\mathbf{xx}}, L_{\mathbf{rr}}$ and $L_{\mathbf{xr}} = L_{\mathbf{rx}}^T$, are needed for solving the dynamic problem. Depending on the solution method (e.g., using the Newton optimization method), the Hessians may be useful in the kinematic stage as well. Finally, the gradients and Hessians of the arclength-related Lagrangian defined in equation 1.2 are arranged as (Koren and Ravve, 2021),

$$L_\mathbf{x} = \frac{\partial L}{\partial \mathbf{x}} = -\frac{\nabla_\mathbf{x} v_{\text{ray}} \sqrt{\mathbf{r} \cdot \mathbf{r}}}{v_{\text{ray}}^2} \quad , \quad L_\mathbf{r} = \frac{\partial L}{\partial \mathbf{r}} = \frac{\mathbf{r}}{v_{\text{ray}} \sqrt{\mathbf{r} \cdot \mathbf{r}}} - \frac{\nabla_\mathbf{r} v_{\text{ray}} \sqrt{\mathbf{r} \cdot \mathbf{r}}}{v_{\text{ray}}^2} \quad ,$$

$$L_{\mathbf{xx}} = \frac{\partial^2 L}{\partial \mathbf{x}^2} = 2\frac{\nabla_\mathbf{x} v_{\text{ray}} \otimes \nabla_\mathbf{x} v_{\text{ray}}}{v_{\text{ray}}^3} - \frac{\nabla_\mathbf{x} \nabla_\mathbf{x} v_{\text{ray}}}{v_{\text{ray}}^2} \quad ,$$

$$L_{\mathbf{xr}} = L_{\mathbf{rx}}^T = \frac{\partial^2 L}{\partial \mathbf{x} \partial \mathbf{r}} = -\frac{\nabla_\mathbf{x} v_{\text{ray}} \otimes \mathbf{r}}{v_{\text{ray}}^2} + 2\frac{\nabla_\mathbf{x} v_{\text{ray}} \otimes \nabla_\mathbf{r} v_{\text{ray}}}{v_{\text{ray}}^3} - \frac{\nabla_\mathbf{x} \nabla_\mathbf{r} v_{\text{ray}}}{v_{\text{ray}}^2} \quad ,$$

$$L_{\mathbf{rr}} = \frac{\mathbf{I} - \mathbf{r} \otimes \mathbf{r}}{v_{\text{ray}}} - \frac{\mathbf{r} \otimes \nabla_\mathbf{r} v_{\text{ray}} + \nabla_\mathbf{r} v_{\text{ray}} \otimes \mathbf{r}}{v_{\text{ray}}^2} + 2\frac{\nabla_\mathbf{r} v_{\text{ray}} \otimes \nabla_\mathbf{r} v_{\text{ray}}}{v_{\text{ray}}^3} - \frac{\nabla_\mathbf{r} \nabla_\mathbf{r} v_{\text{ray}}}{v_{\text{ray}}^2} \quad ,$$

(2.4)

where the derivatives of the Lagrangian are expressed in terms of the derivatives of the ray velocity magnitude (hence, the motivation for this study). Throughout the paper we provide this set of ray velocity derivatives explicitly in terms of the derivatives of the corresponding arclength-related Hamiltonian and we relate it to the reference Hamiltonian. The latter is more convenient for the slowness vector inversion, and the derivatives of the reference Hamiltonian are much simpler than those of the arclength-related one.

## 3. THE REFERENCE HAMILTONIAN

As briefly mentioned in the Introduction, by definition, we call the (vanishing and unitless) Hamiltonian arising from the Christoffel equation (1877),

$$H^{\bar{\tau}} = \det[\mathbf{\Gamma} - \mathbf{I}] \quad , \quad \mathbf{\Gamma} = \mathbf{p} \cdot \tilde{\mathbf{C}} \cdot \mathbf{p} \quad , \tag{3.1}$$

the reference Hamiltonian; we find this Hamiltonian the simplest and most convenient to perform the slowness inversion and it is also used as an auxiliary Hamiltonian in the actual computation of the ray velocity derivatives. Recall that matrix $\mathbf{\Gamma}$ is the Christoffel tensor, $\tilde{\mathbf{C}}$ is the fourth-order density-normalized stiffness tensor, and $\mathbf{I}$ is the identity tensor (matrix). The definition of the reference Hamiltonian is valid for any type of anisotropy, and the scaled time $\bar{\tau}$ is related to the actual running time $\tau$, as (Koren and Ravve, 2021),

$$\alpha_{sc}(\bar{\tau}) = \frac{d\bar{\tau}}{d\tau} = \frac{1}{\mathbf{p} \cdot H_{\mathbf{p}}^{\bar{\tau}}} \quad , \quad \frac{ds}{d\bar{\tau}} = \sqrt{H_{\mathbf{p}}^{\bar{\tau}} \cdot H_{\mathbf{p}}^{\bar{\tau}}} \quad , \quad \mathbf{v}_{ray} = v_{ray} \mathbf{r} = \frac{H_{\mathbf{p}}^{\bar{\tau}}}{\mathbf{p} \cdot H_{\mathbf{p}}^{\bar{\tau}}} \quad , \tag{3.2}$$

where $\alpha_{sc}$ is the scaling factor and $\mathbf{v}_{ray}$ is the ray velocity vector. The gradients and Hessians of the ray velocity magnitude are related to those of the arclength-related Lagrangian, and the latter, in turn, are related to the derivatives of its corresponding arclength-related Hamiltonian. However, the analytic expressions for the gradients and especially for the Hessians of the arclength-related Hamiltonian are unwieldy, and it becomes easier to link the derivatives of the arclength-related Hamiltonian to the derivatives of the reference Hamiltonian. The latter can be computed relatively easily. Note that the components of the Christoffel matrix (tensor) $\mathbf{\Gamma}$ for triclinic media are quadratic functions of the slowness vector components and linear functions of the stiffness tensor components (e.g., Slawinski, 2015),

$$\Gamma_{11} = C_{11}p_1^2 + C_{66}p_2^2 + C_{55}p_3^2 + 2(C_{16}p_1p_2 + C_{56}p_2p_3 + C_{15}p_1p_3),$$
$$\Gamma_{22} = C_{66}p_1^2 + C_{22}p_2^2 + C_{44}p_3^2 + 2(C_{26}p_1p_2 + C_{24}p_2p_3 + C_{46}p_1p_3),$$
$$\Gamma_{33} = C_{55}p_1^2 + C_{44}p_2^2 + C_{33}p_3^2 + 2(C_{45}p_1p_2 + C_{34}p_2p_3 + C_{35}p_1p_3),$$
$$\Gamma_{12} = C_{16}p_1^2 + C_{26}p_2^2 + C_{45}p_3^2 + (C_{12}+C_{66})p_1p_2 + (C_{25}+C_{46})p_2p_3 + (C_{14}+C_{56})p_1p_3,$$
$$\Gamma_{13} = C_{15}p_1^2 + C_{46}p_2^2 + C_{35}p_3^2 + (C_{14}+C_{56})p_1p_2 + (C_{36}+C_{45})p_2p_3 + (C_{13}+C_{55})p_1p_3,$$
$$\Gamma_{23} = C_{56}p_1^2 + C_{24}p_2^2 + C_{34}p_3^2 + (C_{25}+C_{46})p_1p_2 + (C_{23}+C_{44})p_2p_3 + (C_{36}+C_{45})p_1p_3.$$

(3.3)

Remark: The Christoffel equation 3.1 defines the shear-wave Hamiltonian only up to its sign; we discuss this important issue later.

## 4. THE ARCLENGTH-RELATED HAMILTONIAN

In this section, we introduce the arclength-related Hamiltonian $H(\mathbf{x},\mathbf{p})$ and relate its first and second derivatives to those of the reference Hamiltonian, $H^{\bar{\tau}}(\mathbf{x},\mathbf{p})$. The term "arclength-related Hamiltonian" has dual (complementary) meanings. First, with this Hamiltonian, the kinematic ray tracing equation set is,

$$\frac{d\mathbf{p}}{ds} = -H_{\mathbf{x}} \quad , \quad \mathbf{r} = \frac{d\mathbf{x}}{ds} = H_{\mathbf{p}} \quad , \quad \frac{d\tau}{ds} = \mathbf{p} \cdot H_{\mathbf{p}} = \frac{1}{v_{\text{ray}}} \quad , \qquad (4.1)$$

where the corresponding flow parameter is the arclength $s$ and parameter $\tau$ is the current traveltime. Second, as already mentioned in equation 1.5, it is also related to the arclength-related Lagrangian via the Legendre transform, $L(\mathbf{x},\mathbf{r}) = \mathbf{r} \cdot \mathbf{p} - H(\mathbf{x},\mathbf{p})$.

The arclength-related Hamiltonian, $H(\mathbf{x},\mathbf{p})$, can be obtained from the reference Hamiltonian, $H^{\bar{\tau}}(\mathbf{x},\mathbf{p})$, with the scaling operator,

$$H(\mathbf{x},\mathbf{p}) = \frac{H^{\bar{\tau}}(\mathbf{x},\mathbf{p})}{\sqrt{H_{\mathbf{p}}^{\bar{\tau}} \cdot H_{\mathbf{p}}^{\bar{\tau}}}} \quad , \tag{4.2}$$

where,

$$H_{\mathbf{p}}^{\bar{\tau}} \cdot H_{\mathbf{p}}^{\bar{\tau}} = \left(\frac{\partial H^{\bar{\tau}}}{\partial p_x}\right)^2 + \left(\frac{\partial H^{\bar{\tau}}}{\partial p_y}\right)^2 + \left(\frac{\partial H^{\bar{\tau}}}{\partial p_z}\right)^2 \quad . \tag{4.3}$$

The arclength-related Hamiltonian $H(\mathbf{x},\mathbf{p})$, defined with the absolute value of the slowness gradient of the reference Hamiltonian $H_{\mathbf{p}}^{\bar{\tau}}$ in the denominator, is complicated, such that analytical tracking of its Hessians becomes unwieldy. The remedy is to work with $H(\mathbf{x},\mathbf{p})$, but to express its derivatives through those of $H^{\bar{\tau}}(\mathbf{x},\mathbf{p})$. The derivatives include two gradients and three Hessians, and we obtained the corresponding relationships listed below.

The spatial gradient $H_{\mathbf{x}}$ is then written as,

$$H_{\mathbf{x}}(\mathbf{x},\mathbf{p}) = \nabla_{\mathbf{x}}\left[\frac{H^{\bar{\tau}}(\mathbf{x},\mathbf{p})}{\sqrt{H_{\mathbf{p}}^{\bar{\tau}} \cdot H_{\mathbf{p}}^{\bar{\tau}}}}\right] = \frac{H_{\mathbf{x}}^{\bar{\tau}}}{\sqrt{H_{\mathbf{p}}^{\bar{\tau}} \cdot H_{\mathbf{p}}^{\bar{\tau}}}} - \frac{H_{\mathbf{px}}^{\bar{\tau}\,T} \cdot H_{\mathbf{p}}^{\bar{\tau}} H^{\bar{\tau}}}{\left(H_{\mathbf{p}}^{\bar{\tau}} \cdot H_{\mathbf{p}}^{\bar{\tau}}\right)^{3/2}} \quad , \tag{4.4}$$

which leads to,

$$H_{\mathbf{x}}(\mathbf{x},\mathbf{p}) = \frac{H_{\mathbf{x}}^{\bar{\tau}}}{\sqrt{H_{\mathbf{p}}^{\bar{\tau}} \cdot H_{\mathbf{p}}^{\bar{\tau}}}} - \frac{H_{\mathbf{xp}}^{\bar{\tau}} \cdot H_{\mathbf{p}}^{\bar{\tau}}}{\left(H_{\mathbf{p}}^{\bar{\tau}} \cdot H_{\mathbf{p}}^{\bar{\tau}}\right)^{3/2}} H^{\bar{\tau}}(\mathbf{x},\mathbf{p}) \qquad . \tag{4.5}$$

In a similar way we compute the slowness gradient of the arclength-related Hamiltonian $H_{\mathbf{p}}$,

$$H_{\mathbf{p}}(\mathbf{x},\mathbf{p}) = \nabla_{\mathbf{p}}\left[\frac{H^{\bar{\tau}}(\mathbf{x},\mathbf{p})}{\sqrt{H_{\mathbf{p}}^{\bar{\tau}} \cdot H_{\mathbf{p}}^{\bar{\tau}}}}\right] = \frac{H_{\mathbf{p}}^{\bar{\tau}}}{\sqrt{H_{\mathbf{p}}^{\bar{\tau}} \cdot H_{\mathbf{p}}^{\bar{\tau}}}} - \frac{H_{\mathbf{pp}}^{\bar{\tau}} \cdot H_{\mathbf{p}}^{\bar{\tau}}}{\left(H_{\mathbf{p}}^{\bar{\tau}} \cdot H_{\mathbf{p}}^{\bar{\tau}}\right)^{3/2}} H^{\bar{\tau}}(\mathbf{x},\mathbf{p}) \qquad . \tag{4.6}$$

In equations 4.5 and 4.6, the reference Hamiltonian $H^{\bar{\tau}}(\mathbf{x},\mathbf{p})$ in the second terms on the right-hand sides vanishes (since $H^{\bar{\tau}} = 0$). However, we can set it to zero only after we derive the three Hessians of the Hamiltonian (spatial, directional and mixed), otherwise the results will be wrong.

The spatial Hessian of the arclength-related Hamiltonian reads,

$$H_{\mathbf{xx}}(\mathbf{x},\mathbf{p}) = \frac{H_{\mathbf{xx}}^{\bar{\tau}}}{\sqrt{H_{\mathbf{p}}^{\bar{\tau}} \cdot H_{\mathbf{p}}^{\bar{\tau}}}} - \frac{H_{\mathbf{x}}^{\bar{\tau}} \otimes \left(H_{\mathbf{xp}}^{\bar{\tau}} \cdot H_{\mathbf{p}}^{\bar{\tau}}\right) + \left(H_{\mathbf{xp}}^{\bar{\tau}} \cdot H_{\mathbf{p}}^{\bar{\tau}}\right) \otimes H_{\mathbf{x}}^{\bar{\tau}}}{\left(H_{\mathbf{p}}^{\bar{\tau}} \cdot H_{\mathbf{p}}^{\bar{\tau}}\right)^{3/2}} \qquad . \tag{4.7}$$

The slowness Hessian reads,

$$H_{\mathbf{pp}}(\mathbf{x},\mathbf{p}) = \frac{H_{\mathbf{pp}}^{\bar{\tau}}}{\sqrt{H_{\mathbf{p}}^{\bar{\tau}} \cdot H_{\mathbf{p}}^{\bar{\tau}}}} - \frac{H_{\mathbf{p}}^{\bar{\tau}} \otimes \left(H_{\mathbf{pp}}^{\bar{\tau}} \cdot H_{\mathbf{p}}^{\bar{\tau}}\right) + \left(H_{\mathbf{pp}}^{\bar{\tau}} \cdot H_{\mathbf{p}}^{\bar{\tau}}\right) \otimes H_{\mathbf{p}}^{\bar{\tau}}}{\left(H_{\mathbf{p}}^{\bar{\tau}} \cdot H_{\mathbf{p}}^{\bar{\tau}}\right)^{3/2}} \qquad . \tag{4.8}$$

The mixed Hessian reads,

$$H_{\mathbf{xp}}(\mathbf{x},\mathbf{p}) = \frac{H_{\mathbf{xp}}^{\bar{\tau}}}{\sqrt{H_{\mathbf{p}}^{\bar{\tau}} \cdot H_{\mathbf{p}}^{\bar{\tau}}}} - \frac{H_{\mathbf{x}}^{\bar{\tau}} \otimes \left(H_{\mathbf{pp}}^{\bar{\tau}} \cdot H_{\mathbf{p}}^{\bar{\tau}}\right) + \left(H_{\mathbf{xp}}^{\bar{\tau}} \cdot H_{\mathbf{p}}^{\bar{\tau}}\right) \otimes H_{\mathbf{p}}^{\bar{\tau}}}{\left(H_{\mathbf{p}}^{\bar{\tau}} \cdot H_{\mathbf{p}}^{\bar{\tau}}\right)^{3/2}} \qquad . \tag{4.9}$$

Only now we set the Hamiltonian $H^{\bar{\tau}}$ to zero in equations 4.5 and 4.6 and obtain the final results for the spatial and the slowness gradients of the Hamiltonian,

$$H_{\mathbf{p}} = \frac{H_{\mathbf{p}}^{\bar{\tau}}}{\sqrt{H_{\mathbf{p}}^{\bar{\tau}} \cdot H_{\mathbf{p}}^{\bar{\tau}}}} \quad , \quad H_{\mathbf{x}} = \frac{H_{\mathbf{x}}^{\bar{\tau}}}{\sqrt{H_{\mathbf{p}}^{\bar{\tau}} \cdot H_{\mathbf{p}}^{\bar{\tau}}}} \quad . \quad (4.10)$$

Note that for arbitrary vectors $\mathbf{a}$ and $\mathbf{b}$ of lengths $n_a$ and $n_b$, respectively, and matrix $\mathbf{A}$ of size $n_b \times n_a$, the following dual algebraic identities hold,

$$(\mathbf{A} \cdot \mathbf{a}) \otimes \mathbf{b} = \mathbf{A}(\mathbf{a} \otimes \mathbf{b}) \quad , \quad \mathbf{b} \otimes (\mathbf{A} \cdot \mathbf{a}) = (\mathbf{b} \otimes \mathbf{a}) \mathbf{A}^T \quad . \quad (4.11)$$

In both cases, the result is a square matrix $n_b \times n_b$. Applying this rule to equations 4.7, 4.8 and 4.9, we obtain the spatial Hessian,

$$H_{\mathbf{xx}}(\mathbf{x},\mathbf{p}) = \frac{H_{\mathbf{xx}}^{\bar{\tau}}}{\sqrt{H_{\mathbf{p}}^{\bar{\tau}} \cdot H_{\mathbf{p}}^{\bar{\tau}}}} - \frac{\left(H_{\mathbf{x}}^{\bar{\tau}} \otimes H_{\mathbf{p}}^{\bar{\tau}}\right) H_{\mathbf{px}}^{\bar{\tau}} + H_{\mathbf{xp}}^{\bar{\tau}} \left(H_{\mathbf{p}}^{\bar{\tau}} \otimes H_{\mathbf{x}}^{\bar{\tau}}\right)}{\left(H_{\mathbf{p}}^{\bar{\tau}} \cdot H_{\mathbf{p}}^{\bar{\tau}}\right)^{3/2}} \quad , \quad (4.12)$$

the slowness Hessian,

$$H_{\mathbf{pp}}(\mathbf{x},\mathbf{p}) = \frac{H_{\mathbf{pp}}^{\bar{\tau}}}{\sqrt{H_{\mathbf{p}}^{\bar{\tau}} \cdot H_{\mathbf{p}}^{\bar{\tau}}}} - \frac{\left(H_{\mathbf{p}}^{\bar{\tau}} \otimes H_{\mathbf{p}}^{\bar{\tau}}\right) H_{\mathbf{pp}}^{\bar{\tau}} + H_{\mathbf{pp}}^{\bar{\tau}} \left(H_{\mathbf{p}}^{\bar{\tau}} \otimes H_{\mathbf{p}}^{\bar{\tau}}\right)}{\left(H_{\mathbf{p}}^{\bar{\tau}} \cdot H_{\mathbf{p}}^{\bar{\tau}}\right)^{3/2}} \quad , \quad (4.13)$$

and the mixed Hessian,

$$H_{\mathbf{xp}}(\mathbf{x},\mathbf{p}) = \frac{H_{\mathbf{xp}}^{\bar{\tau}}}{\sqrt{H_{\mathbf{p}}^{\bar{\tau}} \cdot H_{\mathbf{p}}^{\bar{\tau}}}} - \frac{\left(H_{\mathbf{x}}^{\bar{\tau}} \otimes H_{\mathbf{p}}^{\bar{\tau}}\right) H_{\mathbf{pp}}^{\bar{\tau}} + H_{\mathbf{xp}}^{\bar{\tau}} \left(H_{\mathbf{p}}^{\bar{\tau}} \otimes H_{\mathbf{p}}^{\bar{\tau}}\right)}{\left(H_{\mathbf{p}}^{\bar{\tau}} \cdot H_{\mathbf{p}}^{\bar{\tau}}\right)^{3/2}} \quad . \quad (4.14)$$

Remark: The relation $H_{\mathbf{p}} = \mathbf{r}$ in equation 4.1 is always consistent for quasi-compressional waves; however, for quasi-shear waves, we may have (or may have not) to change the sign of the reference Hamiltonian, to avoid the case of $H_{\mathbf{p}} = -\mathbf{r}$. Since our derivations and proofs are based on $H_{\mathbf{p}} = \mathbf{r}$, keeping this identity is a must for the proposed approach. The ray velocity derivatives include terms with both odd and even powers of the Hamiltonian components, which means that an improper Hamiltonian sign leads not only to wrong signs, but also to wrong magnitudes of the quasi-shear-wave ray velocity derivatives. The proper sign of the quasi-shear-wave reference Hamiltonian can be established only after performing the slowness inversion (except the cases of qSV and SH waves of polar anisotropy, presented in Part II of this study). This issue is discussed in Appendix E.

## 5. RELATIONS BETWEEN LAGRANGIAN AND HAMILTONIAN DERIVATIVES

As mentioned, the ray velocity and its derivatives are the building stones of the arclength-related Lagrangian $L(\mathbf{x},\mathbf{r})$; however, their explicit forms can only be expressed in terms of the corresponding arclength-related Hamiltonian $H(\mathbf{x},\mathbf{p})$. The gradients of the arclength-related Lagrangian and those of the Hamiltonian are connected by the following relationships (e.g., Koren and Ravve, 2021),

$$\begin{aligned}
L_{\mathbf{r}} &= \mathbf{p} \quad \text{(the generalized momentum equation)}, \\
H_{\mathbf{p}} &= \mathbf{r} \equiv \dot{\mathbf{x}} = d\mathbf{x}/ds, \\
L_{\mathbf{x}} &= -H_{\mathbf{x}} = \dot{\mathbf{p}} = d\mathbf{p}/ds; \quad \text{also:} \ L_{\mathbf{x}} = L_{\mathbf{xr}}\,\mathbf{r}, \\
L_{\mathbf{r}} \cdot \mathbf{r} &= L = \mathbf{p} \cdot \mathbf{r} = v_{\text{ray}}^{-1},
\end{aligned} \quad (5.1)$$

where subscripts $\mathbf{x}, \mathbf{r}, \mathbf{p}$ indicate the vector fields of the corresponding gradients. The Hessians of the arclength-related Lagrangian and those of the Hamiltonian are also connected. The following equation set makes it possible to obtain the Lagrangian Hessians from the Hamiltonian Hessians (Ravve and Koren, 2021),

$$L_{\mathbf{rr}} = H_{\mathbf{pp}}^{-1} - \lambda_{\mathbf{r}} \mathbf{r} \otimes \mathbf{r} \quad , \quad L_{\mathbf{rx}} = -H_{\mathbf{pp}}^{-1} H_{\mathbf{px}} \quad ,$$
$$L_{\mathbf{xr}} = -H_{\mathbf{xp}} H_{\mathbf{pp}}^{-1} \quad , \quad L_{\mathbf{xx}} = H_{\mathbf{xp}} H_{\mathbf{pp}}^{-1} H_{\mathbf{px}} - H_{\mathbf{xx}} \quad ,$$
(5.2)

where symbol $\otimes$ indicates an outer product operation (tensor product) and the double-letter subscript means "the gradient of the gradient", i.e., the Hessian matrix. The reverse transform also exists,

$$H_{\mathbf{pp}}^{-1} = L_{\mathbf{rr}} + \lambda_{\mathbf{r}} \mathbf{r} \otimes \mathbf{r} \quad , \quad H_{\mathbf{px}} = -H_{\mathbf{pp}} L_{\mathbf{rx}} \quad ,$$
$$H_{\mathbf{xp}} = -L_{\mathbf{xr}} H_{\mathbf{pp}} \quad , \quad H_{\mathbf{xx}} = H_{\mathbf{xp}} H_{\mathbf{pp}}^{-1} H_{\mathbf{px}} - L_{\mathbf{xx}} \quad ,$$
(5.3)

where $\lambda_{\mathbf{r}}$ is the eigenvalue of the Hamiltonian's inverse Hessian, $H_{\mathbf{pp}}^{-1}$, corresponding to the eigenvector $\mathbf{r}$. Note that tensors $L_{\mathbf{rr}}$ and $H_{\mathbf{pp}}^{-1}$ have the same eigenvectors: One of them is the ray direction $\mathbf{r}$, while the two others are in the plane normal to the ray (and they are also normal to each other, of course). The eigenvalues, related to the eigenvectors in the normal plane, are the same for $L_{\mathbf{rr}}$ and $H_{\mathbf{pp}}^{-1}$; however, the eigenvalue related to $\mathbf{r}$ vanishes for $L_{\mathbf{rr}}$ and does not vanish for $H_{\mathbf{pp}}^{-1}$. Note also that since $H_{\mathbf{pp}}^{-1}$ is the inverse matrix of $H_{\mathbf{pp}}$, their eigenvectors are identical and their eigenvalues are reciprocals of each other.

There are two more key relationships that will be used in our derivations,

$$H_{\mathbf{x}} = H_{\mathbf{xp}} H_{\mathbf{pp}}^{-1} H_{\mathbf{p}} = H_{\mathbf{xp}} H_{\mathbf{pp}}^{-1} \mathbf{r} \quad , \tag{5.4}$$

proven in Appendix A, and,

$$L_{\mathbf{rr}} = \mathbf{T} H_{\mathbf{pp}}^{-1} \quad , \quad \text{where} \quad \mathbf{T} = \mathbf{I} - \mathbf{r} \otimes \mathbf{r} \quad , \tag{5.5}$$

(Ravve and Koren, 2021).

## 6. SLOWNESS INVERSION: SLOWNESS VECTORS FROM RAY DIRECTION

The combined Lagrangian-Hamiltonian approach used in this study for obtaining the analytic derivatives of the ray velocity magnitude, requires knowing both the ray direction vector $\mathbf{r}$ and the slowness vector $\mathbf{p}$ at the given ray location $\mathbf{x}$. In cases where an initial-value ray tracing is applied, the solution is based on the derivatives of the Hamiltonian $H(\mathbf{x}, \mathbf{p})$ and it provides the values of $\mathbf{x}$ and $\mathbf{p}$ at the ray nodes, where the ray directions $\mathbf{r} = H_{\mathbf{p}}$ and the ray velocity magnitudes $v_{\text{ray}} = (\mathbf{p} \cdot \mathbf{r})^{-1}$ can be directly computed. In cases where a two-point boundary-value ray bending method is used (like in the Eigenray method), the solution is based on an integral over the Lagrangian $L(\mathbf{x}, \mathbf{r})$ which provides the values of $\mathbf{x}$ and $\mathbf{r}$ at the ray nodes. Obtaining all of the corresponding (multiple) slowness vector solutions $\mathbf{p}_i$ (where the index $i$ indicates one of the existing solutions) in general anisotropic media becomes a challenging inverse problem. If the slowness vector components are not known, they should be computed first.

Obtaining the slowness vector from the ray direction vector

The slowness vectors can be obtained from the equation set which manifests that the slowness gradient of the (vanishing) Hamiltonian $H_{\mathbf{p}}^{\bar{\tau}} \equiv \partial H^{\bar{\tau}} / \partial \mathbf{p}$ is collinear with the ray direction $\mathbf{r}$ (Musgrave, 1954a, 1954b, 1970; Fedorov, 1968; Helbig, 1994; Grechka, 2017; Koren and Ravve, 2021, Ravve and Koren, 2021b), written in our notations as,

$$\underbrace{H_{\mathbf{p}}^{\bar{\tau}}\left[\mathbf{m}(\mathbf{x}), \mathbf{p}\right] \times \mathbf{r} = 0}_{\text{use two equations from three}} , \quad \underbrace{H^{\bar{\tau}}\left[\mathbf{m}(\mathbf{x}), \mathbf{p}\right] = 0}_{\text{the third equation}} . \qquad (6.1)$$

The three Cartesian components of the cross-product in the first equation of set 6.1 are linearly dependent, because the mixed product $\mathbf{r} \times H_{\mathbf{p}}^{\bar{\tau}} \cdot \mathbf{r}$ vanishes, so we use only two Cartesian components from the first equation, where the third one is the second equation. The solution for quasi compressional waves is unique, while quasi shear waves may have up to 18 solutions in triclinic media (Grechka, 2017). The numerical example in this study uses the triclinic stiffness parameters suggested by Grechka (2017) and a specific ray direction where all the 19 solutions do exist. Any Hamiltonian can be used for the slowness vector inversion in equation set 6.1; however, for general inhomogeneous anisotropic media, we find the reference Hamiltonian $H^{\bar{\tau}}(\mathbf{x}, \mathbf{p})$ the most suitable.

Note that the other option to formulate the slowness inversion is to arrange equation set 6.1 as,

$$H_{\mathbf{p}}^{\bar{\tau}}\left[\mathbf{m}(\mathbf{x}), \mathbf{p}\right] = \alpha \mathbf{r} , \quad H^{\bar{\tau}}\left[\mathbf{m}(\mathbf{x}), \mathbf{p}\right] = 0 , \qquad (6.2)$$

where $\alpha$ is an unknown scalar value with the units of velocity (positive for compressional waves). In this case we solve a set of four equations with four unknown variables, $\{p_1, p_2, p_3, \alpha\}$. The cost is an additional variable, while the advantage is the revealed sign for the shear-wave Hamiltonian.

## 7. SPATIAL GRADIENT OF THE RAY VELOCITY

Starting from equation 1.6, $v_{ray} = (\mathbf{p} \cdot \mathbf{r})^{-1}$, we compute the spatial gradient of the ray velocity magnitude,

$$\nabla_\mathbf{x} v_{ray} = -\frac{(\nabla_\mathbf{x} \mathbf{p})^T \mathbf{r} + (\nabla_\mathbf{x} \mathbf{r})^T \mathbf{p}}{(\mathbf{p} \cdot \mathbf{r})^2} \quad . \tag{7.1}$$

where we used a general formula for the gradient of a scalar product,

$$\nabla(\mathbf{a} \cdot \mathbf{b}) = (\nabla \mathbf{a})^T \mathbf{b} + (\nabla \mathbf{b})^T \mathbf{a} \quad . \tag{7.2}$$

Recall that in the Eigenray method, the ray location and the ray direction at any given point are assumed independent parameters (degrees of freedom; DoF), and equation 7.1 simplifies to,

$$\nabla_\mathbf{x} v_{ray} = -\frac{(\nabla_\mathbf{x} \mathbf{p})^T \mathbf{r}}{(\mathbf{p} \cdot \mathbf{r})^2} = -v_{ray}^2 (\nabla_\mathbf{x} \mathbf{p})^T \mathbf{r} \quad . \tag{7.3}$$

Thus, we need to compute the spatial gradient of the slowness vector, $\nabla_\mathbf{x} \mathbf{p}$, which is a $3 \times 3$ matrix. Recall the momentum equation, $L_\mathbf{r} = \mathbf{p}$ (equation set 5.1), where $L(\mathbf{x}, \mathbf{r})$ is the arclength-related Lagrangian. Therefore,

$$\nabla_{\mathbf{x}}\mathbf{p} = \nabla_{\mathbf{x}}L_{\mathbf{r}} = L_{\mathbf{rx}} \ , \quad \left(\nabla_{\mathbf{x}}\mathbf{p}\right)^T = L_{\mathbf{xr}} \quad . \tag{7.4}$$

The Hessians of the arclength-related Lagrangian and Hamiltonian are linked. Combining equations 7.3 and 7.4 with the equation from set 5.2, $L_{\mathbf{xr}} = -H_{\mathbf{xp}}H_{\mathbf{pp}}^{-1}$, we obtain,

$$\nabla_{\mathbf{x}}v_{\mathrm{ray}} = +v_{\mathrm{ray}}^2 H_{\mathbf{xp}} H_{\mathbf{pp}}^{-1} \mathbf{r} \quad . \tag{7.5}$$

Next, we introduce identity 5.4 into equation 7.5, and this leads to,

$$\nabla_{\mathbf{x}}v_{\mathrm{ray}} = v_{\mathrm{ray}}^2 H_{\mathbf{x}} \quad . \tag{7.6}$$

Recall that the computation of $H_{\mathbf{x}}$ is simplified using the derivatives of $H^{\bar{\tau}}$ given in equation set 4.10, $H_{\mathbf{x}} = H_{\mathbf{x}}^{\bar{\tau}} / \sqrt{H_{\mathbf{p}}^{\bar{\tau}} \cdot H_{\mathbf{p}}^{\bar{\tau}}}$ , $H_{\mathbf{p}} = H_{\mathbf{p}}^{\bar{\tau}} / \sqrt{H_{\mathbf{p}}^{\bar{\tau}} \cdot H_{\mathbf{p}}^{\bar{\tau}}}$ .

## 8. SPATIAL HESSIAN OF THE RAY VELOCITY

The spatial Hessian of the ray velocity is the gradient of its gradient, so we start from equation 7.6 and compute the gradient of its right-hand side,

$$\nabla_{\mathbf{x}}\nabla_{\mathbf{x}}v_{\mathrm{ray}} = \nabla_{\mathbf{x}}\left[v_{\mathrm{ray}}^2(\mathbf{x},\mathbf{p})H_{\mathbf{x}}(\mathbf{x},\mathbf{p})\right] \quad , \tag{8.1}$$

which is a gradient of a product of a scalar field and a vector field. The gradient of a product of an arbitrary scalar field $c(\mathbf{x})$ and a vector field $\mathbf{d}(\mathbf{x})$ represents a sum of two tensors,

$$\nabla\left[c(\mathbf{x})\mathbf{d}(\mathbf{x})\right] = \mathbf{d}(\mathbf{x}) \otimes \nabla c(\mathbf{x}) + c(\mathbf{x})\nabla \mathbf{d}(\mathbf{x}) \quad . \tag{8.2}$$

Subscript **x** is absent in this identity because it is valid for any gradient, not necessarily spatial.

Applying the rule of equation 8.2 to the right-hand side of equation 8.1, we obtain,

$$\nabla_{\mathbf{x}}\nabla_{\mathbf{x}} v_{\text{ray}} = 2 v_{\text{ray}} H_{\mathbf{x}} \otimes \nabla_{\mathbf{x}} v_{\text{ray}} + v_{\text{ray}}^2 \nabla_{\mathbf{x}} H_{\mathbf{x}} \qquad (8.3)$$

Note, however, that $\nabla_{\mathbf{x}} H_{\mathbf{x}} \neq H_{\mathbf{xx}}$. The reason is that in the Lagrangian formulation, spatial gradient assumes constant ray direction, while in the Hamiltonian formulation, it assumes constant slowness vector. The ray velocity $v_{\text{ray}}(\mathbf{x}, \mathbf{r})$ is an object of the Lagrangian; when computing its spatial derivatives, it is assumed that the ray direction **r** is fixed, while the slowness vector **p** may change as necessary. However, when computing the spatial derivatives of the Hamiltonian, in their conventional notations, the slowness vector is preserved, while the ray direction changes as necessary. This leads to,

$$\nabla_{\mathbf{x}} H_{\mathbf{x}} = \frac{d H_{\mathbf{x}}}{d \mathbf{x}} = \frac{\partial H_{\mathbf{x}}}{\partial \mathbf{x}} + \frac{\partial H_{\mathbf{x}}}{\partial \mathbf{p}} \frac{\partial \mathbf{p}}{\partial \mathbf{x}} = H_{\mathbf{xx}} + H_{\mathbf{xp}} L_{\mathbf{rx}} \qquad (8.4)$$

Since $\mathbf{p} = L_{\mathbf{r}}$, then, for a fixed ray direction, $\partial \mathbf{p} / \partial \mathbf{x} = L_{\mathbf{rx}}$. We then apply a constraint from equation set 5.2, $L_{\mathbf{rx}} = -H_{\mathbf{pp}}^{-1} H_{\mathbf{px}}$, which leads to,

$$\nabla_{\mathbf{x}} H_{\mathbf{x}} = H_{\mathbf{xx}} - H_{\mathbf{xp}} H_{\mathbf{pp}}^{-1} H_{\mathbf{px}} \qquad (8.5)$$

Combining equation 7.6 with 8.3 and 8.5, we obtain,

$$\nabla_{\mathbf{x}} \nabla_{\mathbf{x}} v_{\text{ray}} = 2 v_{\text{ray}}^3 H_{\mathbf{x}} \otimes H_{\mathbf{x}} + v_{\text{ray}}^2 \left( H_{\mathbf{xx}} - H_{\mathbf{xp}} H_{\mathbf{pp}}^{-1} H_{\mathbf{px}} \right) \qquad (8.6)$$

The second term in the brackets accounts for the varying slowness components.

## 9. DIRECTIONAL GRADIENT OF THE RAY VELOCITY MAGNITUDE

To compute the directional gradient and Hessian of the ray velocity magnitude, we need to know both, the ray velocity vector $\mathbf{v}_{ray}$ and the slowness vector $\mathbf{p}$. Given the ray direction $\mathbf{r}$, we compute the slowness vector $\mathbf{p}$ with equation set 6.1 and the ray velocity magnitude $v_{ray}$ with equation 1.6, so that the ray velocity vector becomes,

$$\mathbf{v}_{ray} = v_{ray}\,\mathbf{r} = \frac{\mathbf{r}}{\mathbf{p}\cdot\mathbf{r}} \qquad . \qquad (9.1)$$

The directional gradient of the ray velocity is then delivered by,

$$\nabla_{\mathbf{r}} v_{ray} = -\mathbf{v}_{ray}\times\mathbf{p}\times\mathbf{v}_{ray} = -v_{ray}^{2}\,\mathbf{r}\times\mathbf{p}\times\mathbf{r} \qquad . \qquad (9.2)$$

Note that vector $\mathbf{r}\times\mathbf{p}\times\mathbf{r}$ is the projection of the slowness vector $\mathbf{p}$ onto the plane normal to the ray direction $\mathbf{r}$. The relationships in this section are valid for any type of anisotropic symmetry (proven in Appendix B).

## 10. DIRECTIONAL HESSIAN OF THE RAY VELOCITY MAGNITUDE

The directional Hessian of the ray velocity magnitude can be computed as,

$$\nabla_{\mathbf{r}}\nabla_{\mathbf{r}} v_{ray} = 2 v_{ray}^{3}\,\mathbf{p}\otimes\mathbf{p} - v_{ray}^{2}\,(\mathbf{p}\otimes\mathbf{r}+\mathbf{r}\otimes\mathbf{p}) - v_{ray}\left(v_{ray} L_{\mathbf{rr}} - \mathbf{T}\right) \qquad , \qquad (10.1)$$

where matrix $\mathbf{T}$ is defined in equation 5.5. This relationship is proved in Appendix C. Due to the momentum equation, $L_{\mathbf{r}} = \mathbf{p}$, where $L_{\mathbf{r}}$ is the directional gradient of the Lagrangian $L$, tensor $L_{\mathbf{rr}}$ in equation 10.1,

$$L_{\mathbf{rr}} = \mathbf{p_r} = \nabla_{\mathbf{r}} \mathbf{p} \quad , \tag{10.2}$$

represents the directional Hessian of the arclength-related Lagrangian. This value can be computed through the slowness Hessian of the arclength-related Hamiltonian, listed in equation set 5.5, $L_{\mathbf{rr}} = \mathbf{T} H_{\mathbf{pp}}^{-1}$. Note that $L_{\mathbf{rr}}$ is a singular matrix (its eigenvalue, corresponding to eigenvector $\mathbf{r}$, is zero), while $H_{\mathbf{pp}}$ and its inverse, $H_{\mathbf{pp}}^{-1}$, are regular invertible matrices, with the exception of inflection points along the ray path (Bona and Slawinski, 2003).

Combining equations 5.5, 10.1 and 10.2, we obtain the final relationship for the directional Hessian,

$$\nabla_{\mathbf{r}}\nabla_{\mathbf{r}} v_{\text{ray}} = 2 v_{\text{ray}}^3 \mathbf{p} \otimes \mathbf{p} - v_{\text{ray}}^2 (\mathbf{p} \otimes \mathbf{r} + \mathbf{r} \otimes \mathbf{p}) - v_{\text{ray}} \mathbf{T} \cdot \left( v_{\text{ray}} \mathbf{H}_{\mathbf{pp}}^{-1} - \mathbf{I} \right) \quad . \tag{10.3}$$

## 11. MIXED HESSIAN OF THE RAY VELOCITY MAGNITUDE

The two mixed Hessians, $\nabla_{\mathbf{x}}\nabla_{\mathbf{r}} v_{\text{ray}}$ and $\nabla_{\mathbf{r}}\nabla_{\mathbf{x}} v_{\text{ray}}$ are defined as,

$$\nabla_{\mathbf{x}}\nabla_{\mathbf{r}} v_{\text{ray}} = \left\{ \frac{\partial^2 v_{\text{ray}}}{\partial x_i \partial r_j} \right\} \quad , \quad \nabla_{\mathbf{r}}\nabla_{\mathbf{x}} v_{\text{ray}} = \left\{ \frac{\partial^2 v_{\text{ray}}}{\partial r_i \partial x_j} \right\} \quad , \tag{11.1}$$

where $i, j$ are the indices of the row and the column, respectively. The two tensors are transpose of each other,

$$\nabla_{\mathbf{x}}\nabla_{\mathbf{r}} v_{\text{ray}} = \left( \nabla_{\mathbf{r}}\nabla_{\mathbf{x}} v_{\text{ray}} \right)^T \quad , \tag{11.2}$$

and can be computed as,

$$\nabla_{\mathbf{r}}\nabla_{\mathbf{x}} v_{\text{ray}} = \frac{2}{v_{\text{ray}}} \nabla_{\mathbf{r}} v_{\text{ray}} \otimes \nabla_{\mathbf{x}} v_{\text{ray}} - v_{\text{ray}}^2 \mathbf{T}\mathbf{p_x} \ ,$$

$$\nabla_{\mathbf{x}}\nabla_{\mathbf{r}} v_{\text{ray}} = \frac{2}{v_{\text{ray}}} \nabla_{\mathbf{x}} v_{\text{ray}} \otimes \nabla_{\mathbf{r}} v_{\text{ray}} - v_{\text{ray}}^2 \mathbf{p_x}^T \mathbf{T} \ ,$$
(11.3)

where,

$$\mathbf{p} = L_{\mathbf{r}} \quad \text{and} \quad \mathbf{p_x} = L_{\mathbf{rx}} \ . \tag{11.4}$$

Equation set 11.3 is proven in Appendix D. Tensor $\mathbf{p_x}$ is the spatial gradient of the slowness vector, and it also represents the mixed Hessian of the arclength-related Lagrangian. This matrix (tensor) can be computed through the Hessians of the arclength-related Hamiltonian (equation from set 5.2), $L_{\mathbf{rx}} = -H_{\mathbf{pp}}^{-1} H_{\mathbf{px}}$. The final formulae are,

$$\nabla_{\mathbf{r}}\nabla_{\mathbf{x}} v_{\text{ray}} = \frac{2}{v_{\text{ray}}} \nabla_{\mathbf{r}} v_{\text{ray}} \otimes \nabla_{\mathbf{x}} v_{\text{ray}} + v_{\text{ray}}^2 \mathbf{T} H_{\mathbf{pp}}^{-1} H_{\mathbf{px}} \ ,$$

$$\nabla_{\mathbf{x}}\nabla_{\mathbf{r}} v_{\text{ray}} = \frac{2}{v_{\text{ray}}} \nabla_{\mathbf{x}} v_{\text{ray}} \otimes \nabla_{\mathbf{r}} v_{\text{ray}} + v_{\text{ray}}^2 H_{\mathbf{xp}} H_{\mathbf{pp}}^{-1} \mathbf{T} \ .$$
(11.5)

where matrix (tensor) $\mathbf{T}$ is defined in equation 5.5.

So far, we related the derivatives of the ray velocity magnitude with those of the arclength-related Hamiltonian, and we related the latter with the derivatives of the reference Hamiltonian; these derivatives will be established in the following section.

## 12. SLOWNESS GRADIENT OF THE REFERENCE HAMILTONIAN

The slowness gradient of the reference Hamiltonian reads,

$$H_{\mathbf{p}}^{\bar{\tau}}(\mathbf{x},\mathbf{p}) = \nabla_{\mathbf{p}} H^{\bar{\tau}}(\mathbf{x},\mathbf{p}) \quad . \tag{12.1}$$

The reference Hamiltonian $H^{\bar{\tau}}(\mathbf{x},\mathbf{p})$ for triclinic media is given in equations 6.1 and 6.3. It represents a polynomial in both the slowness vector components and the stiffness tensor components. Computing the derivatives is easy, but the expanded determinant $H^{\bar{\tau}}(\mathbf{x},\mathbf{p}) = \det(\mathbf{\Gamma} - \mathbf{I})$ includes 950 monomials. The direct approach becomes unwieldy, and we expand the derivative of the determinant (e.g., Bellman, 1997; Magnus and Neudecker, 1999). With this approach, the derivative of the determinant (for a square matrix of order $n$) is presented as a sum of $n$ determinants, where in each of them one column (or row) is replaced by its derivative, and the others are untouched,

$$\begin{aligned}\left[H_{\mathbf{p}}^{\bar{\tau}}\right]_i &= \frac{\partial H^{\bar{\tau}}(\mathbf{x},\mathbf{p})}{\partial p_i} = \frac{\partial \det(\mathbf{\Gamma}-\mathbf{I})}{\partial p_i} = \\ &\det\left|\frac{\partial \mathbf{a}_1}{\partial p_i} \quad \mathbf{a}_2 \quad \mathbf{a}_3\right| + \det\left|\mathbf{a}_1 \quad \frac{\partial \mathbf{a}_2}{\partial p_i} \quad \mathbf{a}_3\right| + \det\left|\mathbf{a}_1 \quad \mathbf{a}_2 \quad \frac{\partial \mathbf{a}_3}{\partial p_i}\right| ,\end{aligned} \tag{12.2}$$

where $\mathbf{a}_1$, $\mathbf{a}_2$, $\mathbf{a}_3$ are the columns of matrix $\mathbf{\Gamma} - \mathbf{I}$ (equation 3.3). The relationship yields a single component of the gradient vector $H_{\mathbf{p}}^{\bar{\tau}}$. An advantage of this approach is its easy extension for the second derivatives of the Hamiltonian in a straightforward way.

### 13. SLOWNESS HESSIAN OF THE REFERENCE HAMILTONIAN

Next, we compute the slowness Hessian of the reference Hamiltonian. It represents the gradient of the gradient of the reference Hamiltonian,

$$H^{\bar{\tau}}_{\mathbf{pp}}(\mathbf{x},\mathbf{p}) = \nabla_{\mathbf{p}} H^{\bar{\tau}}_{\mathbf{p}} \qquad (13.1)$$

We use equation 12.2 for the gradient and apply the same rule for the derivative of the determinant,

$$\begin{aligned}
\left[H^{\bar{\tau}}_{\mathbf{pp}}\right]_{ij} &= \frac{\partial^2 H^{\bar{\tau}}(\mathbf{x},\mathbf{p})}{\partial p_i \partial p_j} = \frac{\partial^2 \det(\boldsymbol{\Gamma}-\mathbf{I})}{\partial p_i \partial p_j} = \\
&+ \det\begin{vmatrix} \frac{\partial^2 \mathbf{a}_1}{\partial p_i \partial p_j} & \mathbf{a}_2 & \mathbf{a}_3 \end{vmatrix} + \det\begin{vmatrix} \frac{\partial \mathbf{a}_1}{\partial p_i} & \frac{\partial \mathbf{a}_2}{\partial p_j} & \mathbf{a}_3 \end{vmatrix} + \det\begin{vmatrix} \frac{\partial \mathbf{a}_1}{\partial p_i} & \mathbf{a}_2 & \frac{\partial \mathbf{a}_3}{\partial p_j} \end{vmatrix} \\
&+ \det\begin{vmatrix} \frac{\partial \mathbf{a}_1}{\partial p_j} & \frac{\partial \mathbf{a}_2}{\partial p_i} & \mathbf{a}_3 \end{vmatrix} + \det\begin{vmatrix} \mathbf{a}_1 & \frac{\partial^2 \mathbf{a}_2}{\partial p_i \partial p_j} & \mathbf{a}_3 \end{vmatrix} + \det\begin{vmatrix} \mathbf{a}_1 & \frac{\partial \mathbf{a}_2}{\partial p_i} & \frac{\partial \mathbf{a}_3}{\partial p_j} \end{vmatrix} \\
&+ \det\begin{vmatrix} \frac{\partial \mathbf{a}_1}{\partial p_j} & \mathbf{a}_2 & \frac{\partial \mathbf{a}_3}{\partial p_i} \end{vmatrix} + \det\begin{vmatrix} \mathbf{a}_1 & \frac{\partial \mathbf{a}_2}{\partial p_j} & \frac{\partial \mathbf{a}_3}{\partial p_i} \end{vmatrix} + \det\begin{vmatrix} \mathbf{a}_1 & \mathbf{a}_2 & \frac{\partial^2 \mathbf{a}_3}{\partial p_i \partial p_j} \end{vmatrix} \quad .
\end{aligned} \qquad (13.2)$$

This relationship yields a single component of the Hessian matrix. Equations 12.2 and 13.2 make it possible to compute the slowness gradient and Hessian, respectively, of the arclength Hamiltonian.

## 14. SPATIAL GRADIENT OF THE REFERENCE HAMILTONIAN

The Hamiltonian depends on the medium properties, which, in turn, are position dependent. We introduce vector $\mathbf{m}(\mathbf{x})$ whose components represent spatially varying model parameters defined at any location, along with their gradients and Hessians. Recall that we consider a single spatial location, where the model parameters are the medium properties at this point. In the case of a general (triclinic) anisotropy, these are the stiffness tensor components, while in the case of higher symmetries, the medium properties may be presented by the normalized unitless equivalents of the stiffness components. The Hamiltonian given by equations 3.1 and 3.3 is spatially dependent,

$$H^{\bar{\tau}} = H^{\bar{\tau}}\left[\mathbf{m}(\mathbf{x}), \mathbf{p}\right] \quad . \tag{14.1}$$

The gradient of the Hamiltonian obeys the chain rule. Let $m_i(\mathbf{x})$ be the medium property, then the spatial gradient can be arranged as,

$$H_{\mathbf{x}}^{\bar{\tau}}(\mathbf{x}, \mathbf{p}) = \sum_{i=1}^{n} \frac{\partial H^{\bar{\tau}}}{\partial m_i} \nabla_{\mathbf{x}} m_i \quad , \quad n = 21 \quad . \tag{14.2}$$

Equation 14.2 can be arranged in a more suitable matrix/tensor form, with all summations hidden inside the tensor operations. Let a second-order object $\mathbf{m_x}$ of dimensions $[n, 3]$ be the gradient of the parameter vector $\mathbf{m}$ wrt the spatial coordinates,

$$\nabla_{\mathbf{x}} \mathbf{m} = \mathbf{m_x} = \left\{\frac{\partial m_i}{\partial x_j}\right\} \quad , \tag{14.3}$$

where in the case of a triclinic medium, $n = 21$. We assume that the components of vector $\mathbf{m}$ and matrix $\mathbf{m_x}$ are known or have been computed; actually, these are the medium properties and their derivatives. Vector $\mathbf{m}$ is not a true physical vector; it represents a collection of scalar values. Similarly, matrix $\mathbf{m_x}$ is not a true physical tensor, but a collection of $n$ vectors of length 3.

We now introduce also vector $H_{\mathbf{m}}^{\bar{\tau}}[n]$ which is the gradient of the Hamiltonian wrt the model parameters,

$$H_{\mathbf{m}}^{\bar{\tau}} = \left\{\frac{\partial H^{\bar{\tau}}\left[\mathbf{m}(\mathbf{x}), \mathbf{p}\right]}{\partial m_i}\right\} \quad . \tag{14.4}$$

The derivatives of the Hamiltonian wrt the medium properties are computed similarly to its derivatives wrt the slowness components, equation 12.2,

$$\left[H_{\mathbf{m}}^{\bar{\tau}}\right]_i = \frac{\partial H^{\bar{\tau}}\left[\mathbf{m}(\mathbf{x}),\mathbf{p}\right]}{\partial m_i} = \frac{\partial \det(\mathbf{\Gamma} - \mathbf{I})}{\partial m_i} =$$
$$\det\left|\frac{\partial \mathbf{a}_1}{\partial m_i} \quad \mathbf{a}_2 \quad \mathbf{a}_3\right| + \det\left|\mathbf{a}_1 \quad \frac{\partial \mathbf{a}_2}{\partial m_i} \quad \mathbf{a}_3\right| + \det\left|\mathbf{a}_1 \quad \mathbf{a}_2 \quad \frac{\partial \mathbf{a}_3}{\partial m_i}\right| . \quad (14.5)$$

Combining equations 14.2 – 14.4, we obtain,

$$\underbrace{H_{\mathbf{x}}^{\bar{\tau}}(\mathbf{x},\mathbf{p})}_{1\times 3} = \underbrace{H_{\mathbf{m}}^{\bar{\tau}}}_{1\times n} \underbrace{\mathbf{m}_{\mathbf{x}}}_{n\times 3} , \quad (14.6)$$

where the components of vector $H_{\mathbf{m}}^{\bar{\tau}}$ are computed with equation 14.5. We assume that the components of two-dimensional array $\mathbf{m}_{\mathbf{x}}$ are known or have been computed.

## 15. SPATIAL HESSIAN OF THE REFERENCE HAMILTONIAN

The chain rule can be extended for the Hessian,

$$H_{\mathbf{xx}}^{\bar{\tau}}(\mathbf{x},\mathbf{p}) = \nabla_{\mathbf{x}} H_{\mathbf{x}}^{\bar{\tau}} = \nabla_{\mathbf{x}}\left(\sum_{i=1}^{n} \frac{\partial H^{\bar{\tau}}}{\partial m_i} \nabla_{\mathbf{x}} m_i\right) = \sum_{i=1}^{n} \nabla_{\mathbf{x}}\left(\frac{\partial H^{\bar{\tau}}}{\partial m_i} \nabla_{\mathbf{x}} m_i\right) =$$
$$\sum_{i=1}^{n}\left[\nabla_{\mathbf{x}}\left(\frac{\partial H^{\bar{\tau}}}{\partial m_i}\right) \nabla_{\mathbf{x}} m_i + \frac{\partial H^{\bar{\tau}}}{\partial m_i} \nabla_{\mathbf{x}} \nabla_{\mathbf{x}} m_i\right] = \sum_{i=1}^{n}\sum_{j=1}^{n} \frac{\partial^2 H^{\bar{\tau}}}{\partial m_i \partial m_j} \nabla_{\mathbf{x}} m_i \otimes \nabla_{\mathbf{x}} m_j + \sum_{i=1}^{n} \frac{\partial H^{\bar{\tau}}}{\partial m_i} \nabla_{\mathbf{x}} \nabla_{\mathbf{x}} m_i$$
. (15.1)

Equation 15.1 can be arranged in a more suitable array form, with all summations hidden inside the "tensor" operations. Let a three-dimensional array $\mathbf{m}_{\mathbf{xx}}(n,3,3)$, symmetric for the two last indices, be the Hessian of array $\mathbf{m}$,

Page 30 of 90

$$\mathbf{m_{xx}} = \left\{ \frac{\partial^2 m_i}{\partial x_j \partial x_k} \right\} \quad , \quad m_{xx}(i,j,k) = m_{xx}(i,k,j) \quad . \tag{15.2}$$

We assume that the components of the three-dimensional array $\mathbf{m_{xx}}$ are known or have been computed; actually, these are the Hessians of the medium properties. We emphasize that array $\mathbf{m_{xx}}$ is not a true third-order physical tensor, but a collection of $n$ second-order tensors of dimensions $3 \times 3$, which still makes it possible to apply the tensor multiplication rules.

We now introduce also a symmetric square matrix $H_{\mathbf{mm}}^{\bar{\tau}}(n,n)$, which is the Hessian of the Hamiltonian wrt the model parameters,

$$H_{\mathbf{mm}}^{\bar{\tau}} = \left\{ \frac{\partial^2 H^{\bar{\tau}}[\mathbf{m(x)},\mathbf{p}]}{\partial m_i \partial m_j} \right\} \quad . \tag{15.3}$$

The second derivatives of the Hamiltonian wrt the medium properties are computed similarly to its second derivatives wrt the slowness components, equation 13.2,

$$\begin{aligned}
\left[ H_{\mathbf{mm}}^{\bar{\tau}} \right]_{ij} &= \frac{\partial^2 H^{\bar{\tau}}[\mathbf{m(x)},\mathbf{p}]}{\partial m_i \, \partial m_j} = \frac{\partial^2 \det(\mathbf{\Gamma} - \mathbf{I})}{\partial m_i \, \partial m_j} = \\
&+ \det \left| \frac{\partial^2 \mathbf{a}_1}{\partial m_i \, \partial m_j} \quad \mathbf{a}_2 \quad \mathbf{a}_3 \right| + \det \left| \frac{\partial \mathbf{a}_1}{\partial m_i} \quad \frac{\partial \mathbf{a}_2}{\partial m_j} \quad \mathbf{a}_3 \right| + \det \left| \frac{\partial \mathbf{a}_1}{\partial m_i} \quad \mathbf{a}_2 \quad \frac{\partial \mathbf{a}_3}{\partial m_j} \right| \\
&+ \det \left| \frac{\partial \mathbf{a}_1}{\partial m_j} \quad \frac{\partial \mathbf{a}_2}{\partial m_i} \quad \mathbf{a}_3 \right| + \det \left| \mathbf{a}_1 \quad \frac{\partial^2 \mathbf{a}_2}{\partial m_i \, \partial m_j} \quad \mathbf{a}_3 \right| + \det \left| \mathbf{a}_1 \quad \frac{\partial \mathbf{a}_2}{\partial m_i} \quad \frac{\partial \mathbf{a}_3}{\partial m_j} \right| \\
&+ \det \left| \frac{\partial \mathbf{a}_1}{\partial m_j} \quad \mathbf{a}_2 \quad \frac{\partial \mathbf{a}_3}{\partial m_i} \right| + \det \left| \mathbf{a}_1 \quad \frac{\partial \mathbf{a}_2}{\partial m_j} \quad \frac{\partial \mathbf{a}_3}{\partial m_i} \right| + \det \left| \mathbf{a}_1 \quad \mathbf{a}_2 \quad \frac{\partial^2 \mathbf{a}_3}{\partial m_i \, \partial m_j} \right| \quad ,
\end{aligned} \tag{15.4}$$

Note that according to equation 3.3, the Christoffel matrix components depend on the stiffness components in a linear way. Thus, if the model parameters are the components of the stiffness tensor, the last term on the right-hand side of equation 15.4 vanishes.

With the notations of equations 15.2 and 15.3, equation 15.1 simplifies to,

$$\underbrace{H^{\bar{\tau}}_{\mathbf{xx}}[\mathbf{x(m)},\mathbf{p}]}_{3\times 3} = \underbrace{\mathbf{m}^T_{\mathbf{x}}}_{3\times n} \underbrace{H^{\bar{\tau}}_{\mathbf{mm}}}_{n\times n} \underbrace{\mathbf{m}_{\mathbf{x}}}_{n\times 3} + \underbrace{H^{\bar{\tau}}_{\mathbf{m}}}_{1\times n} \underbrace{\mathbf{m}_{\mathbf{xx}}}_{n\times 3\times 3} \quad , \tag{15.5}$$

where the components of $n\times n$ matrix $H^{\bar{\tau}}_{\mathbf{mm}}$ are computed with equation 15.4. Considering the last item of the right-hand side of this equation, we note that the result of the multiplication of objects with dimensions $(1,n)$ and $(n,3,3)$ is a new object with dimensions $(1,3,3)$, because the adjacent indices collapse, but size 1 means actually that the resulting object is only two-dimensional of the size $(3,3)$. This object is a contribution to the Hessian from the gradient of the reference Hamiltonian $H_{\mathbf{m}}$. The multiplication of a vector and a third-order object is similar to vector-matrix multiplication. Let $\mathbf{a}$ be a vector, $\mathbf{A}$ the third-order object, and $\mathbf{B}$ a matrix resulting from their multiplication (with the vector on the left-hand side). Then,

$$\mathbf{B} = \mathbf{a}\,\mathbf{A} \quad , \qquad B_{jk} = \sum_{i=1}^{n} a_i A_{ijk} \quad . \tag{15.6}$$

We assume that the components of the three-dimensional array $\mathbf{m_{xx}}$ are known or have been computed. In our numerical example, arrays $\mathbf{m}(21)$, $\mathbf{m_x}(21,3)$ and $\mathbf{m_{xx}}(21,3,3)$ are the input data.

## 16. MIXED HESSIAN OF THE REFERENCE HAMILTONIAN

The Hamiltonian has two mixed Hessians, $H_{\mathbf{xp}}^{\bar{\tau}}$ and $H_{\mathbf{px}}^{\bar{\tau}}$, but they are transpose of each other, so we can compute only one of them. In this section, we compute the mixed Hessian $H_{\mathbf{px}}^{\bar{\tau}}$, which can be viewed as a spatial gradient of the slowness gradient,

$$H_{\mathbf{px}}^{\bar{\tau}} = \nabla_{\mathbf{x}} \nabla_{\mathbf{p}} H^{\bar{\tau}} = \nabla_{\mathbf{x}} H_{\mathbf{p}}^{\bar{\tau}} \qquad (16.1)$$

From equation 13.2, we conclude that,

$$\begin{aligned}
\left[H_{\mathbf{pm}}^{\bar{\tau}}\right]_{ij} &= \frac{\partial^2 H^{\bar{\tau}}(\mathbf{x},\mathbf{p})}{\partial p_i \, \partial m_j} = \frac{\partial^2 \det(\mathbf{\Gamma}-\mathbf{I})}{\partial p_i \, \partial m_j} = \\
&+ \det\left|\frac{\partial^2 \mathbf{a}_1}{\partial p_i \, \partial m_j} \quad \mathbf{a}_2 \quad \mathbf{a}_3\right| + \det\left|\frac{\partial \mathbf{a}_1}{\partial p_i} \quad \frac{\partial \mathbf{a}_2}{\partial m_j} \quad \mathbf{a}_3\right| + \det\left|\frac{\partial \mathbf{a}_1}{\partial p_i} \quad \mathbf{a}_2 \quad \frac{\partial \mathbf{a}_3}{\partial m_j}\right| \\
&+ \det\left|\frac{\partial \mathbf{a}_1}{\partial m_j} \quad \frac{\partial \mathbf{a}_2}{\partial p_i} \quad \mathbf{a}_3\right| + \det\left|\mathbf{a}_1 \quad \frac{\partial^2 \mathbf{a}_2}{\partial p_i \, \partial m_j} \quad \mathbf{a}_3\right| + \det\left|\mathbf{a}_1 \quad \frac{\partial \mathbf{a}_2}{\partial p_i} \quad \frac{\partial \mathbf{a}_3}{\partial m_j}\right| \\
&+ \det\left|\frac{\partial \mathbf{a}_1}{\partial m_j} \quad \mathbf{a}_2 \quad \frac{\partial \mathbf{a}_3}{\partial p_i}\right| + \det\left|\mathbf{a}_1 \quad \frac{\partial \mathbf{a}_2}{\partial m_j} \quad \frac{\partial \mathbf{a}_3}{\partial p_i}\right| + \det\left|\mathbf{a}_1 \quad \mathbf{a}_2 \quad \frac{\partial^2 \mathbf{a}_3}{\partial p_i \, \partial m_j}\right|
\end{aligned} \qquad (16.2)$$

Thus, we found the derivatives of the mixed Hessian of the reference Hamiltonian wrt the slowness vector of length 3 and the medium properties vector of length up to $n=21$, $\partial H_{\mathbf{pm}}^{\bar{\tau}}$. This set can be considered a matrix of dimension $[3,n]$. This matrix, $H_{\mathbf{pm}}^{\bar{\tau}}$, where $\mathbf{m}$ is a set of the spatially varying model parameters, is not a physical tensor, but each column of this matrix, with a fixed $m_i$, is a true physical tensor.

Recall that the second-order object $\mathbf{m_x}$ of dimensions $[n,3]$ is the gradient of the parameter vector $\mathbf{m}$ wrt the spatial coordinates, with all components available (e.g., the derivatives are computed numerically). We multiply the two matrices of dimensions $[3,n]$ and $[n,3]$, and obtain the mixed Hessian, whose dimensions are $3\times 3$, of the reference Hamiltonian,

$$H_{\mathbf{px}}^{\bar{\tau}} = H_{\mathbf{pm}}^{\bar{\tau}}\, \mathbf{m_x} \qquad . \tag{16.3}$$

## 17. GRADIENT OF THE RAY VELOCITY WRT THE MEDIUM PROPERTIES

In this section we derive the first partial derivatives of the ray velocity magnitude along fixed ray elements wrt the medium properties (the Fréchet derivatives). These derivatives construct the traveltime derivatives, which are, for example, the components of the traveltime sensitivity matrix used in traveltime tomography for updating the model parameters. The ray trajectories and hence the ray directions $\mathbf{r}$ are assumed fixed, where each partial derivative describes the ray velocity change wrt a given medium's property $m_i$, (index $i$ refers to the model parameter type at a given point) where all the other properties are fixed. It is also assumed that the slowness vectors $\mathbf{p}$ for the wave mode(s) under consideration are known (have been computed).

Again, starting from equation 1.6, $v_{\mathrm{ray}} = (\mathbf{p}\cdot\mathbf{r})^{-1}$, we obtain,

$$\frac{\partial v_{\mathrm{ray}}}{\partial m_i} = -\frac{\partial \mathbf{p}/\partial m_i}{(\mathbf{p}\cdot\mathbf{r})^2}\cdot\mathbf{r} = -v_{\mathrm{ray}}^2 \frac{\partial \mathbf{p}}{\partial m_i}\cdot\mathbf{r} \qquad . \tag{17.1}$$

Thus, to compute the ray velocity magnitude derivative wrt a specific model parameter, we need first to establish the corresponding derivative of the slowness vector.

Consider the general equation set for the slowness inversion (valid for any anisotropic symmetry),

$$H_\mathbf{p}(\mathbf{m},\mathbf{p}) \times \mathbf{r} = 0 \ , \quad H(\mathbf{m},\mathbf{p}) = 0 \quad , \tag{17.2}$$

where $H$ is the arclength-related Hamiltonian, and $\mathbf{m} = \mathbf{m}_o(\mathbf{x}) + \delta\mathbf{m}(\mathbf{x})$ is a set of model parameters; $\mathbf{m}_o$ is a background model. Consider a fixed point and a fixed ray direction; thus, both $\mathbf{x}$ and $\mathbf{r}$ are kept constant, but the ray velocity may vary due to a change of the parameters at the given point. For the arclength-related Hamiltonian, we replace the first equation of set 17.2 by $H_\mathbf{p} = \mathbf{r}$, which means that the slowness gradient of this Hamiltonian is not only collinear to the normalized ray direction vector $\mathbf{r}$, but also equal to this vector.

The medium property $m_i$ slightly changes, and as a result of this change, the slowness vector varies, such that both the updated Hamiltonian, $H(\mathbf{m},\mathbf{p}) = 0$, and its slowness gradient, $H_\mathbf{p}(\mathbf{m},\mathbf{p}) = \mathbf{r}$, remain constant. This means that the total derivatives of each of them vanish,

$$\frac{dH_\mathbf{p}}{dm_i} = \frac{\partial H_\mathbf{p}}{\partial m_i} + \frac{\partial H_\mathbf{p}}{\partial \mathbf{p}} \cdot \frac{\partial \mathbf{p}}{\partial m_i} = 0 \quad , \quad \frac{dH}{dm_i} = \frac{\partial H}{\partial m_i} + \frac{\partial H}{\partial \mathbf{p}} \cdot \frac{\partial \mathbf{p}}{\partial m_i} = 0 \quad , \tag{17.3}$$

and the equation set can be arranged as,

$$H_\mathbf{pp} \cdot \frac{\partial \mathbf{p}}{\partial m_i} = -\frac{\partial H_\mathbf{p}}{\partial m_i} \quad , \quad \frac{\partial H}{\partial m_i} + H_\mathbf{p} \cdot \frac{\partial \mathbf{p}}{\partial m_i} = 0 \quad . \tag{17.4}$$

Note that vector $\partial\mathbf{p}/\partial m_i$ (of length 3) can be computed from the first equation of set 17.4,

$$\frac{\partial \mathbf{p}}{\partial m_i} = -H_\mathbf{pp}^{-1} \cdot \frac{\partial H_\mathbf{p}}{\partial m_i} \quad . \tag{17.5}$$

Next, we introduce this result into the second equation of set 17.4,

$$\frac{\partial H}{\partial m_i} = H_{\mathbf{p}} H_{\mathbf{pp}}^{-1} \cdot \frac{\partial H_{\mathbf{p}}}{\partial m_i} \qquad . \qquad (17.6)$$

Recall that the ray direction, $H_{\mathbf{p}} = \mathbf{r}$, is the eigenvector of the inverse Hessian $H_{\mathbf{pp}}^{-1}$, and the corresponding eigenvalue is $\lambda_{\mathbf{r}}$,

$$\mathbf{r} H_{\mathbf{pp}}^{-1} = H_{\mathbf{pp}}^{-1} \mathbf{r} = \lambda_{\mathbf{r}} \mathbf{r} \qquad . \qquad (17.7)$$

This leads to,

$$\frac{\partial H}{\partial m_i} = \lambda_{\mathbf{r}} \frac{\partial H_{\mathbf{p}}}{\partial m_i} \cdot \mathbf{r} \qquad . \qquad (17.8)$$

Next, we introduce equation 17.5 into 17.1,

$$\frac{\partial v_{\text{ray}}}{\partial m_i} = v_{\text{ray}}^2 \mathbf{r} H_{\mathbf{pp}}^{-1} \cdot \frac{\partial H_{\mathbf{p}}}{\partial m_i} = v_{\text{ray}}^2 \lambda_{\mathbf{r}} \frac{\partial H_{\mathbf{p}}}{\partial m_i} \cdot \mathbf{r} \qquad . \qquad (17.9)$$

Finally, we combine equations 17.8 and 17.9 to obtain the derivative of the ray velocity wrt the model parameter,

$$\frac{\partial v_{\text{ray}}}{\partial m_i} = v_{\text{ray}}^2 \frac{\partial H(\mathbf{m},\mathbf{p})}{\partial m_i} \qquad . \qquad (17.10)$$

We emphasize that equation 17.10 is only valid for the arclength-related Hamiltonian. As mentioned above, the reference Hamiltonian is easier to deal with. The relationship between the two Hamiltonians is given in equation 4.2, and it leads to,

$$\frac{\partial H(\mathbf{m},\mathbf{p})}{\partial m_i} = \frac{1}{\sqrt{H_{\mathbf{p}}^{\bar{\tau}} \cdot H_{\mathbf{p}}^{\bar{\tau}}}} \frac{\partial H^{\bar{\tau}}(\mathbf{m},\mathbf{p})}{\partial m_i} - H_{\mathbf{p}}^{\bar{\tau}}(\mathbf{m},\mathbf{p}) \cdot \frac{\partial H_{\mathbf{p}}^{\bar{\tau}}(\mathbf{m},\mathbf{p})}{\partial m_i} \frac{H^{\bar{\tau}}(\mathbf{m},\mathbf{p})}{\left(H_{\mathbf{p}}^{\bar{\tau}} \cdot H_{\mathbf{p}}^{\bar{\tau}}\right)^{3/2}} \qquad . \quad (17.11)$$

Since the Hamiltonian vanishes, the second item on the right-hand side of equation 17.11 vanishes, and this relationship simplifies to,

$$\frac{\partial H(\mathbf{m},\mathbf{p})}{\partial m_i} = \frac{1}{\sqrt{H_{\mathbf{p}}^{\bar{\tau}} \cdot H_{\mathbf{p}}^{\bar{\tau}}}} \frac{\partial H^{\bar{\tau}}(\mathbf{m},\mathbf{p})}{\partial m_i} \qquad . \quad (17.12)$$

However, we still need expression 17.11 since it will be used to further derive the second derivatives of the ray velocity. We arrange equations 17.10 and 17.12 in a concise vector form,

$$\nabla_{\mathbf{m}} v_{\text{ray}} = v_{\text{ray}}^2 \, H_{\mathbf{m}}(\mathbf{m},\mathbf{p}) \quad \text{where} \quad H_{\mathbf{m}}(\mathbf{m},\mathbf{p}) = \frac{H_{\mathbf{m}}^{\bar{\tau}}(\mathbf{m},\mathbf{p})}{\sqrt{H_{\mathbf{p}}^{\bar{\tau}} \cdot H_{\mathbf{p}}^{\bar{\tau}}}} \qquad . \quad (17.13)$$

Note that both Hamiltonians depend on the medium properties and the slowness vector, thus operators $H_{\mathbf{m}}(\mathbf{m},\mathbf{p})$ and $H_{\mathbf{m}}^{\bar{\tau}}(\mathbf{m},\mathbf{p})$ result in model-related gradients, where the slowness vector $\mathbf{p}$ is assumed fixed, $\partial H^{\bar{\tau}}/\partial \mathbf{m}\big|_{\mathbf{p}=\text{const}}$, $\partial H/\partial \mathbf{m}\big|_{\mathbf{p}=\text{const}}$. However, the resulting value, $\nabla_{\mathbf{m}} v_{\text{ray}}$, is the model-related ray velocity gradient, computed for the fixed ray direction $\mathbf{r}$ (and actually, for the fixed location as well, but due to the direct dependence on $\mathbf{m}$, dependence on $\mathbf{x}$ becomes pointless). This gradient shows how the ray velocity varies with the change of the medium properties, while the ray direction is kept constant and the slowness components modify as necessary in the updated medium. The same is true for the model-related ray velocity Hessian, $\nabla_{\mathbf{m}} \nabla_{\mathbf{m}} v_{\text{ray}}$, established in the next section.

## 18. HESSIAN OF THE RAY VELOCITY WRT THE MEDIUM PROPERTIES

We start from the resulting relationship 17.13 for a gradient component, which leads to a Hessian component,

$$\frac{\partial^2 v_{\text{ray}}}{\partial m_i \, \partial m_j} = 2 v_{\text{ray}} \frac{\partial v_{\text{ray}}}{\partial m_j} \frac{\partial H}{\partial m_i} + v_{\text{ray}}^2 \frac{d}{dm_j} \frac{\partial H}{\partial m_i} \quad , \tag{18.1}$$

where operator $d/dm_j$ denotes the full derivative, resulting from both explicit dependency of $\partial H / \partial m_i$ on parameter $m_j$ and implicit dependency through the slowness components,

$$\frac{d}{dm_j} \frac{\partial H}{\partial m_j} = \frac{\partial^2 H}{\partial m_i \, \partial m_j} + \frac{\partial^2 H}{\partial m_i \partial \mathbf{p}} \frac{\partial \mathbf{p}}{\partial m_j} = \frac{\partial^2 H}{\partial m_i \, \partial m_j} + \frac{\partial H_{\mathbf{p}}}{\partial m_i} \frac{\partial \mathbf{p}}{\partial m_j} \quad . \tag{18.2}$$

The (arclength-related) Hamiltonian $H(\mathbf{m},\mathbf{p})$ and its slowness gradient $H_{\mathbf{p}}$ are explicit functions of the medium properties, and their derivatives wrt these parameters can be computed directly. The slowness vector derivative, $\partial \mathbf{p} / \partial m_j$ is given in equation 17.5, which leads to,

$$\frac{d}{dm_j} \frac{\partial H}{\partial m_j} = \frac{\partial^2 H}{\partial m_i \, \partial m_j} + \frac{\partial^2 H}{\partial m_i \partial \mathbf{p}} \frac{\partial \mathbf{p}}{\partial m_j} = \frac{\partial^2 H}{\partial m_i \, \partial m_j} - \frac{\partial H_{\mathbf{p}}}{\partial m_i} H_{\mathbf{pp}}^{-1} \cdot \frac{\partial H_{\mathbf{p}}}{\partial m_j} \quad . \tag{18.3}$$

Finally, we introduce equation 18.3 into equation 18.1, and the Hessian of the ray velocity, wrt the model parameters, simplifies to,

$$\frac{\partial^2 v_{\text{ray}}}{\partial m_i \, \partial m_j} = 2 v_{\text{ray}}^3 \frac{\partial H}{\partial m_i} \frac{\partial H}{\partial m_j} + v_{\text{ray}}^2 \frac{\partial^2 H}{\partial m_i \, \partial m_j} - v_{\text{ray}}^2 \frac{\partial H_{\mathbf{p}}}{\partial m_i} \cdot H_{\mathbf{pp}}^{-1} \cdot \frac{\partial H_{\mathbf{p}}}{\partial m_j} \quad . \tag{18.4}$$

Recall that the arclength-related Hamiltonian has the units of slowness. We still need to convert the Hessian components $\partial^2 H / (\partial m_i \partial m_j)$ into the Hessian components of the reference Hamiltonian, $\partial H^{\bar{\tau}} / \partial m_i$ and $\partial^2 H^{\bar{\tau}} / (\partial m_i \partial m_j)$. Applying equation 17.11, we obtain,

$$\frac{\partial^2 H(\mathbf{x},\mathbf{p})}{\partial m_i \partial m_j} = \frac{\dfrac{\partial^2 H^{\bar{\tau}}(\mathbf{x},\mathbf{p})}{\partial m_i \partial m_j}}{\sqrt{H^{\bar{\tau}}_{\mathbf{p}} \cdot H^{\bar{\tau}}_{\mathbf{p}}}} - \frac{\dfrac{\partial H^{\bar{\tau}}}{\partial m_i}\left(H^{\bar{\tau}}_{\mathbf{p}} \cdot \dfrac{\partial H^{\bar{\tau}}_{\mathbf{p}}}{\partial m_j}\right) + \dfrac{\partial H^{\bar{\tau}}}{\partial m_j}\left(H^{\bar{\tau}}_{\mathbf{p}} \cdot \dfrac{\partial H^{\bar{\tau}}_{\mathbf{p}}}{\partial m_i}\right)}{\left(H^{\bar{\tau}}_{\mathbf{p}} \cdot H^{\bar{\tau}}_{\mathbf{p}}\right)^{3/2}} . \qquad (18.5)$$

Equations 18.4 and 18.5 can be arranged in a concise tensor form,

$$\nabla_{\mathbf{m}}\nabla_{\mathbf{m}} v_{\text{ray}} = 2 v_{\text{ray}}^3 H_{\mathbf{m}} \otimes H_{\mathbf{m}} + v_{\text{ray}}^2 H_{\mathbf{mm}} - v_{\text{ray}}^2 H_{\mathbf{mp}} \cdot H_{\mathbf{pp}}^{-1} \cdot H_{\mathbf{pm}} , \qquad (18.6)$$

and,

$$H_{\mathbf{mm}}(\mathbf{x},\mathbf{p}) = \frac{H^{\bar{\tau}}_{\mathbf{mm}}}{\sqrt{H^{\bar{\tau}}_{\mathbf{p}} \cdot H^{\bar{\tau}}_{\mathbf{p}}}} - \frac{H^{\bar{\tau}}_{\mathbf{m}} \otimes \left(H^{\bar{\tau}}_{\mathbf{mp}} \cdot H^{\bar{\tau}}_{\mathbf{p}}\right) + \left(H^{\bar{\tau}}_{\mathbf{mp}} \cdot H^{\bar{\tau}}_{\mathbf{p}}\right) \otimes H^{\bar{\tau}}_{\mathbf{m}}}{\left(H^{\bar{\tau}}_{\mathbf{p}} \cdot H^{\bar{\tau}}_{\mathbf{p}}\right)^{3/2}} . \qquad (18.7)$$

One more auxiliary relationship is needed to complete the computations,

$$H_{\mathbf{mp}}(\mathbf{x},\mathbf{p}) = \frac{H^{\bar{\tau}}_{\mathbf{mp}}}{\sqrt{H^{\bar{\tau}}_{\mathbf{p}} \cdot H^{\bar{\tau}}_{\mathbf{p}}}} - \frac{H^{\bar{\tau}}_{\mathbf{m}} \otimes \left(H^{\bar{\tau}}_{\mathbf{pp}} \cdot H^{\bar{\tau}}_{\mathbf{p}}\right) + \left(H^{\bar{\tau}}_{\mathbf{mp}} \cdot H^{\bar{\tau}}_{\mathbf{p}}\right) \otimes H^{\bar{\tau}}_{\mathbf{p}}}{\left(H^{\bar{\tau}}_{\mathbf{p}} \cdot H^{\bar{\tau}}_{\mathbf{p}}\right)^{3/2}} . \qquad (18.8)$$

Applying the auxiliary algebraic identities of equation 4.11, we arrange equation 18.7 as,

$$H_{\mathbf{mm}}(\mathbf{x},\mathbf{p}) = \frac{H^{\bar{\tau}}_{\mathbf{mm}}}{\sqrt{H^{\bar{\tau}}_{\mathbf{p}} \cdot H^{\bar{\tau}}_{\mathbf{p}}}} - \frac{\left(H^{\bar{\tau}}_{\mathbf{m}} \otimes H^{\bar{\tau}}_{\mathbf{p}}\right) \cdot H^{\bar{\tau}}_{\mathbf{pm}} + H^{\bar{\tau}}_{\mathbf{mp}}\left(H^{\bar{\tau}}_{\mathbf{p}} \otimes H^{\bar{\tau}}_{\mathbf{m}}\right)}{\left(H^{\bar{\tau}}_{\mathbf{p}} \cdot H^{\bar{\tau}}_{\mathbf{p}}\right)^{3/2}} , \qquad (18.9)$$

and equation 18.8 as,

$$H_{\mathbf{mp}}(\mathbf{x},\mathbf{p}) = \frac{H_{\mathbf{mp}}^{\bar{\tau}}}{\sqrt{H_{\mathbf{p}}^{\bar{\tau}} \cdot H_{\mathbf{p}}^{\bar{\tau}}}} - \frac{\left(H_{\mathbf{m}}^{\bar{\tau}} \otimes H_{\mathbf{p}}^{\bar{\tau}}\right) \cdot H_{\mathbf{pp}}^{\bar{\tau}} + H_{\mathbf{mp}}^{\bar{\tau}} \left(H_{\mathbf{p}}^{\bar{\tau}} \otimes H_{\mathbf{p}}^{\bar{\tau}}\right)}{\left(H_{\mathbf{p}}^{\bar{\tau}} \cdot H_{\mathbf{p}}^{\bar{\tau}}\right)^{3/2}} \quad . \tag{18.10}$$

## 19. COMPUTATIONAL TEST 1

Input data at a reference point

An example of a medium with triclinic symmetry is given by Grechka (2017). It is presented below in the Voigt form,

$$\mathbf{C} = \begin{bmatrix} +45.597 & +0.935 & +0.895 & +0.103 & +0.074 & +0.070 \\ +0.935 & +14.442 & +0.887 & -0.083 & +0.018 & -0.059 \\ +0.895 & +0.887 & +44.303 & -0.049 & -0.040 & -0.026 \\ +0.103 & -0.083 & -0.049 & +0.459 & +0.090 & +0.52 \\ +0.074 & +0.018 & -0.040 & +0.90 & +0.374 & +0.101 \\ +0.070 & -0.059 & -0.026 & +0.52 & +0.101 & +0.900 \end{bmatrix} \quad . \tag{19.1}$$

However, it becomes more suitable to present the 21 stiffness components as a vector, rather than a tensor or a matrix. Table 1 gives the correspondence between two indices of the Voigt matrix and a single index of a vector. For each stiffness component, we introduce the normalized spatial gradient and the normalized spatial Hessian, whose units are $\text{km}^{-1}$ and $\text{km}^{-2}$, respectively.

$$\overline{\nabla_{\mathbf{x}} C_i} = \nabla_{\mathbf{x}} C_i / C_i^o \quad , \qquad \overline{\nabla_{\mathbf{x}} \nabla_{\mathbf{x}} C_i} = \nabla_{\mathbf{x}} \nabla_{\mathbf{x}} C_i / C_i^o \quad . \tag{19.2}$$

Thus, in the proximity of a reference node,

$$\frac{C_i}{C_i^o} = 1 + \overline{\nabla_{\mathbf{x}} C_i} \cdot \mathbf{x} + \frac{1}{2} \mathbf{x} \cdot \overline{\nabla_{\mathbf{x}} \nabla_{\mathbf{x}} C_i} \cdot \mathbf{x} \quad . \tag{19.3}$$

For each medium property (component of the triclinic stiffness tensor), we provide the normalized (relative) spatial gradient in Table 1 and the normalized spatial Hessian in Table 2. Normally, these derivatives are computed numerically, given the parameter field, but in this example, we consider a single reference point; therefore, the spatial derivatives of the model parameters are assumed input data. Table 1 represents a normalized (divided by $m_i$) matrix $\mathbf{m_x}(21,3)$, and Table 2 is a normalized 3D array $\mathbf{m_{xx}}(21,3,3)$, symmetric wrt the last two indices (and therefore stored as $21 \times 6$).

Table 1. Elastic properties of the triclinic model and their spatial gradients.

| Component, km$^2$/s$^2$ | | | Relative gradient, km$^{-1}$ | | |
|---|---|---|---|---|---|
| $C_i$ | $C_{ij}$ | value | $x_1$ | $x_2$ | $x_3$ |
| 1 | $C_{11}$ | +45.597 | +0.4647 | +0.1201 | +0.2126 |
| 2 | $C_{12}$ | +0.935 | +0.2993 | +0.0965 | +0.1592 |
| 3 | $C_{13}$ | +0.895 | +0.1517 | −0.0708 | +0.2103 |
| 4 | $C_{14}$ | +0.103 | +0.0062 | +0.3696 | −0.0399 |
| 5 | $C_{15}$ | +0.074 | −0.0308 | −0.0730 | +0.0514 |
| 6 | $C_{16}$ | +0.070 | +0.3953 | +0.2842 | +0.3468 |
| 7 | $C_{22}$ | +14.442 | −0.0786 | +0.0767 | +0.3028 |
| 8 | $C_{23}$ | +0.887 | +0.1143 | +0.1624 | +0.3564 |
| 9 | $C_{24}$ | −0.083 | +0.1668 | −0.0605 | +0.4459 |
| 10 | $C_{25}$ | +0.018 | +0.3221 | +0.3742 | +0.3664 |
| 11 | $C_{26}$ | −0.059 | +0.4380 | +0.1161 | +0.4072 |
| 12 | $C_{33}$ | +44.303 | +0.1007 | +0.2677 | +0.0293 |
| 13 | $C_{34}$ | −0.049 | +0.1362 | +0.1640 | +0.1998 |
| 14 | $C_{35}$ | −0.040 | −0.0412 | +0.2176 | +0.0890 |
| 15 | $C_{36}$ | −0.026 | +0.2618 | +0.2117 | +0.4530 |
| 16 | $C_{44}$ | +0.459 | −0.0788 | +0.4569 | +0.0150 |
| 17 | $C_{45}$ | +0.090 | +0.0019 | +0.3758 | +0.0294 |
| 18 | $C_{46}$ | +0.052 | +0.1136 | +0.2295 | +0.0055 |
| 19 | $C_{55}$ | +0.374 | +0.1294 | +0.1885 | +0.4181 |
| 20 | $C_{56}$ | +0.101 | +0.1619 | +0.4534 | +0.2185 |
| 21 | $C_{66}$ | +0.450 | +0.1679 | +0.3936 | +0.4576 |

Table 2. Hessians of elastic properties for triclinic model

| Component | | Normalized spatial Hessian components, km$^{-2}$ | | | | | |
|---|---|---|---|---|---|---|---|
| $C_i$ | $C_{ij}$ | $x_1 x_1$ | $x_1 x_2$ | $x_1 x_3$ | $x_2 x_2$ | $x_2 x_3$ | $x_3 x_3$ |
| 1 | $C_{11}$ | +0.1312 | +0.0956 | +0.2023 | –0.1279 | +0.0801 | +0.1125 |
| 2 | $C_{12}$ | +0.1241 | +0.1150 | –0.1544 | +0.0905 | –0.1351 | –0.0607 |
| 3 | $C_{13}$ | +0.1420 | –0.1518 | +0.1108 | –0.1373 | +0.1648 | +0.0142 |
| 4 | $C_{14}$ | +0.0562 | +0.1298 | –0.0895 | +0.1630 | –0.1255 | +0.0348 |
| 5 | $C_{15}$ | –0.1311 | –0.0842 | +0.0514 | +0.1220 | +0.0792 | –0.1346 |
| 6 | $C_{16}$ | +0.1942 | +0.1832 | +0.1981 | +0.1257 | –0.1114 | +0.1634 |
| 7 | $C_{22}$ | –0.1536 | +0.0233 | +0.1011 | –0.1488 | –0.0711 | +0.0823 |
| 8 | $C_{23}$ | +0.1053 | +0.1124 | +0.0564 | –0.1196 | –0.1291 | +0.1066 |
| 9 | $C_{24}$ | +0.0877 | –0.1008 | +0.1240 | +0.1321 | –0.1190 | –0.1572 |
| 10 | $C_{25}$ | +0.1209 | +0.1554 | +0.1365 | +0.0240 | +0.1355 | –0.0960 |
| 11 | $C_{26}$ | +0.1370 | +0.1255 | +0.0062 | –0.1371 | +0.1220 | +0.1342 |
| 12 | $C_{33}$ | +0.0050 | +0.0289 | +0.0233 | –0.1458 | +0.0630 | +0.0432 |
| 13 | $C_{34}$ | +0.1362 | +0.1640 | +0.1998 | +0.1265 | –0.1481 | –0.0455 |
| 14 | $C_{35}$ | –0.0412 | +0.1176 | +0.0890 | –0.0902 | +0.0239 | –0.1030 |
| 15 | $C_{36}$ | +0.1418 | +0.1917 | +0.1530 | –0.1499 | +0.2370 | +0.0723 |
| 16 | $C_{44}$ | –0.0788 | +0.1571 | +0.0150 | +0.0914 | +0.1200 | –0.1650 |
| 17 | $C_{45}$ | +0.1020 | +0.1762 | +0.0184 | +0.0912 | +0.0631 | +0.1271 |
| 18 | $C_{46}$ | +0.1136 | +0.2295 | +0.0055 | +0.1199 | –0.1235 | +0.1080 |
| 19 | $C_{55}$ | +0.1062 | +0.0380 | +0.2011 | +0.0731 | +0.1655 | +0.0887 |
| 20 | $C_{56}$ | +0.0810 | +0.1234 | +0.1185 | –0.1452 | –0.1300 | –0.0988 |
| 21 | $C_{66}$ | +0.1679 | +0.1936 | +0.1576 | +0.2149 | +0.1425 | –0.1241 |

Normally, the accuracy of the resolving parameters does not exceed 3-4 significant decimal digits; in these examples we keep the accuracy up to eight digits to allow reproducibility of the benchmark problems. We accept the ray velocity direction used by Grechka (2017),

$$\mathbf{r} = \begin{bmatrix} 0.54812444 & 0.55112512 & 0.62914283 \end{bmatrix} \quad . \tag{19.4}$$

Inversion for the slowness vector

We compute analytically the reference Hamiltonian with equations 3.1 and 3.3, and solve equation set 6.1 to invert for the slowness vector. In Table 3, the solutions are sorted in the order of decreasing ray velocity magnitudes. The first solution is for a compressional wave, and the others are for shear waves. For each solution, the table includes the Cartesian slowness components, $p_k$, the magnitudes of the phase and ray velocities, $v_{\text{phs}}$ and $v_{\text{ray}}$ and the angle $\Delta\vartheta$ between the phase and ray directions. We compute the ray velocity derivatives for the first three solutions (out of the 19): compressional and shear with the highest and second-highest ray velocities; we call them $P$, $S_A$ and $S_B$. For both shear waves, the correct sign before the reference Hamiltonian proved to be minus (see Appendix D).

Table 4 includes the minor, major and mean curvatures of the slowness surface. The last column includes the coefficient $\alpha$ (equation 6.2) that defines the sign of the shear-wave reference Hamiltonian $H^{\bar{\tau}}$. The slowness surface is the constant (in our case, vanishing) Hamiltonian. Both, the reference and the arclength-related Hamiltonians can be used, but the arclength-related Hamiltonian $H$ is more convenient because $H_{\mathbf{p}} \cdot H_{\mathbf{p}} = 1$; it becomes clear from the formulae for computing the curvature, listed below. For a fixed location $\mathbf{x}$, the Hamiltonian can be considered an implicit function of the three slowness components,

Table 3. Inverted slowness vector for example 1

| # | $p_1$, s/km | $p_2$, s/km | $p_3$, s/km | $v_{\text{phs}}$, km/s | $v_{\text{ray}}$, km/s | $\Delta\vartheta$, deg |
|---|---|---|---|---|---|---|
| 1 | 0.13555828 | 0.25145731 | 0.14025204 | 3.1422707 | 3.3208711 | 18.876378 |
| 2 | 0.14145161 | 0.26175272 | 0.15494351 | 2.9810194 | 3.1321140 | 17.869186 |
| 3 | 0.15324689 | 0.26250236 | 0.14535503 | 2.9679970 | 3.1238375 | 18.174221 |
| 4 | 0.14621400 | 0.27368703 | 0.14881941 | 2.9058187 | 3.0806395 | 19.394976 |
| 5 | .020462473 | 1.3739451 | .028826692 | 0.72759035 | 1.2713463 | 55.089296 |
| 6 | 1.3261367 | .069294564 | .042152101 | 0.75266367 | 1.2632681 | 53.429836 |
| 7 | .082601563 | 1.4915586 | .021901670 | 0.66934197 | 1.1349570 | 53.860664 |
| 8 | .017962805 | 1.5077454 | .071948739 | 0.6624412 | 1.1285813 | 54.057851 |
| 9 | .026411709 | 1.5436997 | .036288735 | 0.64752077 | 1.1260255 | 54.896878 |
| 10 | .025072060 | .058899768 | 1.3518256 | 0.73891242 | 1.1152063 | 48.503128 |
| 11 | 1.4998649 | .056791944 | 0.14900379 | 0.66299032 | 1.0557915 | 51.100538 |
| 12 | 1.6495928 | 0.20673786 | 0.20673786 | 0.60145923 | 0.96994027 | 51.676618 |
| 13 | 1.7319248 | 0.12539465 | .023070089 | 0.57583402 | 0.96811716 | 53.501776 |
| 14 | 0.15535027 | .053104697 | 1.4831328 | 0.67015508 | 0.95463482 | 45.412193 |
| 15 | 1.1315446 | 0.86296430 | .015418487 | 0.70266966 | 0.90454434 | 39.029513 |
| 16 | .012640499 | 0.22331913 | 1.6154554 | 0.61317076 | 0.87232820 | 45.338822 |
| 17 | .016424568 | 0.1147058 | 1.7137075 | 0.58220075 | 0.86927289 | 47.951763 |
| 18 | .007662205 | 0.74119655 | 1.2320229 | 0.69549985 | 0.84188530 | 34.297604 |
| 19 | 0.90607175 | .043187752 | 1.0649368 | 0.71484731 | 0.84002605 | 31.681327 |

Table 4. Ratio between $H_{\mathbf{p}}^{\bar{\tau}}$ and $\mathbf{r}$ vs. curvatures of the slowness surface for example 1

| # | Curvature, km/s | | | Ratio |
|---|---|---|---|---|
|   | Minor | Major | Mean | $\alpha$, km/s |
| 1  | +61.795621  | +100.49936 | +81.147488 | +0.12995592 |
| 2  | −139.14815  | +93.142899 | −23.002627 | −0.07025568 |
| 3  | −184.87913  | +76.061900 | −54.408616 | −0.06537122 |
| 4  | −1060.1390  | +62.168078 | −498.98546 | −0.01406616 |
| 5  | −24.293369  | −5.6765690 | −14.984969 | −14.3592906 |
| 6  | −15.626902  | −3.1727608 | −9.3998313 | −79.0563923 |
| 7  | −4.7748834  | +16.982291 | +6.1037036 | +28.2306563 |
| 8  | −3.8466348  | +18.203224 | +7.1782947 | +23.8592781 |
| 9  | +5.2220325  | +11.396992 | +8.3095121 | +20.7148975 |
| 10 | −14.381405  | −4.5066681 | −9.4440365 | −68.782846 |
| 11 | −5.9287231  | +7.4664983 | +0.7688876 | +193.30857 |
| 12 | −0.71235548 | +23.417681 | +11.352663 | +218.41032 |
| 13 | +1.03549108 | +20.138339 | +10.586915 | +194.78277 |
| 14 | −5.17687070 | +9.1493832 | +1.9862563 | +192.63753 |
| 15 | −0.33563937 | +29.765275 | +15.050457 | +1066.0033 |
| 16 | −0.79302121 | +32.195990 | +15.701484 | +188.14619 |
| 17 | +1.49472127 | +26.710559 | +14.102640 | +150.09615 |
| 18 | +0.33667408 | +37.726961 | +19.031818 | +834.74199 |
| 19 | +0.46726311 | +12.200988 | +6.3341253 | +3074.6920 |

$$H(p_1, p_2, p_3) = 0 \quad , \tag{19.5}$$

The minor and major curvatures of an implicit surface are the two nonzero eigenvalues of the matrix,

$$\mathbf{M} = \left( \mathbf{I} - \frac{H_{\mathbf{p}} \otimes \nabla H_{\mathbf{p}}}{H_{\mathbf{p}} \cdot H_{\mathbf{p}}} \right) \cdot \frac{H_{\mathbf{pp}}}{\sqrt{H_{\mathbf{p}} \cdot H_{\mathbf{p}}}} \quad , \tag{19.6}$$

and for $H_{\mathbf{p}} = \pm \mathbf{r}$, the matrix simplifies to,

$$\mathbf{M} = (\mathbf{I} - \mathbf{r} \otimes \mathbf{r}) H_{\mathbf{pp}} = \mathbf{T} H_{\mathbf{pp}} \qquad . \tag{19.7}$$

Multiplication by matrix $\mathbf{T}$ nullifies the eigenvalue related to the eigenvector $\mathbf{r}$ and does not alter the two other eigenvalues and the three eigenvectors. The eigenvectors of matrix $\mathbf{M}$, corresponding to the nonzero eigenvalues, are the principal directions in the tangent plane, normal each other and to the gradient $H_{\mathbf{p}}$. The latter, in turn, is the third eigenvector, collinear with the ray direction $\mathbf{r}$ and corresponding to the zero eigenvalue. To obtain the mean curvature only, there is no need to solve the eigenvalue problem,

$$k_{\text{mean}} = \frac{k_{\min} + k_{\max}}{2} = \frac{\operatorname{tr} \mathbf{M}}{2} \qquad . \tag{19.8}$$

Note that for all 19 solutions of the inverse problem, the signs of the coefficient $\alpha$ in equation 6.2 and the mean curvature, listed in the last two columns of Table 4, are identical. This is, however, not a must, as shown in the other example, Section 22.

Gradients and Hessians of the reference Hamiltonian

We compute the slowness gradients and Hessians of the reference Hamiltonian using equations 12.2 and 13.2, respectively; its spatial gradients with equations 14.5 and 14.6, the spatial Hessians with equations 15.4 and 15.5, and the mixed Hessians with equations 16.2 and 16.3. The results are,

- the slowness gradient,

$$H_{\mathbf{p}}^{\bar{\tau}}(P) = \begin{bmatrix} 7.1232017 \cdot 10^{-2} & 7.1621973 \cdot 10^{-2} & 8.1760837 \cdot 10^{-2} \end{bmatrix} \text{ km/s} ,$$

$$H_{\mathbf{p}}^{\bar{\tau}}(S_A) = \begin{bmatrix} 3.8508853 \cdot 10^{-2} & 3.8719668 \cdot 10^{-2} & 4.4200855 \cdot 10^{-2} \end{bmatrix} \text{ km/s} , \quad (19.9)$$

$$H_{\mathbf{p}}^{\bar{\tau}}(S_B) = \begin{bmatrix} 3.5831561 \cdot 10^{-2} & 3.6027719 \cdot 10^{-2} & 4.1127832 \cdot 10^{-2} \end{bmatrix} \text{ km/s} ,$$

- the slowness Hessian,

$$H_{\mathbf{pp}}^{\bar{\tau}}(P) = \begin{bmatrix} -1.1819972 \cdot 10^{-1} & -8.8313085 & -1.3030042 \cdot 10^{+1} \\ -8.8313085 & -5.6423112 \cdot 10^{-1} & -1.0655657 \cdot 10^{+1} \\ -1.3030042 \cdot 10^{+1} & -1.0655657 \cdot 10^{+1} & -2.4011328 \cdot 10^{-1} \end{bmatrix} (\text{km/s})^2 ,$$

$$H_{\mathbf{pp}}^{\bar{\tau}}(S_A) = \begin{bmatrix} +1.3973083 \cdot 10^{-1} & -9.6178410 & +8.8403500 \cdot 10^{-1} \\ -9.6178410 & -7.9895182 \cdot 10^{-2} & +3.4468498 \\ +8.8403500 \cdot 10^{-1} & +3.4468498 & +6.276404 \cdot 10^{-1} \end{bmatrix} (\text{km/s})^2 , \quad (19.10)$$

$$H_{\mathbf{pp}}^{\bar{\tau}}(S_B) = \begin{bmatrix} +4.5439111 \cdot 10^{-1} & +2.5317289 & -2.3447280 \cdot 10^{-1} \\ +2.5317289 & -1.4788900 \cdot 10^{-1} & -1.1774930 \cdot 10^{+1} \\ -2.3447280 \cdot 10^{-1} & -1.1774930 \cdot 10^{+1} & +1.6699778 \cdot 10^{-1} \end{bmatrix} (\text{km/s})^2 ,$$

- the spatial gradient,

$$H_{\mathbf{x}}^{\bar{\tau}}(P) = \begin{bmatrix} 2.0292344 \cdot 10^{-3} & 3.1343522 \cdot 10^{-3} & 3.9306528 \cdot 10^{-3} \end{bmatrix} \text{ km}^{-1} ,$$

$$H_{\mathbf{x}}^{\bar{\tau}}(S_A) = \begin{bmatrix} 1.2277450 \cdot 10^{-3} & 1.7432780 \cdot 10^{-3} & 2.1761336 \cdot 10^{-3} \end{bmatrix} \text{ km}^{-1} , \quad (19.11)$$

$$H_{\mathbf{x}}^{\bar{\tau}}(S_B) = \begin{bmatrix} 1.2017822 \cdot 10^{-3} & 1.6136573 \cdot 10^{-3} & 2.1131523 \cdot 10^{-3} \end{bmatrix} \text{ km}^{-1} ,$$

- the spatial Hessian,

$$H^{\bar{\tau}}_{\mathbf{xx}}(P) = \begin{bmatrix} +7.2522389\cdot 10^{-4} & -9.0927236\cdot 10^{-3} & -1.1994543\cdot 10^{-2} \\ -9.0927236\cdot 10^{-3} & -1.3759214\cdot 10^{-2} & -1.7000955\cdot 10^{-2} \\ -1.1994543\cdot 10^{-2} & -1.7000955\cdot 10^{-2} & -1.1904140\cdot 10^{-2} \end{bmatrix} \text{km}^{-2} \ ,$$

$$H^{\bar{\tau}}_{\mathbf{xx}}(S_A) = \begin{bmatrix} +6.6642191\cdot 10^{-3} & -1.3858581\cdot 10^{-3} & -8.8621517\cdot 10^{-3} \\ -1.3858581\cdot 10^{-3} & -7.5007740\cdot 10^{-4} & -1.7354923\cdot 10^{-3} \\ -8.8621517\cdot 10^{-3} & -1.7354923\cdot 10^{-3} & -1.0995600\cdot 10^{-2} \end{bmatrix} \text{km}^{-2} \ , \quad (19.12)$$

$$H^{\bar{\tau}}_{\mathbf{xx}}(S_B) = \begin{bmatrix} -4.4251729\cdot 10^{-4} & +3.3951758\cdot 10^{-3} & +1.4315441\cdot 10^{-3} \\ +3.3951758\cdot 10^{-3} & -5.7939013\cdot 10^{-3} & -8.2410466\cdot 10^{-3} \\ +1.4315441\cdot 10^{-3} & -8.2410466\cdot 10^{-3} & +2.6664265\cdot 10^{-3} \end{bmatrix} \text{km}^{-2} \ ,$$

- the mixed Hessian,

$$H^{\bar{\tau}}_{\mathbf{px}}(P) = \begin{bmatrix} +1.8112914\cdot 10^{-2} & -3.5206988\cdot 10^{-1} & -3.4982070\cdot 10^{-1} \\ -3.5096106\cdot 10^{-1} & -2.9061771\cdot 10^{-1} & -1.5876352\cdot 10^{-1} \\ -3.0335156\cdot 10^{-1} & -2.2712095\cdot 10^{-1} & -5.9803835\cdot 10^{-1} \end{bmatrix} \text{s}^{-1} \ ,$$

$$H^{\bar{\tau}}_{\mathbf{px}}(S_A) = \begin{bmatrix} +1.2808909\cdot 10^{-1} & -7.1417659\cdot 10^{-2} & -3.6938443\cdot 10^{-1} \\ -3.0040773\cdot 10^{-1} & -1.5095586\cdot 10^{-2} & -1.4296670\cdot 10^{-1} \\ -1.4853915\cdot 10^{-3} & +7.3969663\cdot 10^{-2} & +1.5029337\cdot 10^{-1} \end{bmatrix} \text{s}^{-1} \ , \quad (19.13)$$

$$H^{\bar{\tau}}_{\mathbf{px}}(S_B) = \begin{bmatrix} -6.6815737\cdot 10^{-3} & +3.8371842\cdot 10^{-2} & +1.1460374\cdot 10^{-1} \\ +6.7151471\cdot 10^{-3} & -2.1783209\cdot 10^{-1} & +2.6220593\cdot 10^{-2} \\ +1.2915470\cdot 10^{-1} & -1.1076245\cdot 10^{-1} & -4.7146619\cdot 10^{-1} \end{bmatrix} \text{s}^{-1} \ ,$$

Gradients and Hessians of the arclength-related Hamiltonian

Applying relationships of Section 4, we convert the gradients and Hessians of the reference Hamiltonian into those of the arclength-related Hamiltonian. The results are,

- the slowness gradient of the arclength Hamiltonian is the same for all wave modes and equal to the ray direction $\mathbf{r}$,

$$H_{\mathbf{p}}(P, S_A, S_B) = [0.54812444 \quad 0.55112512 \quad 0.62914283] = \mathbf{r} \quad , \qquad (19.14)$$

- the slowness Hessian,

$$H_{\mathbf{pp}}(P) = \begin{bmatrix} +109.84663 & +37.729064 & +18.828577 \\ +37.729064 & +96.213771 & +31.234835 \\ +18.828577 & +31.234835 & +125.62845 \end{bmatrix} \text{km/s} \quad ,$$

$$H_{\mathbf{pp}}(S_A) = \begin{bmatrix} +74.824798 & -75.726044 & +32.702021 \\ -75.726044 & +48.240324 & +55.444553 \\ +32.702021 & +55.444553 & -40.839958 \end{bmatrix} \text{km/s} \quad , \qquad (19.15)$$

$$H_{\mathbf{pp}}(S_B) = \begin{bmatrix} -18.150556 & +77.272371 & +36.616808 \\ +77.272371 & +100.62455 & -66.489855 \\ +36.616808 & -66.489855 & +127.91723 \end{bmatrix} \text{km/s} \quad ,$$

- the inverse of this matrix,

$$H_{\mathbf{pp}}^{-1}(P) = \begin{bmatrix} +1.0562259 \cdot 10^{-2} & -3.9464905 \cdot 10^{-3} & -6.0180896 \cdot 10^{-4} \\ -3.9464905 \cdot 10^{-3} & +1.2780661 \cdot 10^{-2} & -2.5861583 \cdot 10^{-3} \\ -6.0180896 \cdot 10^{-4} & -2.5861583 \cdot 10^{-3} & +8.6931698 \cdot 10^{-3} \end{bmatrix} \text{s/km} \quad ,$$

$$H_{\mathbf{pp}}^{-1}(S_A) = \begin{bmatrix} +1.0745347 \cdot 10^{-2} & +2.7256217 \cdot 10^{-3} & +1.2304504 \cdot 10^{-2} \\ +2.7256217 \cdot 10^{-3} & +8.7877394 \cdot 10^{-3} & +1.4112787 \cdot 10^{-2} \\ +1.2304504 \cdot 10^{-2} & +1.4112787 \cdot 10^{-2} & +4.5264324 \cdot 10^{-3} \end{bmatrix} \text{s/km} \quad , \qquad (19.16)$$

$$H_{\mathbf{pp}}^{-1}(S_B) = \begin{bmatrix} -5.9163758 \cdot 10^{-3} & +8.6246585 \cdot 10^{-3} & +6.1765808 \cdot 10^{-3} \\ +8.6246585 \cdot 10^{-3} & +2.5641719 \cdot 10^{-3} & -1.1360162 \cdot 10^{-3} \\ +6.1765808 \cdot 10^{-3} & -1.1360162 \cdot 10^{-3} & +5.4589971 \cdot 10^{-3} \end{bmatrix} \text{s/km} \quad ,$$

- the spatial gradient,

$$H_{\mathbf{x}}(P) = \begin{bmatrix} 1.5614790 \cdot 10^{-2} & 2.4118579 \cdot 10^{-2} & 3.0246046 \cdot 10^{-2} \end{bmatrix} \text{ s/km}^2 ,$$

$$H_{\mathbf{x}}(S_A) = \begin{bmatrix} 1.7475385 \cdot 10^{-2} & 2.4813340 \cdot 10^{-2} & 3.0974488 \cdot 10^{-2} \end{bmatrix} \text{ s/km}^2 , \quad (19.17)$$

$$H_{\mathbf{x}}(S_B) = \begin{bmatrix} 1.8383965 \cdot 10^{-2} & 2.4684524 \cdot 10^{-2} & 3.2325424 \cdot 10^{-2} \end{bmatrix} \text{ s/km}^2 ,$$

- the spatial Hessian,

$$H_{\mathbf{xx}}(P) = \begin{bmatrix} +9.5539445 \cdot 10^{-2} & +5.9108444 \cdot 10^{-2} & +7.3589490 \cdot 10^{-2} \\ +5.9108444 \cdot 10^{-2} & +7.8243216 \cdot 10^{-2} & +1.0628065 \cdot 10^{-1} \\ +7.3589490 \cdot 10^{-2} & +1.0628065 \cdot 10^{-1} & +2.1351968 \cdot 10^{-1} \end{bmatrix} \text{ s/km}^3 ,$$

$$H_{\mathbf{xx}}(S_A) = \begin{bmatrix} +1.4275799 \cdot 10^{-1} & +1.4512470 \cdot 10^{-2} & -3.7248733 \cdot 10^{-2} \\ +1.4512470 \cdot 10^{-2} & -1.0020997 \cdot 10^{-2} & +4.1648252 \cdot 10^{-2} \\ -3.7248733 \cdot 10^{-2} & +4.1648252 \cdot 10^{-2} & +8.1216863 \cdot 10^{-3} \end{bmatrix} \text{ s/km}^3 , \quad (19.18)$$

$$H_{\mathbf{xx}}(S_B) = \begin{bmatrix} -5.2493707 \cdot 10^{-2} & +6.8683405 \cdot 10^{-2} & +4.3385941 \cdot 10^{-2} \\ +6.8683405 \cdot 10^{-2} & +3.8777515 \cdot 10^{-2} & +4.0186250 \cdot 10^{-2} \\ +4.3385941 \cdot 10^{-2} & +4.0186250 \cdot 10^{-2} & +2.5772346 \cdot 10^{-1} \end{bmatrix} \text{ s/km}^3 ,$$

- the mixed Hessian,

$$H_{\mathbf{px}}(P) = \begin{bmatrix} +3.2958795 & +1.8197670 & +3.1287060 \\ +3.1143439 \cdot 10^{-1} & +2.0676191 & +4.3174630 \\ +1.0599475 & +3.0971727 & +1.6357372 \end{bmatrix} \text{km}^{-1} \;,$$

$$H_{\mathbf{px}}(S_A) = \begin{bmatrix} +3.7354925 & +6.3932346 \cdot 10^{-1} & -1.7430933 \\ -2.7377385 & +9.0397583 \cdot 10^{-1} & +8.1723673 \cdot 10^{-1} \\ +1.498525 \cdot 10^{-1} & +7.9639035 \cdot 10^{-2} & +2.5859398 \end{bmatrix} \text{km}^{-1} \;, \quad (19.19)$$

$$H_{\mathbf{px}}(S_B) = \begin{bmatrix} -1.2048035 & +1.4363292 & +2.8521662 \\ +1.1333531 & +3.9419000 \cdot 10^{-1} & +5.2677282 \\ +3.0249022 & +2.3885996 & -1.8804907 \end{bmatrix} \text{km}^{-1} \;.$$

<u>Gradients and Hessians of the ray velocity</u>

Computing the spatial gradient and Hessian of the ray velocity magnitude is explained in Sections 7 and 8, the directional gradient and Hessian - in Sections 9 and 10, and the mixed Hessian in Section 11. Recall that we consider a quasi-compressional wave and two shear waves (from 18 shear solutions total), with the highest and second-highest ray velocity magnitudes,

$$v_{\text{ray}}(P) = 3.3208711 \text{ km/s} \;, \quad \begin{array}{l} v_{\text{ray}}(S1) = 3.1321140 \text{ km/s} \;, \\ v_{\text{ray}}(S2) = 3.1238375 \text{ km/s} \;. \end{array} \quad (19.20)$$

The results are,

- the spatial gradients of the ray velocity,

$$\nabla_{\mathbf{x}} v_{\text{ray}}(P) = \begin{bmatrix} +1.7220279 \cdot 10^{-1} & +2.6598414 \cdot 10^{-1} & +3.3355897 \cdot 10^{-1} \end{bmatrix} \text{s}^{-1} \;,$$
$$\nabla_{\mathbf{x}} v_{\text{ray}}(S_A) = \begin{bmatrix} +1.7143594 \cdot 10^{-1} & +2.4342230 \cdot 10^{-1} & +3.0386401 \cdot 10^{-1} \end{bmatrix} \text{s}^{-1} \;, \quad (19.21)$$
$$\nabla_{\mathbf{x}} v_{\text{ray}}(S_B) = \begin{bmatrix} +1.7939737 \cdot 10^{-1} & +2.4088049 \cdot 10^{-1} & +3.1544315 \cdot 10^{-1} \end{bmatrix} \text{s}^{-1} \;,$$

- the spatial Hessians of the ray velocity,

$$\nabla_{\mathbf{x}}\nabla_{\mathbf{x}} v_{\text{ray}}(P) = \begin{bmatrix} -1.6067446 \cdot 10^{-1} & +6.7148368 \cdot 10^{-2} & +1.5383046 \cdot 10^{-1} \\ +6.7148368 \cdot 10^{-2} & -2.3482982 \cdot 10^{-2} & +3.8049694 \cdot 10^{-3} \\ +1.5383046 \cdot 10^{-1} & +3.8049694 \cdot 10^{-3} & +4.4258402 \cdot 10^{-2} \end{bmatrix} 1/(\text{km s}) \quad ,$$

$$\nabla_{\mathbf{x}}\nabla_{\mathbf{x}} v_{\text{ray}}(S_A) = \begin{bmatrix} -1.7347332 \cdot 10^{-1} & +5.0560836 \cdot 10^{-2} & +1.4939360 \cdot 10^{-1} \\ +5.0560836 \cdot 10^{-2} & -2.3741840 \cdot 10^{-1} & +1.3151670 \cdot 10^{-2} \\ +1.4939360 \cdot 10^{-1} & +1.3151670 \cdot 10^{-2} & +4.3038140 \cdot 10^{-2} \end{bmatrix} 1/(\text{km s}) \quad ,$$

$$\nabla_{\mathbf{x}}\nabla_{\mathbf{x}} v_{\text{ray}}(S_B) = \begin{bmatrix} -1.8224066 \cdot 10^{-1} & +5.9695079 \cdot 10^{-2} & +1.7338675 \cdot 10^{-1} \\ +5.9695079 \cdot 10^{-2} & -2.6115926 \cdot 10^{-1} & +1.6666942 \cdot 10^{-2} \\ +1.7338675 \cdot 10^{-1} & +1.6666942 \cdot 10^{-2} & +6.3539163 \cdot 10^{-2} \end{bmatrix} 1/(\text{km s}) \quad ,$$

(19.22)

- the directional gradients of the ray velocity

$$\nabla_{\mathbf{r}} v_{\text{ray}}(P) = \begin{bmatrix} +3.2528882 \cdot 10^{-1} & -9.4290215 \cdot 10^{-1} & +5.4257681 \cdot 10^{-1} \end{bmatrix} \text{km}/\text{s} \quad ,$$
$$\nabla_{\mathbf{r}} v_{\text{ray}}(S_A) = \begin{bmatrix} +3.2912842 \cdot 10^{-1} & -8.4164359 \cdot 10^{-1} & +4.5052980 \cdot 10^{-1} \end{bmatrix} \text{km}/\text{s} \quad , \quad (19.23)$$
$$\nabla_{\mathbf{r}} v_{\text{ray}}(S_B) = \begin{bmatrix} +2.1681322 \cdot 10^{-1} & -8.3996740 \cdot 10^{-1} & +5.4691318 \cdot 10^{-1} \end{bmatrix} \text{km}/\text{s} \quad ,$$

- the directional Hessians of the ray velocity,

$$\nabla_{\mathbf{r}}\nabla_{\mathbf{r}}v_{\text{ray}}(P) = \begin{bmatrix} +1.9333529 & -7.8716336 \cdot 10^{-1} & -1.5118686 \\ -7.8716336 \cdot 10^{-1} & +3.7657751 & -1.1142901 \\ -1.5118686 & -1.1142901 & +1.4308811 \end{bmatrix} \text{km/s} \quad,$$

$$\nabla_{\mathbf{r}}\nabla_{\mathbf{r}}v_{\text{ray}}(S_A) = \begin{bmatrix} +1.8754254 & -7.8803278 \cdot 10^{-1} & -1.4667420 \\ -7.8803278 \cdot 10^{-1} & +3.5568527 & -1.0914648 \\ -1.4667420 & -1.0914648 & +1.5178764 \end{bmatrix} \text{km/s} \quad, (19.24)$$

$$\nabla_{\mathbf{r}}\nabla_{\mathbf{r}}v_{\text{ray}}(S_B) = \begin{bmatrix} +2.0643234 & -7.7448787 \cdot 10^{-1} & -1.4646589 \\ -7.7448787 \cdot 10^{-1} & +3.5567363 & -1.1058277 \\ -1.4646589 & -1.1058277 & +1.3754454 \end{bmatrix} \text{km/s} \quad,$$

- and the mixed Hessian of the ray velocity,

$$\nabla_{\mathbf{r}}\nabla_{\mathbf{x}}v_{\text{ray}}(P) = \begin{bmatrix} +3.0267021 \cdot 10^{-1} & +7.7427806 \cdot 10^{-3} & +4.8190430 \cdot 10^{-2} \\ -3.2247291 \cdot 10^{-1} & -1.7374209 \cdot 10^{-1} & +5.2464846 \cdot 10^{-2} \\ +1.8790615 \cdot 10^{-2} & +1.4545127 \cdot 10^{-1} & -8.7943538 \cdot 10^{-2} \end{bmatrix} \text{s}^{-1} \quad,$$

$$\nabla_{\mathbf{r}}\nabla_{\mathbf{x}}v_{\text{ray}}(S_A) = \begin{bmatrix} +2.8071700 \cdot 10^{-1} & +1.8910353 \cdot 10^{-2} & +4.7558410 \cdot 10^{-2} \\ -3.0200560 \cdot 10^{-1} & -1.5892682 \cdot 10^{-1} & +5.1091963 \cdot 10^{-2} \\ +1.9987545 \cdot 10^{-2} & +1.2274370 \cdot 10^{-1} & -8.6190271 \cdot 10^{-2} \end{bmatrix} \text{s}^{-1} \quad, (19.25)$$

$$\nabla_{\mathbf{r}}\nabla_{\mathbf{x}}v_{\text{ray}}(S_B) = \begin{bmatrix} +2.7383543 \cdot 10^{-1} & -4.3757573 \cdot 10^{-3} & +3.6219836 \cdot 10^{-2} \\ -3.0192001 \cdot 10^{-1} & -1.5802643 \cdot 10^{-1} & +4.9214288 \cdot 10^{-2} \\ +2.5907962 \cdot 10^{-2} & +1.4224241 \cdot 10^{-1} & -7.4667000 \cdot 10^{-2} \end{bmatrix} \text{s}^{-1} \quad.$$

Next, we compute the gradients and Hessians of the ray velocity wrt the medium properties, $\nabla_{\mathbf{m}}$ and $\nabla_{\mathbf{m}}\nabla_{\mathbf{m}}v_{\text{ray}}$, and the auxiliary objects that need to be precomputed to obtain these gradients and Hessians. To avoid long vectors and large matrices, let's assume, for example that we are looking for the derivatives wrt only four density-normalized stiffness tensor components: $\mathbf{m} = \{C_{11} \quad C_{22} \quad C_{13} \quad C_{44}\}$. The other stiffness components are assumed fixed. The "upgrade" of the test from four parameters to twenty-one is straightforward. To establish $\nabla_{\mathbf{m}}$ and $\nabla_{\mathbf{m}}\nabla_{\mathbf{m}} v_{\text{ray}}$, we compute the following values,

- the model-related gradients of the reference Hamiltonian,

$$H_{\mathbf{m}}^{\bar{\tau}}(P) = \begin{bmatrix} +9.0042207 \cdot 10^{-5} & +2.0345354 \cdot 10^{-4} & +1.1492761 \cdot 10^{-4} & +1.0494571 \cdot 10^{-3} \end{bmatrix} (\text{km/s})^{-2},$$

$$H_{\mathbf{m}}^{\bar{\tau}}(S_A) = \begin{bmatrix} +5.6365232 \cdot 10^{-5} & -1.3562183 \cdot 10^{-4} & +8.1580786 \cdot 10^{-5} & +1.8508376 \cdot 10^{-5} \end{bmatrix} (\text{km/s})^{-2},$$

$$H_{\mathbf{m}}^{\bar{\tau}}(S_B) = \begin{bmatrix} +6.5020628 \cdot 10^{-5} & -1.2952212 \cdot 10^{-4} & +6.4502530 \cdot 10^{-5} & +5.9571028 \cdot 10^{-4} \end{bmatrix} (\text{km/s})^{-2},$$

(19.26)

- the model-related Hessians of the reference Hamiltonian,

$$H_{\mathbf{mm}}^{\bar{\tau}}(P) = \begin{bmatrix} 0 & 0 & -2.798578 \cdot 10^{-5} & -1.8357794 \cdot 10^{-4} \\ 0 & +5.5971560 \cdot 10^{-5} & 0 & +4.3095661 \cdot 10^{-5} \\ -2.7985780 \cdot 10^{-5} & 0 & 0 & -4.3176653 \cdot 10^{-5} \\ -1.8357794 \cdot 10^{-4} & +4.3095661 \cdot 10^{-5} & -4.3176653 \cdot 10^{-5} & 0 \end{bmatrix} (\text{km/s})^{-4},$$

$$H_{\mathbf{mm}}^{\bar{\tau}}(S_A) = \begin{bmatrix} 0 & 0 & -3.3020085 \cdot 10^{-7} & +3.6575709 \cdot 10^{-5} \\ 0 & +6.6040169 \cdot 10^{-7} & 0 & -5.9059236 \cdot 10^{-5} \\ -3.3020085 \cdot 10^{-7} & 0 & 0 & +1.8005321 \cdot 10^{-5} \\ +3.6575709 \cdot 10^{-5} & -5.9059236 \cdot 10^{-5} & +1.8005321 \cdot 10^{-5} & 0 \end{bmatrix} (\text{km/s})^{-4},$$

$$H_{\mathbf{mm}}^{\bar{\tau}}(S_B) = \begin{bmatrix} 0 & 0 & -3.3064158 \cdot 10^{-6} & +9.1837521 \cdot 10^{-5} \\ 0 & +6.6128316 \cdot 10^{-6} & 0 & -6.5246235 \cdot 10^{-5} \\ -3.3064158 \cdot 10^{-6} & 0 & 0 & -5.6412911 \cdot 10^{-5} \\ +9.1837521 \cdot 10^{-5} & -6.5246235 \cdot 10^{-5} & -5.6412911 \cdot 10^{-5} & 0 \end{bmatrix} (\text{km/s})^{-4},$$

(19.27)

- the model-related gradients of the arclength-related Hamiltonian,

$$H_{\mathbf{m}}(P) = \begin{bmatrix} +6.9286728\cdot 10^{-4} & +1.5655580\cdot 10^{-3} & +8.8435839\cdot 10^{-4} & +8.0754853\cdot 10^{-3} \end{bmatrix} (\text{km/s})^{-3},$$

$$H_{\mathbf{m}}(S_A) = \begin{bmatrix} +8.0228723\cdot 10^{-4} & -1.9304038\cdot 10^{-3} & +1.1611985\cdot 10^{-3} & +2.6344315\cdot 10^{-4} \end{bmatrix} (\text{km/s})^{-3},$$

$$H_{\mathbf{m}}(S_B) = \begin{bmatrix} +9.9463697\cdot 10^{-4} & -1.9813326\cdot 10^{-3} & +9.8671149\cdot 10^{-4} & +9.1127306\cdot 10^{-4} \end{bmatrix} (\text{km/s})^{-3},$$

(19.28)

- the model-related Hessians of the arclength-related Hamiltonian,

$$H_{\mathbf{mm}}(P) = \begin{bmatrix} +1.9316689\cdot 10^{-4} & +2.2864973\cdot 10^{-4} & +1.1767691\cdot 10^{-5} & +2.9318832\cdot 10^{-4} \\ +2.2864973\cdot 10^{-4} & +4.7776726\cdot 10^{-4} & +2.4792321\cdot 10^{-4} & +1.7637944\cdot 10^{-3} \\ +1.1767691\cdot 10^{-5} & +2.4792321\cdot 10^{-4} & +2.6507562\cdot 10^{-4} & +1.6184596\cdot 10^{-3} \\ +2.9318832\cdot 10^{-4} & +1.7637944\cdot 10^{-3} & +1.6184596\cdot 10^{-3} & +1.3522533\cdot 10^{-2} \end{bmatrix} (\text{km/s})^{-5},$$

$$H_{\mathbf{mm}}(S_A) = \begin{bmatrix} +1.7804213\cdot 10^{-4} & -2.7198404\cdot 10^{-4} & +6.9522805\cdot 10^{-5} & +2.1679589\cdot 10^{-4} \\ -2.7198404\cdot 10^{-4} & +2.8749120\cdot 10^{-4} & +4.7788876\cdot 10^{-5} & -5.8262593\cdot 10^{-5} \\ +6.9522805\cdot 10^{-5} & +4.7788876\cdot 10^{-5} & -1.5811755\cdot 10^{-4} & -2.4368816\cdot 10^{-4} \\ +2.1679589\cdot 10^{-4} & -5.8262593\cdot 10^{-5} & -2.4368816\cdot 10^{-4} & -2.1872015\cdot 10^{-4} \end{bmatrix} (\text{km/s})^{-5},$$

$$H_{\mathbf{mm}}(S_B) = \begin{bmatrix} -6.6522865\cdot 10^{-5} & -1.8043291\cdot 10^{-5} & +8.3964057\cdot 10^{-5} & +1.4101125\cdot 10^{-3} \\ -1.8043291\cdot 10^{-5} & +4.3701438\cdot 10^{-4} & -4.1737027\cdot 10^{-4} & -2.3879390\cdot 10^{-3} \\ +8.3964057\cdot 10^{-5} & -4.1737027\cdot 10^{-4} & +3.3240908\cdot 10^{-4} & +9.7952971\cdot 10^{-4} \\ +1.4101125\cdot 10^{-3} & -2.3879390\cdot 10^{-3} & +9.7952971\cdot 10^{-4} & +5.6801289\cdot 10^{-3} \end{bmatrix} (\text{km/s})^{-5},$$

(19.29)

- the model-related gradients of the ray velocity,

$$\nabla_{\mathbf{m}} v_{\text{ray}}(P) = \begin{bmatrix} +7.6410683\cdot 10^{-3} & +1.7265263\cdot 10^{-2} & +9.7528675\cdot 10^{-3} & +8.9057943\cdot 10^{-2} \end{bmatrix} (\text{km/s})^{-1},$$

$$\nabla_{\mathbf{m}} v_{\text{ray}}(S_A) = \begin{bmatrix} +7.8705485\cdot 10^{-3} & -1.8937528\cdot 10^{-2} & +1.1391518\cdot 10^{-2} & +2.5844136\cdot 10^{-3} \end{bmatrix} (\text{km/s})^{-1},$$

$$\nabla_{\mathbf{m}} v_{\text{ray}}(S_B) = \begin{bmatrix} +9.7060263\cdot 10^{-3} & -1.9334558\cdot 10^{-2} & +9.6286866\cdot 10^{-3} & +8.8925312\cdot 10^{-2} \end{bmatrix} (\text{km/s})^{-1},$$

(19.30)

- for model-related Hessians of the ray velocity,

$$\nabla_{\mathbf{m}}\nabla_{\mathbf{m}} v_{\text{ray}}(P) = \begin{bmatrix} -2.0556231 \cdot 10^{-4} & -1.8372495 \cdot 10^{-4} & +5.4808073 \cdot 10^{-5} & +5.1314962 \cdot 10^{-4} \\ -1.8372495 \cdot 10^{-4} & -3.6063794 \cdot 10^{-4} & -1.7293516 \cdot 10^{-4} & -1.1951304 \cdot 10^{-4} \\ +5.4808073 \cdot 10^{-5} & -1.7293516 \cdot 10^{-4} & -2.6533219 \cdot 10^{-4} & -7.8999324 \cdot 10^{-4} \\ +5.1314962 \cdot 10^{-4} & -1.1951304 \cdot 10^{-4} & -7.8999324 \cdot 10^{-4} & -4.7887107 \cdot 10^{-3} \end{bmatrix} (\text{km}/\text{s})^{-3}$$

$$\nabla_{\mathbf{m}}\nabla_{\mathbf{m}} v_{\text{ray}}(S_A) = \begin{bmatrix} -2.1472900 \cdot 10^{-4} & +2.1390164 \cdot 10^{-4} & +5.3453083 \cdot 10^{-5} & +2.2838676 \cdot 10^{-5} \\ +2.1390164 \cdot 10^{-4} & -4.1649702 \cdot 10^{-4} & +1.9147937 \cdot 10^{-4} & -8.2522857 \cdot 10^{-5} \\ +5.3453083 \cdot 10^{-5} & +1.9147937 \cdot 10^{-4} & -3.0772757 \cdot 10^{-4} & +6.6224391 \cdot 10^{-5} \\ +2.2838676 \cdot 10^{-5} & -8.2522857 \cdot 10^{-5} & +6.6224391 \cdot 10^{-5} & +5.8754274 \cdot 10^{-4} \end{bmatrix} (\text{km}/\text{s})^{-3}$$

$$\nabla_{\mathbf{m}}\nabla_{\mathbf{m}} v_{\text{ray}}(S_B) = \begin{bmatrix} -2.6517448 \cdot 10^{-4} & +2.0915783 \cdot 10^{-4} & +5.4738655 \cdot 10^{-5} & +4.9249128 \cdot 10^{-4} \\ +2.0915783 \cdot 10^{-4} & -4.1840506 \cdot 10^{-4} & +2.0924354 \cdot 10^{-4} & +3.0226709 \cdot 10^{-4} \\ +5.4738655 \cdot 10^{-5} & +2.0924354 \cdot 10^{-4} & -2.6271071 \cdot 10^{-4} & -7.8962743 \cdot 10^{-4} \\ +4.9249128 \cdot 10^{-4} & +3.0226709 \cdot 10^{-4} & -7.8962743 \cdot 10^{-4} & -4.6702001 \cdot 10^{-3} \end{bmatrix} (\text{km}/\text{s})^{-3}$$

(19.31)

## 20. VALIDATION WITH NUMERICAL TESTS

We assumed that each component of the stiffness tensor varies quadratically in a proximity of the reference point, where all of them are defined, along with their spatial gradients and Hessians. Thus, we can compute the ray velocity values (invert for the slowness vector and then apply equation 1.4) in a stencil of neighbor points, where the medium properties differ slightly from those at the central (reference) point and establish the spatial gradient and Hessian by the second-order differences. Similarly, we can compute the ray velocity magnitude for slightly different ray directions and establish numerically the directional gradient and Hessian. Finally, we can change the location and direction simultaneously for numerical computation of the mixed Hessian.

In all cases, we present the (signed) relative error,

$$E = \frac{N - A}{A} \quad . \tag{20.1}$$

For the first derivatives, diagonal and off-diagonal second derivatives, the second-order central difference approximation read, respectively,

$$\frac{\partial f}{\partial x} = \frac{f(x+\Delta) - f(x-\Delta)}{2\Delta} + O(\Delta^2)$$

$$\frac{\partial^2 f}{\partial x^2} = \frac{f(x+\Delta) - 2f(x) + f(x-\Delta)}{\Delta^2} + O(\Delta^2)$$

$$\frac{\partial^2 f}{\partial x_1 \partial x_2} = \frac{f(x_1-\Delta_1, x_2-\Delta_2) - f(x_1+\Delta_1, x_2-\Delta_2) - f(x_1-\Delta_1, x_2+\Delta_2) + f(x_1+\Delta_1, x_2+\Delta_2)}{4\Delta_1 \Delta_2}$$

(20.2)

We used the spatial resolution $\Delta = 5 \cdot 10^{-5}$ km and the directional (angular) resolution $10^{-5}$ radian for all cases except the mixed Hessian of the ray velocity, where the angular resolution was $10^{-4}$ radian. Note that the goal is not to achieve the best numerical accuracy (for this, the fourth-order differences will work). The aim of this section is to make sure that our analytic relationships are correct. The computed relative numerical errors are as follows,

- for the spatial gradients of the ray velocity,

$$\begin{aligned}
E\left[\nabla_{\mathbf{x}} v_{\text{ray}}(P)\right] &= \begin{bmatrix} +4.84 \cdot 10^{-8} & +1.27 \cdot 10^{-6} & +4.24 \cdot 10^{-7} \end{bmatrix} , \\
E\left[\nabla_{\mathbf{x}} v_{\text{ray}}(S_A)\right] &= \begin{bmatrix} +9.19 \cdot 10^{-8} & +7.07 \cdot 10^{-7} & +5.45 \cdot 10^{-9} \end{bmatrix} , \\
E\left[\nabla_{\mathbf{x}} v_{\text{ray}}(S_B)\right] &= \begin{bmatrix} +1.25 \cdot 10^{-7} & -7.39 \cdot 10^{-7} & -1.78 \cdot 10^{-7} \end{bmatrix} ,
\end{aligned} \tag{20.3}$$

- for the spatial Hessian of the ray velocity,

$$E\left[\nabla_{\mathbf{x}}\nabla_{\mathbf{x}}v_{\text{ray}}(P)\right] = \begin{bmatrix} -4.18 \cdot 10^{-5} & -8.12 \cdot 10^{-5} & +3.95 \cdot 10^{-5} \\ -8.12 \cdot 10^{-5} & +4.25 \cdot 10^{-6} & +1.70 \cdot 10^{-3} \\ +3.95 \cdot 10^{-5} & +1.70 \cdot 10^{-3} & -6.97 \cdot 10^{-4} \end{bmatrix} ,$$

$$E\left[\nabla_{\mathbf{x}}\nabla_{\mathbf{x}}v_{\text{ray}}(S_A)\right] = \begin{bmatrix} +2.61 \cdot 10^{-4} & +2.89 \cdot 10^{-4} & -8.94 \cdot 10^{-5} \\ +2.89 \cdot 10^{-4} & +1.54 \cdot 10^{-4} & -1.41 \cdot 10^{-3} \\ -8.94 \cdot 10^{-5} & -1.41 \cdot 10^{-3} & +5.15 \cdot 10^{-4} \end{bmatrix} , \quad (20.4)$$

$$E\left[\nabla_{\mathbf{x}}\nabla_{\mathbf{x}}v_{\text{ray}}(S_B)\right] = \begin{bmatrix} -4.65 \cdot 10^{-5} & +1.06 \cdot 10^{-4} & -2.86 \cdot 10^{-5} \\ +1.06 \cdot 10^{-4} & +1.23 \cdot 10^{-5} & -8.04 \cdot 10^{-5} \\ -2.86 \cdot 10^{-5} & -8.04 \cdot 10^{-5} & -1.07 \cdot 10^{-3} \end{bmatrix} ,$$

- for the directional gradients of the ray velocity (with the angular resolution $10^{-5}$ rad),

$$E\left[\nabla_{\mathbf{r}}v_{\text{ray}}(P)\right] = \begin{bmatrix} +1.26 \cdot 10^{-9} & +1.27 \cdot 10^{-9} & +1.26 \cdot 10^{-9} \end{bmatrix} ,$$

$$E\left[\nabla_{\mathbf{r}}v_{\text{ray}}(S_A)\right] = \begin{bmatrix} -2.94 \cdot 10^{-9} & +1.32 \cdot 10^{-9} & +4.03 \cdot 10^{-9} \end{bmatrix} , \quad (20.5)$$

$$E\left[\nabla_{\mathbf{r}}v_{\text{ray}}(S_B)\right] = \begin{bmatrix} -5.70 \cdot 10^{-10} & +2.63 \cdot 10^{-9} & +3.74 \cdot 10^{-9} \end{bmatrix} ,$$

- for the directional Hessians of the ray velocity,

$$E\left[\nabla_{\mathbf{r}}\nabla_{\mathbf{r}}v_{\text{ray}}(P)\right] = \begin{bmatrix} +4.96 \cdot 10^{-4} & +1.06 \cdot 10^{-3} & +7.01 \cdot 10^{-5} \\ +1.06 \cdot 10^{-3} & +2.82 \cdot 10^{-4} & +1.83 \cdot 10^{-4} \\ +7.01 \cdot 10^{-5} & +1.83 \cdot 10^{-4} & +1.89 \cdot 10^{-4} \end{bmatrix} ,$$

$$E\left[\nabla_{\mathbf{r}}\nabla_{\mathbf{r}}v_{\text{ray}}(S_A)\right] = \begin{bmatrix} -9.93 \cdot 10^{-4} & -1.99 \cdot 10^{-3} & -1.72 \cdot 10^{-4} \\ -1.99 \cdot 10^{-3} & -6.30 \cdot 10^{-4} & -5.49 \cdot 10^{-4} \\ -1.72 \cdot 10^{-4} & -5.49 \cdot 10^{-4} & -4.91 \cdot 10^{-4} \end{bmatrix} , \quad (20.6)$$

$$E\left[\nabla_{\mathbf{r}}\nabla_{\mathbf{r}}v_{\text{ray}}(S_B)\right] = \begin{bmatrix} -4.64 \cdot 10^{-4} & -1.61 \cdot 10^{-3} & +1.75 \cdot 10^{-4} \\ -1.61 \cdot 10^{-3} & -2.66 \cdot 10^{-4} & +2.31 \cdot 10^{-4} \\ +1.75 \cdot 10^{-4} & +2.31 \cdot 10^{-4} & +3.25 \cdot 10^{-4} \end{bmatrix} ,$$

- for the mixed Hessians of the ray velocity,

$$E\left[\nabla_{\mathbf{r}}\nabla_{\mathbf{x}}v_{\text{ray}}(P)\right] = \begin{bmatrix} +1.77 \cdot 10^{-5} & +2.50 \cdot 10^{-6} & -2.11 \cdot 10^{-4} \\ -3.93 \cdot 10^{-5} & +5.65 \cdot 10^{-6} & +7.74 \cdot 10^{-6} \\ +6.07 \cdot 10^{-5} & -3.42 \cdot 10^{-5} & +1.11 \cdot 10^{-5} \end{bmatrix},$$

$$E\left[\nabla_{\mathbf{r}}\nabla_{\mathbf{x}}v_{\text{ray}}(S_A)\right] = \begin{bmatrix} -4.24 \cdot 10^{-5} & -1.50 \cdot 10^{-5} & +3.20 \cdot 10^{-4} \\ -1.14 \cdot 10^{-4} & -2.54 \cdot 10^{-5} & -1.35 \cdot 10^{-5} \\ -1.72 \cdot 10^{-4} & +9.63 \cdot 10^{-5} & -3.25 \cdot 10^{-5} \end{bmatrix}, \quad (20.7)$$

$$E\left[\nabla_{\mathbf{r}}\nabla_{\mathbf{x}}v_{\text{ray}}(S_B)\right] = \begin{bmatrix} -5.34 \cdot 10^{-5} & -7.51 \cdot 10^{-6} & +4.15 \cdot 10^{-4} \\ +8.80 \cdot 10^{-4} & -2.53 \cdot 10^{-5} & -1.01 \cdot 10^{-6} \\ -2.57 \cdot 10^{-4} & +5.29 \cdot 10^{-5} & -7.81 \cdot 10^{-5} \end{bmatrix}.$$

- for the model-related gradients of the ray velocity,

$$\begin{aligned} E\left[\nabla_{\mathbf{m}}v_{\text{ray}}(P)\right] &= \begin{bmatrix} +4.40 \cdot 10^{-7} & -4.40 \cdot 10^{-10} & +4.28 \cdot 10^{-7} & +8.58 \cdot 10^{-10} \end{bmatrix}, \\ E\left[\nabla_{\mathbf{m}}v_{\text{ray}}(S_A)\right] &= \begin{bmatrix} +4.51 \cdot 10^{-7} & -1.31 \cdot 10^{-9} & +4.19 \cdot 10^{-7} & -3.04 \cdot 10^{-8} \end{bmatrix}, \quad (20.8) \\ E\left[\nabla_{\mathbf{m}}v_{\text{ray}}(S_B)\right] &= \begin{bmatrix} +4.49 \cdot 10^{-7} & -2.90 \cdot 10^{-9} & +4.29 \cdot 10^{-7} & -8.13 \cdot 10^{-10} \end{bmatrix}, \end{aligned}$$

- for the model-related Hessians of the ray velocity,

$$E\left[\nabla_{\mathbf{m}}\nabla_{\mathbf{m}}v_{\text{ray}}(P)\right]=\begin{bmatrix}+3.44\cdot10^{-6} & +2.78\cdot10^{-6} & +4.54\cdot10^{-6} & +2.58\cdot10^{-6}\\ +2.78\cdot10^{-6} & -6.56\cdot10^{-5} & +2.74\cdot10^{-6} & -2.87\cdot10^{-5}\\ +4.54\cdot10^{-6} & +2.74\cdot10^{-6} & +3.38\cdot10^{-6} & +2.50\cdot10^{-6}\\ +2.58\cdot10^{-6} & -2.87\cdot10^{-5} & +2.50\cdot10^{-6} & -1.29\cdot10^{-5}\end{bmatrix}(\text{km}/\text{s})^{-3},$$

$$E\left[\nabla_{\mathbf{m}}\nabla_{\mathbf{m}}v_{\text{ray}}(S_A)\right]=\begin{bmatrix}+3.62\cdot10^{-6} & +3.10\cdot10^{-6} & +4.41\cdot10^{-6} & -3.19\cdot10^{-6}\\ +3.10\cdot10^{-6} & +1.42\cdot10^{-4} & +2.75\cdot10^{-6} & +9.51\cdot10^{-5}\\ +4.41\cdot10^{-6} & +2.75\cdot10^{-6} & +3.38\cdot10^{-6} & -7.00\cdot10^{-7}\\ -3.19\cdot10^{-6} & +9.51\cdot10^{-5} & -7.00\cdot10^{-7} & -2.27\cdot10^{-4}\end{bmatrix}(\text{km}/\text{s})^{-3},$$

$$E\left[\nabla_{\mathbf{m}}\nabla_{\mathbf{m}}v_{\text{ray}}(S_B)\right]=\begin{bmatrix}+2.96\cdot10^{-3} & +6.63\cdot10^{-4} & -3.88\cdot10^{-4} & +8.94\cdot10^{-4}\\ +6.63\cdot10^{-4} & -1.68\cdot10^{-4} & -1.38\cdot10^{-3} & +1.90\cdot10^{-3}\\ -3.88\cdot10^{-4} & -1.38\cdot10^{-3} & -3.68\cdot10^{-3} & -2.10\cdot10^{-3}\\ +8.94\cdot10^{-4} & +1.90\cdot10^{-3} & -2.10\cdot10^{-3} & -1.54\cdot10^{-3}\end{bmatrix}(\text{km}/\text{s})^{-3},$$

(20.9)

## 21. DISCUSSION: DO WE NEED TO COMPUTE DERIVATIVES ANALYTICALLY?

As we show in the numerical test section, the numerical difference schemes are simple and deliver the ray velocity derivatives directly, without the need to compute first analytically the derivatives of the reference and arclength-related Hamiltonians. However, any finite difference approach requires computing the function at the neighbor points (at the nodes of the stencil). The simplest, second-order central difference scheme for the first and second derivatives requires a spatial stencil with 15 nodes (eight at the vertices of the box, six at the face centers and one at the central node), and angular stencil with eight additional nodes and a number of stencil nodes to approximate the derivatives wrt the medium properties. At each node, the slowness vector inversion should be performed to establish the ray velocity. However, the low efficiency of the numerical approach is not the most essential issue. The slowness inversion is an iterative procedure resulting in an approximate solution. While the slowness components and the ray velocity magnitude can be obtained with a very good accuracy, and the ray velocity first derivative can be still estimated with a reasonable accuracy as well, this is not so for the second derivatives of the ray velocity, especially for the shear waves with non-convex ray (group) velocity surfaces. Approximating derivatives with finite differences is not a "healthy" operation in itself, and in particular, in this problem where we apply an approximation to another approximation. The choice of the finite difference resolution (spatial, $\Delta x_i$, directional $\Delta \theta_{\text{ray}}$ and $\Delta \psi_{\text{ray}}$, and medium properties, $\Delta m_i$) becomes challenging as well. The absolute error of the second-order finite difference scheme is $O(\Delta^2)$ but this is only the case when there are no round-off errors due to a limited number of machine digits. In reality, keeping $\Delta$ very small requires subtracting very close values (and even closer for the second derivatives), resulting in a considerable accuracy decrease. On the other hand, increasing $\Delta$ obviously reduces the "theoretic" accuracy of the scheme. Moreover, the most essential problem is that by increasing $\Delta$ (spatial, directional or related to medium properties) we can switch to "an alien event", which may be a different leaf of the shear-wave ray velocity surface leading to completely wrongful values. As a result, each parameter or pair of parameters related to the values of the off-diagonal second derivatives has a narrow "window" of its feasible resolution range $\Delta_{\min} \leq \Delta \leq \Delta_{\max}$. For different components of the Hessian matrix, these windows may not necessarily overlap, and different resolutions should be applied when computing these components. In the most "accidental" case, the resolution window may not exist at all, or the result may become dependent on the sequence of equal priority operations of the difference scheme. An example of such dependency is the numerator of the divided difference operator in the third equation of set 20.2, which can be arranged as,

$$f(x_1-\Delta_1, x_2-\Delta_2) - f(x_1+\Delta_1, x_2-\Delta_2) - f(x_1-\Delta_1, x_2+\Delta_2) + f(x_1+\Delta_1, x_2+\Delta_2) ,$$
$$\text{or} \tag{21.1}$$
$$f(x_1-\Delta_1, x_2-\Delta_2) - f(x_1-\Delta_1, x_2+\Delta_2) - f(x_1+\Delta_1, x_2-\Delta_2) + f(x_1+\Delta_1, x_2+\Delta_2) .$$

Thus, the analytical derivatives of the ray velocity are essentially more accurate and stable than their finite difference approximations. The accuracy of the analytical derivatives depends only on the accuracy of the slowness inversion and the spatial derivatives of the medium (model) properties, which, although in real case applications are computed numerically, in this study they are considered the input data.

In summary, the proposed analytical derivatives are computed in four steps: a) Slowness inversion, b) Derivatives of the reference Hamiltonian, c) Derivatives of the arclength-related Hamiltonian, and d) Derivatives of the ray velocity. In Part II of this study, we formulate the derivatives for models with a higher anisotropic symmetry (polar anisotropy – TTI) where we adjust the reference Hamiltonian in the first two steps and considerably improve the computational performance of the proposed approach. The two other steps remain the same, as they are generic, governed by identical relationships for any anisotropic media.

## 22. COMPUTATIONAL TEST 2

In this test we do not establish the derivatives of the ray velocity magnitude; we only perform the slowness inversion and compare the signs of coefficient $\alpha$ in equation 6.2 and the mean curvature. Consider another triclinic medium, introduced by Igel et al. (1995) and later used by Saenger and Bohlen (2004) and Köhn et al. (2015). Its density-normalized stiffness matrix in the Voigt notation reads,

$$\mathbf{C} = \begin{bmatrix} +5.0000 & +1.7500 & +1.2500 & -2.5000 & +0.0500 & +0.1500 \\ +1.7500 & +4.0000 & +0.7500 & +0.1000 & -0.0500 & -0.0750 \\ +1.2500 & +0.7500 & +3.0000 & +0.5000 & +0.2000 & +0.1200 \\ -2.5000 & +0.1000 & +0.5000 & +2.5000 & +0.1750 & +0.2625 \\ +0.0500 & -0.0500 & +0.2000 & +0.1750 & +2.0000 & -0.5000 \\ +0.1500 & -0.0750 & +0.1200 & +0.2625 & -0.5000 & +1.5000 \end{bmatrix}. \quad (22.1)$$

Assume the ray direction vector,

$$\mathbf{r} = \frac{[5 \quad 6 \quad 8]}{5\sqrt{5}} = [0.4472136 \quad 0.53665631 \quad 0.71554175] \quad . \quad (22.2)$$

The slowness inversion for these data leads to a qP solution and six qS solutions, total seven, listed in Table 5. In Table 6, we list the minor, major, and mean eigenvalues, along with the coefficient $\alpha$ for all solutions. The inversion solutions are sorted by descending ray velocity magnitude. The first solution corresponds to a compressional wave, the others – to shear waves.

Table 5. Inverted slowness vector for example 2

| # | $p_1$, s/km | $p_2$, s/km | $p_3$, s/km | $v_{\text{phs}}$, km/s | $v_{\text{ray}}$, km/s | $\Delta\vartheta$, deg |
|---|---|---|---|---|---|---|
| 1 | 0.34085086 | 0.23796514 | 0.27892780 | 1.99757467 | 2.08453574 | 16.607983 |
| 2 | 0.45478451 | 0.17682680 | 0.32080596 | 1.71243639 | 1.89454546 | 25.327521 |
| 3 | 0.53132424 | 0.11778196 | 0.31926646 | 1.58488933 | 1.88938649 | 32.982374 |
| 4 | 0.27891578 | 0.59341395 | 0.45936882 | 1.24906955 | 1.29551828 | 15.388950 |
| 5 | 0.25878957 | 0.65852196 | 0.42718589 | 1.20991361 | 1.29064979 | 20.373135 |
| 6 | 0.42767426 | 0.71341951 | 0.28412679 | 1.13768610 | 1.28629372 | 27.813832 |
| 7 | 0.95822553 | −0.24471372 | 1.14712665 | 0.66024503 | 0.89443731 | 42.424408 |

Table 6. Ratio between $H_{\mathbf{p}}^{\bar{\tau}}$ and **r** vs. curvatures of the slowness surface for example 2

| # | Curvature, km/s | | | Ratio |
|---|---|---|---|---|
| | Minor | Major | Mean | $\alpha$, km/s |
| 1 | +5.8756786 | +1.38104105 | +3.62835984 | +0.60480963 |
| 2 | −1.4790732 | +1.39905354 | −0.04000981 | −0.75725244 |
| 3 | −1.39432774 | −0.65548554 | −1.02490664 | −1.26735659 |
| 4 | −5.23881096 | +1.03870922 | −2.10005087 | +0.75965662 |
| 5 | +2.07729872 | −1.00756251 | +0.53486811 | +1.68395842 |
| 6 | +3.00571428 | +0.13393940 | +1.56982684 | +7.47070975 |
| 7 | +6.14183727 | +0.31905916 | +3.23044821 | +90.3238680 |

The case, where the mean curvature of the slowness surface and coefficient $\alpha$ relating the slowness gradient of the reference Hamiltonian, $H_{\mathbf{p}}^{\bar{\tau}}$, and the ray direction vector, **r**, have opposite signs (solution # 4), is highlighted in yellow. Recall that the sign of the Hamiltonian has to be changed when $\alpha$ proves to be negative.

## 23. CONCLUSIONS

In this paper we have presented an original and generic approach for computing all types of first and second partial derivatives of the ray velocity at any point along a ray path in general anisotropic elastic media. The derivatives are obtained for both quasi-compressional and quasi-shear waves. This set of partial derivatives construct the traveltime sensitivity kernels (matrices) which provide a deep insight to the sensitivity and reliability of seismic traveltimes and amplitudes to changes in the ray locations and directions, and to perturbations of the anisotropic model parameters. These derivatives are intensively used in different key modeling and inversion methods, such as ray bending and seismic tomography.

We start with an arclength-related Lagrangian representing the reciprocal ray velocity magnitude, which, in turn, is a function of the spatially varying anisotropic model parameters and the ray direction at each ray node. Since there is no explicit form for the ray velocity (and hence the arclength-related Lagrangian) as a function of the ray direction components, we reduce the problem to computing the derivatives of the corresponding arclength-related Hamiltonian that explicitly depends on the slowness vector components (rather than the ray direction components). We then connect the arclength-related Hamiltonian (vanishing, with the units of slowness) to a simpler, reference Hamiltonian (vanishing, unitless, related to a scaled traveltime flow parameter) obtained directly from the Christoffel equation. The reference Hamiltonian is used to invert for the slowness vector components and for the actual computation of the required ray velocity gradients and Hessians.

This part (Part I) provides the derivatives (gradients and Hessians) of the ray velocity magnitude for general anisotropic media, and it represents the theoretical basis for providing the same partial derivatives for the higher symmetry anisotropic media. The advantage of the proposed scheme is that the final derivative expressions are solely based on the derivatives of the reference

Hamiltonian which can be explicitly expressed wrt the slowness components for any type of anisotropic media. Hence, for obtaining the ray velocity derivatives for a specific anisotropic medium (e.g., monoclinic, orthorhombic, or polar anisotropic, where all of them may be tilted), only the reference Hamiltonian and its derivatives should be adjusted. In the next part (Part II), we provide the explicit derivatives for polar anisotropic media (TTI), for the coupled qP and qSV waves and for SH waves. The correctness of the derivations has been tested and confirmed by checking the consistency between the proposed analytical derivatives and the corresponding numerical ones.

## ACKNOWLEDGEMENT

The authors are grateful to Emerson for the financial and technical support of this study and for the permission to publish its results. Gratitude is extended to Anne-Laure Tertois, Alexey Stovas, Yuriy Ivanov, Yury Kligerman, Michael Slawinski, Mikhail Kochetov, and Beth Orshalimy for valuable remarks and comments that helped to improve the content and the style of this paper.

## APPENDIX A. PROPERTY OF THE ARCLENGTH-RELATED HAMILTONIAN

In this appendix, we prove equation 5.4, a property of the arclength-related Hamiltonian, used in our derivations, that connects its spatial gradient, $H_\mathbf{x}$, to the slowness gradient, $H_\mathbf{p} = \mathbf{r}$,

$$H_\mathbf{x} = H_\mathbf{xp} H_\mathbf{pp}^{-1} H_\mathbf{p} \qquad . \tag{A1}$$

Consider the inversion equation set 3.1, where for the arclength related Hamiltonian, the collinearity of the slowness gradient and the ray direction has been replaced by their equality,

$$H_{\mathbf{p}}(\mathbf{x},\mathbf{p}) = \mathbf{r} , \quad H(\mathbf{x},\mathbf{p}) = 0 . \tag{A2}$$

Since both the Hamiltonian and its slowness gradient are constant, the corresponding full derivatives wrt the location vector vanish,

$$\frac{dH_{\mathbf{p}}(\mathbf{x},\mathbf{p})}{d\mathbf{x}} = \frac{\partial H_{\mathbf{p}}}{\partial \mathbf{x}} + \frac{\partial H_{\mathbf{p}}}{\partial \mathbf{p}} \frac{\partial \mathbf{p}}{\partial \mathbf{x}} = 0 ,$$
$$\frac{dH(\mathbf{x},\mathbf{p})}{d\mathbf{x}} = \frac{\partial H}{\partial \mathbf{x}} + \frac{\partial H}{\partial \mathbf{p}} \frac{\partial \mathbf{p}}{\partial \mathbf{x}} , \tag{A3}$$

or, equivalently,

$$H_{\mathbf{px}} + H_{\mathbf{pp}} \cdot \frac{\partial \mathbf{p}}{\partial \mathbf{x}} = 0 , \quad H_{\mathbf{x}} + H_{\mathbf{p}} \cdot \frac{\partial \mathbf{p}}{\partial \mathbf{x}} = 0 . \tag{A3}$$

From the first equation of set A3, we compute the matrix $\partial \mathbf{p}/\partial \mathbf{x}$

$$\frac{\partial \mathbf{p}}{\partial \mathbf{x}} = -H_{\mathbf{pp}}^{-1} H_{\mathbf{px}} . \tag{A4}$$

Next, we introduce equation A4 into the second equation of set A3,

$$H_{\mathbf{x}} = H_{\mathbf{p}} H_{\mathbf{pp}}^{-1} H_{\mathbf{px}} . \tag{A5}$$

In this study, we use the tensor algebra convention that does not distinguish between the row and column vectors: a one-dimensional object is assumed equal to its transpose. Applying this rule to equation A5, we obtain,

$$H_{\mathbf{x}} = H_{\mathbf{x}}^T = \left( H_{\mathbf{p}} \cdot H_{\mathbf{pp}}^{-1} H_{\mathbf{px}} \right)^T = H_{\mathbf{xp}} \cdot H_{\mathbf{pp}}^{-1} H_{\mathbf{p}} = H_{\mathbf{xp}} \cdot H_{\mathbf{pp}}^{-1} \mathbf{r} . \tag{A6}$$

## APPENDIX B. DIRECTIONAL GRADIENT OF THE RAY VELOCITY

Partial derivatives $\partial v_{\text{ray}} / \partial \mathbf{r}$ characterize a change in the ray velocity magnitude due to infinitesimal change of a ray direction component $r_i$ while the two other components remain fixed. However, changing a single component of the ray direction ruins the normalization $\mathbf{r} \cdot \mathbf{r} = 1$. The remedy is to keep the direction fixed after a change in a single component and adjust all three components, keeping the vector at a unit length. As a result, the gradient vector obtains a correction,

$$\nabla_{\mathbf{r}} v_{\text{ray}} = \mathbf{T} \frac{\partial v_{\text{ray}}}{\partial \mathbf{r}} \quad , \tag{B1}$$

where the transformation matrix $\mathbf{T} = \mathbf{I} - \mathbf{r} \otimes \mathbf{r}$ is defined in equation 5.5. The proof of equation B1 is given by Ravve and Koren (2019) and Koren and Ravve (2021). The key idea is that we distinguish between the normalized, $\nabla_{\mathbf{r}} v_{\text{ray}}$, and the non-normalized, $\partial v_{\text{ray}} / \partial \mathbf{r}$, directional gradients of the ray velocity. The non-normalized gradient reads,

$$\frac{\partial v_{\text{ray}}}{\partial \mathbf{r}} = \frac{\partial}{\partial \mathbf{r}} \frac{1}{\mathbf{p} \cdot \mathbf{r}} = -\frac{(\nabla_{\mathbf{r}} \mathbf{p})^T \mathbf{r} + (\nabla_{\mathbf{r}} \mathbf{r})^T \mathbf{p}}{(\mathbf{p} \cdot \mathbf{r})^2} = -v_{\text{ray}}^2 \left( \mathbf{p}_{\mathbf{r}}^T \cdot \mathbf{r} + \mathbf{I} \mathbf{p} \right) \quad , \tag{B2}$$

where,

$$\mathbf{p} = L_{\mathbf{r}} \quad , \quad \mathbf{p}_{\mathbf{r}} = L_{\mathbf{rr}} \quad , \quad L_{\mathbf{rr}} = L_{\mathbf{rr}}^T \quad , \quad L_{\mathbf{rr}} \cdot \mathbf{r} = 0 \quad . \tag{B3}$$

Thus, $L_{\mathbf{rr}}$ is a symmetric matrix (tensor), and the ray direction $\mathbf{r}$ is one of its eigenvectors, with the corresponding eigenvalue zero. Equation B2 simplifies to,

$$\frac{\partial v_{\text{ray}}}{\partial \mathbf{r}} = -v_{\text{ray}}^2 \mathbf{p} \quad . \tag{B4}$$

Combining equations B1 and B4, we obtain the normalized directional gradient,

$$\nabla_{\mathbf{r}} v_{\text{ray}} = -v_{\text{ray}}^2 (\mathbf{I} - \mathbf{r} \otimes \mathbf{r}) \mathbf{p} = -v_{\text{ray}}^2 \mathbf{p} + (\mathbf{r} \otimes \mathbf{r}) \mathbf{p} \quad . \tag{B5}$$

Next, we apply a general algebraic formula,

$$(\mathbf{a} \otimes \mathbf{b}) \mathbf{c} = (\mathbf{b} \cdot \mathbf{c}) \mathbf{a} \quad , \tag{B6}$$

and obtain,

$$\nabla_{\mathbf{r}} v_{\text{ray}} = -v_{\text{ray}}^2 \mathbf{p} + v_{\text{ray}}^2 (\mathbf{p} \cdot \mathbf{r}) \mathbf{r} = -v_{\text{ray}}^2 \mathbf{p} + v_{\text{ray}} \mathbf{r} = \mathbf{v}_{\text{ray}} - v_{\text{ray}}^2 \mathbf{p} \quad . \tag{B7}$$

It follows from equation B7 that the directional gradient of the ray velocity is normal to the ray, $\nabla_{\mathbf{r}} v_{\text{ray}} \cdot \mathbf{r} = 0$. Equation B7 can be arranged as,

$$\nabla_{\mathbf{r}} v_{\text{ray}} = -v_{\text{ray}}^2 \left( \mathbf{p} - \frac{\mathbf{r}}{v_{\text{ray}}} \right) = -v_{\text{ray}}^2 \left[ \mathbf{p} - \mathbf{r} (\mathbf{p} \cdot \mathbf{r}) \right] \quad . \tag{B8}$$

The second item in brackets is the component of the slowness vector along the ray direction. The whole expression in brackets is the difference between the full slowness vector and the ray-tangent component; it represents the slowness component in a plane normal to the ray,

$$\mathbf{p} - \mathbf{r} (\mathbf{p} \cdot \mathbf{r}) = \mathbf{r} \times \mathbf{p} \times \mathbf{r} \quad . \tag{B9}$$

Combining equations B8 and B9, we obtain the directional gradient of the ray velocity,

$$\nabla_{\mathbf{r}} v_{\text{ray}} = -v_{\text{ray}}^2 \, \mathbf{r} \times \mathbf{p} \times \mathbf{r} = -\mathbf{v}_{\text{ray}} \times \mathbf{p} \times \mathbf{v}_{\text{ray}} \qquad (B10)$$

**APPENDIX C. DIRECTIONAL HESSIAN OF THE RAY VELOCITY**

Like the directional gradient, the directional Hessian of the ray velocity magnitude is also computed in two stages, obtaining first the non-normalized and then the normalized objects. The non-normalized directional Hessian reads,

$$\frac{\partial^2 v_{\text{ray}}}{\partial \mathbf{r}^2} = -\frac{\partial}{\partial \mathbf{r}}\left(v_{\text{ray}}^2 \, \mathbf{p}\right) = -2v_{\text{ray}} \mathbf{p} \otimes \frac{\partial v_{\text{ray}}}{\partial \mathbf{r}} - v_{\text{ray}}^2 \frac{\partial \mathbf{p}}{\partial \mathbf{r}} \qquad (C1)$$

Combining equations B4 and C1, and taking into account that,

$$L_{\mathbf{r}} = \mathbf{p} \, , \quad L_{\mathbf{rr}} = \frac{\partial \mathbf{p}}{\partial \mathbf{r}} \, , \qquad (C2)$$

we obtain,

$$\frac{\partial^2 v_{\text{ray}}}{\partial \mathbf{r}^2} = 2v_{\text{ray}}^3 \mathbf{p} \otimes \mathbf{p} - v_{\text{ray}}^2 L_{\mathbf{rr}} \qquad (C3)$$

Next, we apply equation from set 5.2, $L_{\mathbf{rr}} = H_{\mathbf{pp}}^{-1} - \lambda_{\mathbf{r}} \mathbf{r} \otimes \mathbf{r} = \mathbf{T} H_{\mathbf{pp}}^{-1}$, where $\lambda_{\mathbf{r}}$ is the eigenvalue of the inverse Hessian of the arclength-related Hamiltonian, $H_{\mathbf{pp}}^{-1}$ corresponding to the eigenvector $\mathbf{r}$.

The normalization of the Hessian differs from the normalization of the gradient. To obtain the normalized Hessian, we need not only the non-normalized Hessian, but also the non-normalized gradient, $\partial v_{ray} / \partial \mathbf{r}$. Still, it is a linear operator, and the components of the matrix in this operator depend on the ray direction alone,

$$\nabla_\mathbf{r} \nabla_\mathbf{r} v_{ray} = \hat{\mathbf{E}} \frac{\partial v_{ray}}{\partial \mathbf{r}} + \mathbf{T} \frac{\partial^2 v_{ray}}{\partial \mathbf{r}^2} \mathbf{T} \quad , \tag{C4}$$

where $\mathbf{T}$ is the second-order symmetric transformation tensor defined in equation 1.9, and $\hat{\mathbf{E}}$ is a third-order super-symmetric tensor (i.e., its three indices can be swapped in any order), defined as,

$$\hat{\mathbf{E}} = -\mathbf{T} \otimes \mathbf{r} - \mathbf{r} \otimes \mathbf{T} - (\mathbf{T} \otimes \mathbf{r})^{T\{1,3,2\}} \quad . \tag{C5}$$

Notation $T\{1,3,2\}$ means that the third-order tensor is transposed, the first index is unchanged, while the second and the third are swapped.

Remark: The linear operator normalizing the directional Hessian in equation C4, along with the definition of its gradient-related third-order tensor $\hat{\mathbf{E}}$ in equation C5, are equivalent to those of equations 42-47 in Ravve and Koren (2019), but this operator is presented here in a compact form, involving only physical vectors and tensors.

Taking into account that $\mathbf{T} = \mathbf{I} - \mathbf{r} \otimes \mathbf{r}$, equation C5 simplifies to,

$$\hat{\mathbf{E}} = 3 \mathbf{r} \otimes \mathbf{r} \otimes \mathbf{r} - \mathbf{I} \otimes \mathbf{r} - \mathbf{r} \otimes \mathbf{I} - (\mathbf{I} \otimes \mathbf{r})^{T\{1,3,2\}} \quad , \tag{C6}$$

where $\mathbf{I}$ is the second-order identity tensor (matrix), resulting in,

$$\hat{E}_{ijk} = 3 r_i r_j r_k - \delta_{ij} r_k - \delta_{jk} r_i - \delta_{ki} r_j \quad , \tag{C7}$$

where $\delta_{lm}$ is the Kronecker delta.

Due to its complete symmetry (obvious from equation C7), tensor $\hat{\mathbf{E}}$ has only ten distinct components (out of the twenty-seven), defined, for example, by three orientation parameters (e.g., Euler's angles) and seven rotational invariants (e.g., the non-negative eigenvalues).

The first term on the right-hand side of equation C4 leads to a symmetric matrix,

$$\hat{\mathbf{E}} \frac{\partial v_{\text{ray}}}{\partial \mathbf{r}} = -v_{\text{ray}}^2 \hat{\mathbf{E}} \mathbf{p} = -v_{\text{ray}}^2 \left[ 3\mathbf{r} \otimes \mathbf{r} (\mathbf{p} \cdot \mathbf{r}) - \mathbf{I} (\mathbf{p} \cdot \mathbf{r}) - \mathbf{p} \otimes \mathbf{r} - \mathbf{r} \otimes \mathbf{p} \right]$$
$$= v_{\text{ray}}^2 (\mathbf{p} \otimes \mathbf{r} + \mathbf{r} \otimes \mathbf{p}) - 2v_{\text{ray}} \mathbf{r} \otimes \mathbf{r} + v_{\text{ray}} \mathbf{T} \quad . \tag{C8}$$

The second term on the right-hand side of equation C4, with the use of equations B4 and C3, simplifies to,

$$\mathbf{T} \frac{\partial^2 v_{\text{ray}}}{\partial \mathbf{r}^2} \mathbf{T} = 2v_{\text{ray}}^3 \mathbf{T} (\mathbf{p} \otimes \mathbf{p}) \mathbf{T} - v_{\text{ray}}^2 \mathbf{T} L_{\mathbf{rr}} \mathbf{T} =$$
$$2v_{\text{ray}}^3 (\mathbf{I} - \mathbf{r} \otimes \mathbf{r}) \mathbf{p} \otimes \mathbf{p} (\mathbf{I} - \mathbf{r} \otimes \mathbf{r}) - v_{\text{ray}}^2 (\mathbf{I} - \mathbf{r} \otimes \mathbf{r}) L_{\mathbf{rr}} (\mathbf{I} - \mathbf{r} \otimes \mathbf{r}) = 2v_{\text{ray}}^3 \mathbf{p} \otimes \mathbf{p} \tag{C9}$$
$$-2v_{\text{ray}}^2 (\mathbf{p} \otimes \mathbf{r} + \mathbf{r} \otimes \mathbf{p}) + 2v_{\text{ray}} \mathbf{r} \otimes \mathbf{r} - v_{\text{ray}}^2 \left[ L_{\mathbf{rr}} - \mathbf{r} \otimes (\mathbf{r} L_{\mathbf{rr}}) - (L_{\mathbf{rr}} \mathbf{r}) \otimes \mathbf{r} + (\mathbf{r} L_{\mathbf{rr}} \mathbf{r}) \mathbf{r} \otimes \mathbf{r} \right] \quad ,$$

which is a symmetric matrix as well. Recall that the ray direction is the eigenvector of the symmetric matrix $L_{\mathbf{rr}}$, with the corresponding zero eigenvalue, $\mathbf{r} L_{\mathbf{rr}} = L_{\mathbf{rr}} \mathbf{r} = 0$. Thus, only the first item in the square brackets on the right-hand side of equation C9 does not vanish. Combining equations C4, C8 and C9, we obtain the final expression for the normalized directional Hessian of the ray velocity,

$$\nabla_{\mathbf{r}} \nabla_{\mathbf{r}} v_{\text{ray}} = 2v_{\text{ray}}^3 \mathbf{p} \otimes \mathbf{p} - v_{\text{ray}}^2 (\mathbf{p} \otimes \mathbf{r} + \mathbf{r} \otimes \mathbf{p}) - v_{\text{ray}} (v_{\text{ray}} L_{\mathbf{rr}} - \mathbf{T}) \quad , \tag{C10}$$

where $L_{\mathbf{rr}}$ can be computed with the equation set 5.5.

For isotropic media, where $\mathbf{p} = \mathbf{r}/v_{\text{ray}}$ and $L_{\mathbf{rr}} = \mathbf{T}/v_{\text{ray}}$, this directional Hessian vanishes.

### APPENDIX D. MIXED HESSIAN OF THE RAY VELOCITY MAGNITUDE

We start from the non-normalized mixed Hessians of the ray velocity magnitude,

$$\frac{\partial^2 v_{\text{ray}}}{\partial \mathbf{x} \partial \mathbf{r}} = \frac{2}{v_{\text{ray}}} \frac{\partial v_{\text{ray}}}{\partial \mathbf{x}} \otimes \frac{\partial v_{\text{ray}}}{\partial \mathbf{r}} - v_{\text{ray}}^2 \mathbf{p}_{\mathbf{x}}^T \quad,$$
$$\frac{\partial^2 v_{\text{ray}}}{\partial \mathbf{r} \partial \mathbf{x}} = \frac{2}{v_{\text{ray}}} \frac{\partial v_{\text{ray}}}{\partial \mathbf{r}} \otimes \frac{\partial v_{\text{ray}}}{\partial \mathbf{x}} - v_{\text{ray}}^2 \mathbf{p}_{\mathbf{x}} \quad. \tag{D1}$$

The normalized mixed Hessians are,

$$\nabla_{\mathbf{x}} \nabla_{\mathbf{r}} v_{\text{ray}} = \frac{\partial^2 v_{\text{ray}}}{\partial \mathbf{x} \partial \mathbf{r}} \mathbf{T} \quad, \qquad \nabla_{\mathbf{r}} \nabla_{\mathbf{x}} v_{\text{ray}} = \mathbf{T} \frac{\partial^2 v_{\text{ray}}}{\partial \mathbf{r} \partial \mathbf{x}} \quad. \tag{D2}$$

This leads to,

$$\nabla_{\mathbf{x}} \nabla_{\mathbf{r}} v_{\text{ray}} = \frac{2}{v_{\text{ray}}} \nabla_{\mathbf{x}} v_{\text{ray}} \otimes \nabla_{\mathbf{r}} v_{\text{ray}} - v_{\text{ray}}^2 L_{\mathbf{xr}} \mathbf{T} \quad,$$
$$\nabla_{\mathbf{r}} \nabla_{\mathbf{x}} v_{\text{ray}} = \frac{2}{v_{\text{ray}}} \nabla_{\mathbf{r}} v_{\text{ray}} \otimes \nabla_{\mathbf{x}} v_{\text{ray}} - v_{\text{ray}}^2 \mathbf{T} L_{\mathbf{rx}} \quad, \tag{D3}$$

where,

$$\nabla_{\mathbf{x}} v_{\text{ray}} = v_{\text{ray}}^2 H_{\mathbf{x}} \quad, \quad L_{\mathbf{xr}} = -H_{\mathbf{xp}} H_{\mathbf{pp}}^{-1} \quad, \quad L_{\mathbf{rx}} = -H_{\mathbf{pp}}^{-1} H_{\mathbf{px}} \quad. \tag{D4}$$

Note that,

$$(\mathbf{T} \cdot \mathbf{p}) \otimes H_{\mathbf{x}} = \mathbf{T} \cdot (\mathbf{p} \otimes H_{\mathbf{x}}) \quad \text{and} \quad H_{\mathbf{x}} \otimes (\mathbf{T} \cdot \mathbf{p}) = (H_{\mathbf{x}} \otimes \mathbf{p}) \cdot \mathbf{T} \quad , \tag{D5}$$

equation set D3 can be arranged as,

$$\begin{aligned} \nabla_{\mathbf{x}} \nabla_{\mathbf{r}} v_{\text{ray}} &= -v_{\text{ray}}^2 \left( 2 v_{\text{ray}} H_{\mathbf{x}} \otimes \mathbf{p} + L_{\mathbf{xr}} \right) \mathbf{T} \quad , \\ \nabla_{\mathbf{r}} \nabla_{\mathbf{x}} v_{\text{ray}} &= -v_{\text{ray}}^2 \mathbf{T} \left( 2 v_{\text{ray}} \mathbf{p} \otimes H_{\mathbf{x}} + L_{\mathbf{rx}} \right) \quad . \end{aligned} \tag{D6}$$

As mentioned, the two mixed Hessians of the ray velocity are transpose of each other, $\nabla_{\mathbf{r}} \nabla_{\mathbf{x}} v_{\text{ray}} = \left( \nabla_{\mathbf{x}} \nabla_{\mathbf{r}} v_{\text{ray}} \right)^T$.

**APPENDIX E. THE SIGN OF THE REFERENCE HAMILTONIAN**

In a general case, the Hamiltonian has to be defined as,

$$\begin{aligned} H^{\bar{\tau}} &= +\det[\mathbf{\Gamma} - \mathbf{I}] \quad \text{quasi-compressional wave} \quad , \\ H^{\bar{\tau}} &= \pm \det[\mathbf{\Gamma} - \mathbf{I}] \quad \text{quasi-shear wave} \quad , \end{aligned} \tag{E1}$$

otherwise (if we do not pick the right sign in the second equation of this set) the shear-wave direction of the slowness gradient $H_{\mathbf{p}}^{\bar{\tau}}$ may become opposite to the ray direction $\mathbf{r}$ (which is legitimate, but inconvenient). To explain where this rule comes from and when the upper or lower sign should be used in the second equation of set E1, consider a particular case of isotropic media; due to continuity of the Christoffel matrix, the conclusions will be valid for anisotropic media as well. For isotropic media, the Hamiltonian reduces to,

$$H^{\bar{\tau}} = \left( C_{11} p^2 - 1 \right) \left( C_{44} p^2 - 1 \right)^2 \quad , \tag{E2}$$

or alternatively,

$$H^{\bar{\tau}} = \left(p^2 v_P^2 - 1\right)\left(p^2 v_S^2 - 1\right)^2 \quad, \quad \text{where} \quad v_S = v_P \sqrt{1-f} \quad . \tag{E3}$$

Of course, for isotropic media, we might apply the Hamiltonian $H^{\bar{\tau}} = p^2 v_P^2 - 1$ for compressional waves and $H^{\bar{\tau}} = p^2 v_S^2 - 1$ for shear waves, but our goal now is to explore what happens to the Hamiltonian that we use in the case of "infinitesimal anisotropy".

For compressional waves, $p^2 = v_P^2$, and the slowness gradient becomes,

$$H_{\mathbf{p}}^{\bar{\tau}} = \frac{2\left(v_P^2 - v_S^2\right)^2 \mathbf{r}}{v_P^3} = 2 v_P f^2 \mathbf{r}, \quad \text{and} \quad H_{\mathbf{p}} = \mathbf{r} \quad . \tag{E4}$$

Factor $2 v_P f^2$ is positive; thus, for compressional waves, the sign is plus in the second equation of set E1. However, there is no definite conclusion for shear waves, where in the isotropic case, the gradient of the reference Hamiltonian vanishes, $H_{\mathbf{p}}^{\bar{\tau}} = 0$, while for the arclength-related Hamiltonian, the limits for the gradient are different from the left and from the right, $H_{\mathbf{p}} = \pm \mathbf{r}$ depending on whether the slowness magnitude $p$ approaches $v_S^{-1}$ from above or from below,

$$H_{\mathbf{p}} = \mathrm{sgn}\left(p^2 v_S^2 - 1\right) \mathbf{r} \quad , \quad \lim_{p \to (1/v_S)^-} H_{\mathbf{p}} = -\mathbf{r} \quad , \quad \lim_{p \to (1/v_S)^+} H_{\mathbf{p}} = +\mathbf{r} \quad . \tag{E5}$$

Thus, for the shear wave, the sign of the Hamiltonian is undefined, and to pick the right sign, we need to check $H_{\mathbf{p}}^{\bar{\tau}}$ that may prove to be either collinear or "anti-parallel" with $\mathbf{r}$. Another option is to modify the slowness inversion set as in equation set 6.2, $H_{\mathbf{p}}^{\bar{\tau}}(\mathbf{x},\mathbf{p}) = \alpha \mathbf{r}, H^{\bar{\tau}}(\mathbf{x},\mathbf{p}) = 0$.

Set 6.2 consists of four equations and four unknown variables: three slowness components and a scalar coefficient $\alpha$. If in the shear-wave solution we obtain a negative alpha, we pick the minus sign in the second equation of set E1. Then the slowness gradient of the arclength Hamiltonian is $H_{\mathbf{p}} = +\mathbf{r}$ for all cases.

# APPENDIX F. COMPUTING DIRECTIONAL DERIVATIVES WITH FINITE DIFFERENCES

To compute the directional derivatives, we follow the approach suggested by Ravve and Koren (2019). The ray velocity magnitude, $v_{\text{ray}}(\mathbf{x},\mathbf{r})$ can be computed numerically for any direction. The direction is normalized, $\mathbf{r} \cdot \mathbf{r} = 1$. In this Appendix, we compute the first and second directional derivatives, and the mixed derivatives of the ray velocity,

$$\frac{\partial v_{\text{ray}}}{\partial r_i} \quad , \quad \frac{\partial^2 v_{\text{ray}}}{\partial r_i \, \partial r_j} \quad \text{and} \quad \frac{\partial^2 v_{\text{ray}}}{\partial x_i \, \partial r_j} \quad , \quad i, j = 1, 2, 3 \quad , \tag{F1}$$

approximating them by finite differences. Regular partial derivative means that one of the arguments varies while the other stays unchanged. This is not the case for the normalized ray (group) velocity direction, because a change of only one direction component ruins the normalization, $\mathbf{r} \cdot \mathbf{r} \neq 1$. The remedy is to introduce the two components of the polar angle – zenith $\theta_{\text{ray}}$ and azimuth $\psi_{\text{ray}}$, related to the components $r_i$, to compute the finite differences of the ray velocity,

$$v_{\text{ray}}(\mathbf{r}) = v_{\text{ray}}\left[\theta_{\text{ray}}(\mathbf{r}), \psi_{\text{ray}}(\mathbf{r})\right] \quad , \tag{F2}$$

with respect to zenith and azimuth, and to convert then the derivatives of the ray velocity wrt the polar angle components to those wrt the Cartesian components of the ray direction, applying the chain rule. The components of the polar angle are,

$$\theta_{ray} = \arccos \frac{r_3}{\sqrt{r_1^2 + r_2^2 + r_3^2}} \quad , \quad \psi_{ray} = \arctan(r_2, r_1) \quad , \tag{F3}$$

where the inverse tangent of two arguments means,

$$\cos\psi_{ray} = \frac{r_1}{\sqrt{r_1^2 + r_2^2}} \quad , \quad \sin\psi_{ray} = \frac{r_2}{\sqrt{r_1^2 + r_2^2}} \quad . \tag{F4}$$

The ray direction is normalized, $r_1^2 + r_2^2 + r_3^2 = 1$, but this normalization should be taken into account only after the derivatives are computed. Applying normalization in equation F3 before computing derivatives leads to wrongful results.

The Cartesian components of the ray velocity direction are related to the polar angle components $\theta_{ray}$ and $\psi_{ray}$,

$$r_1 = |\mathbf{r}|\sin\theta_{ray}\cos\psi_{ray} \; , \quad r_2 = |\mathbf{r}|\sin\theta_{ray}\sin\psi_{ray} \; , \quad r_3 = |\mathbf{r}|\cos\theta_{ray} \; , \quad |\mathbf{r}| = 1 \quad . \tag{F5}$$

Equation F4 may be also arranged as,

$$\begin{aligned} \psi_{ray} &= \arctan\frac{r_2}{r_1} \quad \text{for} \quad r_1 > 0 \; , \\ \psi_{ray} &= \arctan\frac{r_2}{r_1} + \pi \quad \text{for} \quad r_1 < 0 \; . \end{aligned} \tag{F6}$$

For computing derivatives, the constant additive factor $\pi$ does not matter, and therefore we can assume that for any signs of the horizontal direction components $r_1$ and $r_2$, the first equation of set F6 is applied.

The first derivatives

The vector form of the chain rule for the first derivatives reads,

$$\frac{\partial v_{\text{ray}}}{\partial \mathbf{r}} = \frac{\partial v_{\text{ray}}}{\partial \theta_{\text{ray}}} \frac{\partial \theta_{\text{ray}}}{\partial \mathbf{r}} + \frac{\partial v_{\text{ray}}}{\partial \psi_{\text{ray}}} \frac{\partial \psi_{\text{ray}}}{\partial \mathbf{r}} \quad . \tag{F7}$$

The derivatives of zenith $\theta_{\text{ray}}$ are,

$$\begin{aligned}
\frac{\partial \theta_{\text{ray}}}{\partial r_1} &= +\frac{1}{\sqrt{r_1^2 + r_2^2}} \frac{r_1 r_3}{r_1^2 + r_2^2 + r_3^2} = +\cos\psi_{\text{ray}} \cos\theta_{\text{ray}} \quad , \\
\frac{\partial \theta_{\text{ray}}}{\partial r_2} &= +\frac{1}{\sqrt{r_1^2 + r_2^2}} \frac{r_2 r_3}{r_1^2 + r_2^2 + r_3^2} = +\sin\psi_{\text{ray}} \cos\theta_{\text{ray}} \quad , \\
\frac{\partial \theta_{\text{ray}}}{\partial r_3} &= -\frac{\sqrt{r_1^2 + r_2^2}}{r_1^2 + r_2^2 + r_3^2} = -\sin\theta_{\text{ray}} \quad ,
\end{aligned} \tag{F8}$$

The derivatives of azimuth $\psi_{\text{ray}}$ are,

$$\begin{aligned}
\frac{\partial \psi_{\text{ray}}}{\partial r_1} &= -\frac{r_2}{r_1^2 + r_2^2} = -\frac{\sin\psi_{\text{ray}}}{\sin\theta_{\text{ray}}} \quad , \\
\frac{\partial \psi_{\text{ray}}}{\partial r_2} &= +\frac{r_1}{r_1^2 + r_2^2} = +\frac{\cos\psi_{\text{ray}}}{\sin\theta_{\text{ray}}} \quad .
\end{aligned} \tag{F9}$$

<u>The second derivatives</u>

The tensor form of the chain rule for the second derivatives reads,

$$\begin{aligned}
\frac{\partial^2 v_{\text{ray}}}{\partial \mathbf{r}^2} &= \frac{\partial^2 v_{\text{ray}}}{\partial \theta_{\text{ray}}^2} \frac{\partial \theta_{\text{ray}}}{\partial \mathbf{r}} \otimes \frac{\partial \theta_{\text{ray}}}{\partial \mathbf{r}} + \frac{\partial^2 v_{\text{ray}}}{\partial \psi_{\text{ray}}^2} \frac{\partial \psi_{\text{ray}}}{\partial \mathbf{r}} \otimes \frac{\partial \psi_{\text{ray}}}{\partial \mathbf{r}} + \frac{\partial^2 v_{\text{ray}}}{\partial \theta_{\text{ray}} \partial \psi_{\text{ray}}} \times \\
&\times \left( \frac{\partial \theta_{\text{ray}}}{\partial \mathbf{r}} \otimes \frac{\partial \psi_{\text{ray}}}{\partial \mathbf{r}} + \frac{\partial \psi_{\text{ray}}}{\partial \mathbf{r}} \otimes \frac{\partial \theta_{\text{ray}}}{\partial \mathbf{r}} \right) + \frac{\partial v_{\text{ray}}}{\partial \theta_{\text{ray}}} \frac{\partial^2 \theta_{\text{ray}}}{\partial \mathbf{r}^2} + \frac{\partial v_{\text{ray}}}{\partial \psi_{\text{ray}}} \frac{\partial^2 \psi_{\text{ray}}}{\partial \mathbf{r}^2} \quad .
\end{aligned} \tag{F10}$$

The "pure" second derivatives of zenith $\theta_{\text{ray}}$ are,

$$\frac{\partial^2 \theta_{\text{ray}}}{\partial r_1^2} = -\frac{r_3}{\left(r_1^2 + r_2^2\right)^{3/2}} \frac{2r_1^2 + r_1^2 r_2^2 - r_2^4 - r_2^2 r_3^2}{\left(r_1^2 + r_2^2 + r_3^2\right)^2} = \left(\cos 2\theta_{\text{ray}} \cos^2 \psi_{\text{ray}} - \cos 2\psi_{\text{ray}}\right) \cot \theta_{\text{ray}},$$

$$\frac{\partial^2 \theta_{\text{ray}}}{\partial r_2^2} = -\frac{r_3}{\left(r_1^2 + r_2^2\right)^{3/2}} \frac{2r_2^2 + r_1^2 r_2^2 - r_1^4 - r_1^2 r_3^2}{\left(r_1^2 + r_2^2 + r_3^2\right)^2} = \left(\cos 2\theta_{\text{ray}} \sin^2 \psi_{\text{ray}} + \cos 2\psi_{\text{ray}}\right) \cot \theta_{\text{ray}}, \quad \text{(F11)}$$

$$\frac{\partial^2 \theta_{\text{ray}}}{\partial r_3^2} = +\frac{2\sqrt{r_1^2 + r_2^2}\, r_3}{\left(r_1^2 + r_2^2 + r_3^2\right)^2} = \sin 2\theta_{\text{ray}}.$$

The mixed second derivatives of zenith $\theta_{\text{ray}}$ are,

$$\frac{\partial^2 \theta_{\text{ray}}}{\partial r_1 \partial r_2} = -\frac{r_1 r_2 r_3}{\left(r_1^2 + r_2^2\right)^{3/2}} \frac{3\left(r_1^2 + r_2^2\right) + r_3^2}{\left(r_1^2 + r_2^2 + r_3^2\right)^2} = -\frac{2 - \cos 2\theta_{\text{ray}}}{2} \cot \theta_{\text{ray}} \sin 2\psi_{\text{ray}}$$

$$\frac{\partial^2 \theta_{\text{ray}}}{\partial r_1 \partial r_3} = +\frac{r_1}{\sqrt{r_1^2 + r_2^2}} \frac{r_1^2 + r_2^2 - r_3^2}{\left(r_1^2 + r_2^2 + r_3^2\right)^2} = -\cos 2\theta_{\text{ray}} \cos \psi_{\text{ray}} \quad \text{(F12)}$$

$$\frac{\partial^2 \theta_{\text{ray}}}{\partial r_2 \partial r_3} = +\frac{r_2}{\sqrt{r_1^2 + r_2^2}} \frac{r_1^2 + r_2^2 - r_3^2}{\left(r_1^2 + r_2^2 + r_3^2\right)^2} = -\cos 2\theta_{\text{ray}} \sin \psi_{\text{ray}}$$

The second derivatives of azimuth $\psi_{\text{ray}}$ are,

$$\frac{\partial^2 \psi_{\text{ray}}}{\partial r_1^2} = +\frac{2r_1 r_2}{\left(r_1^2 + r_2^2\right)^2} = +\frac{\sin 2\psi_{\text{ray}}}{\sin^2 \theta_{\text{ray}}},$$

$$\frac{\partial^2 \psi_{\text{ray}}}{\partial r_2^2} = -\frac{2r_1 r_2}{\left(r_1^2 + r_2^2\right)^2} = -\frac{\sin 2\psi_{\text{ray}}}{\sin^2 \theta_{\text{ray}}}, \quad \text{(F13)}$$

$$\frac{\partial^2 \psi_{\text{ray}}}{\partial r_1 \partial r_2} = -\frac{r_1^2 - r_2^2}{\left(r_1^2 + r_2^2\right)^2} = -\frac{\cos 2\psi_{\text{ray}}}{\sin^2 \theta_{\text{ray}}}.$$

The finite differences will be applied to approximate the derivatives of the ray velocity with respect to the polar angles only. The first finite differences read,

$$\frac{\partial v_{\text{ray}}}{\partial \theta_{\text{ray}}} = \frac{v_{\text{ray}}\left(\theta_{\text{ray}} + \Delta\theta_{\text{ray}}, \psi_{\text{ray}}\right) - v_{\text{ray}}\left(\theta_{\text{ray}} - \Delta\theta_{\text{ray}}, \psi_{\text{ray}}\right)}{2\Delta\theta_{\text{ray}}},$$
$$\frac{\partial v_{\text{ray}}}{\partial \psi_{\text{ray}}} = \frac{v_{\text{ray}}\left(\theta_{\text{ray}}, \psi_{\text{ray}} + \Delta\psi_{\text{ray}}\right) - v_{\text{ray}}\left(\theta_{\text{ray}}, \psi_{\text{ray}} - \Delta\psi_{\text{ray}}\right)}{2\Delta\psi_{\text{ray}}}.$$
(F14)

The second "pure" finite differences read,

$$\frac{\partial^2 v_{\text{ray}}}{\partial \theta_{\text{ray}}^2} = \frac{v_{\text{ray}}\left(\theta_{\text{ray}} + \Delta\theta_{\text{ray}}, \psi_{\text{ray}}\right) - 2v_{\text{ray}}\left(\theta_{\text{ray}}, \psi_{\text{ray}}\right) + v_{\text{ray}}\left(\theta_{\text{ray}} - \Delta\theta_{\text{ray}}, \psi_{\text{ray}}\right)}{\Delta\theta_{\text{ray}}^2},$$
$$\frac{\partial^2 v_{\text{ray}}}{\partial \psi_{\text{ray}}^2} = \frac{v_{\text{ray}}\left(\theta_{\text{ray}}, \psi_{\text{ray}} + \Delta\psi_{\text{ray}}\right) - 2v_{\text{ray}}\left(\theta_{\text{ray}}, \psi_{\text{ray}}\right) + v_{\text{ray}}\left(\theta_{\text{ray}}, \psi_{\text{ray}} - \Delta\psi_{\text{ray}}\right)}{\Delta\psi_{\text{ray}}^2}.$$
(F15)

The second mixed finite difference reads,

$$\frac{\partial^2 v_{\text{ray}}}{\partial \theta_{\text{ray}} \partial \psi_{\text{ray}}} = \frac{1}{4\Delta\theta_{\text{ray}} \Delta\psi_{\text{ray}}} \times$$
$$\left[ v_{\text{ray}}\left(\theta_{\text{ray}} + \Delta\theta_{\text{ray}}, \psi_{\text{ray}} + \Delta\psi_{\text{ray}}\right) - v_{\text{ray}}\left(\theta_{\text{ray}} + \Delta\theta_{\text{ray}}, \psi_{\text{ray}} - \Delta\psi_{\text{ray}}\right) \right.$$
$$\left. - v_{\text{ray}}\left(\theta_{\text{ray}} - \Delta\theta_{\text{ray}}, \psi_{\text{ray}} + \Delta\psi_{\text{ray}}\right) + v_{\text{ray}}\left(\theta_{\text{ray}} - \Delta\theta_{\text{ray}}, \psi_{\text{ray}} - \Delta\psi_{\text{ray}}\right) \right].$$
(F16)

<u>The mixed "location-direction" derivatives of the ray velocity</u>

The mixed derivative is approximated by finite differences as,

$$\nabla_{\mathbf{x}} \nabla_{\mathbf{r}} v_{\text{ray}} = \left\{ \frac{\partial^2 v_{\text{ray}}}{\partial x_i \partial r_j} \right\} = \nabla_{\mathbf{r}}\left[\nabla_{\mathbf{x}} v_{\text{ray}}(\mathbf{x}, \mathbf{r})\right] = \nabla_{\mathbf{r}} \begin{bmatrix} \dfrac{v_{\text{ray}}(x_1 + \Delta) - v_{\text{ray}}(x_1 - \Delta)}{2\Delta} \\ \dfrac{v_{\text{ray}}(x_2 + \Delta) - v_{\text{ray}}(x_2 - \Delta)}{2\Delta} \\ \dfrac{v_{\text{ray}}(x_3 + \Delta) - v_{\text{ray}}(x_3 - \Delta)}{2\Delta} \end{bmatrix}.$$
(F17)

The items inside the brackets form a column. After computing the directional finite differences, each item evolves in a row, and all of them become a matrix. Applying the chain rule of equation F7, we obtain each component of this matrix,

$$\nabla_{\mathbf{r}} \left[ \frac{v_{\text{ray}}(x_i + \Delta) - v_{\text{ray}}(x_i - \Delta)}{2\Delta} \right] =$$
$$\frac{\partial}{\partial \theta_{\text{ray}}} \frac{v_{\text{ray}}(x_i + \Delta) - v_{\text{ray}}(x_i - \Delta)}{2\Delta} \cdot \frac{\partial \theta_{\text{ray}}}{\partial \mathbf{r}} + \frac{\partial}{\partial \psi_{\text{ray}}} \frac{v_{\text{ray}}(x_i + \Delta) - v_{\text{ray}}(x_i - \Delta)}{2\Delta} \frac{\partial \psi_{\text{ray}}}{\partial \mathbf{r}} \quad . \tag{F18}$$

This yields the components of the mixed Hessian,

$$\left[ \nabla_{\mathbf{x}} \nabla_{\mathbf{r}} v_{\text{ray}} \right]_{ij} = \left\{ \frac{\partial^2 v_{\text{ray}}}{\partial x_i \partial r_j} \right\} = \frac{1}{4\Delta_{\text{dst}} \Delta_{\text{ang}}} \left( A_i \frac{\partial \theta_{\text{ray}}}{\partial r_j} + B_i \frac{\partial \psi_{\text{ray}}}{\partial r_j} \right) \quad , \tag{F19}$$

where,

$$A_i = \left[ v_{\text{ray}}(x_i + \Delta_{\text{dst}}, \theta_{\text{ray}} + \Delta_{\text{ang}}, \psi_{\text{ray}}) - v_{\text{ray}}(x_i - \Delta_{\text{dst}}, \theta_{\text{ray}} + \Delta_{\text{ang}}, \psi_{\text{ray}}) \right.$$
$$\left. - v_{\text{ray}}(x_i + \Delta_{\text{dst}}, \theta_{\text{ray}} - \Delta_{\text{ang}}, \psi_{\text{ray}}) + v_{\text{ray}}(x_i + \Delta_{\text{dst}}, \theta_{\text{ray}} - \Delta_{\text{ang}}, \psi_{\text{ray}}) \right] \quad ,$$
and
$$B_i = \left[ v_{\text{ray}}(x_i + \Delta_{\text{dst}}, \theta_{\text{ray}}, \psi_{\text{ray}} + \Delta_{\text{ang}}) - v_{\text{ray}}(x_i - \Delta_{\text{dst}}, \theta_{\text{ray}}, \psi_{\text{ray}} + \Delta_{\text{ang}}) \right. \tag{F20}$$
$$\left. - v_{\text{ray}}(x_i + \Delta_{\text{dst}}, \theta_{\text{ray}}, \psi_{\text{ray}} - \Delta_{\text{ang}}) + v_{\text{ray}}(x_i + \Delta_{\text{dst}}, \theta_{\text{ray}}, \psi_{\text{ray}} - \Delta_{\text{ang}}) \right] \quad .$$

Thus, to approximate the directional gradient and Hessian and the mixed Hessian of the ray velocity, we need two first and three second derivatives, approximated by finite differences,

$$\frac{\partial v_{\text{ray}}}{\partial \theta_{\text{ray}}}, \frac{\partial v_{\text{ray}}}{\partial \psi_{\text{ray}}}, \frac{\partial^2 v_{\text{ray}}}{\partial \theta_{\text{ray}}^2}, \frac{\partial^2 v_{\text{ray}}}{\partial \psi_{\text{ray}}^2}, \frac{\partial^2 v_{\text{ray}}}{\partial \theta_{\text{ray}} \partial \psi_{\text{ray}}} \quad . \tag{F21}$$

The stencil for directional derivatives is shown in Figure 1. Each of the first derivatives requires the function values at two nodes. The pure second derivatives require in addition the central node value. These nodes are shown in blue. The mixed second derivative requires four additional node values, where the corresponding nodes are shown in red.

The advantage of the proposed analytical approach is that it requires solving the nonlinear equation set (that yields the slowness vector components) only at the central node. With the numerical approach, we solve this set at all nodes of the stencil. For the second-order central differences approximating the first and second derivatives of the ray velocity with respect to the polar angle components, this stencil includes nine nodes, as shown in the figure. For more accurate fourth-order finite differences, the stencil includes twenty-five nodes. In addition, the numerical azimuthal derivatives become unstable at or near spherical grid singularities $\theta_{\text{ray}} = 0$ and $\theta_{\text{ray}} = \pi$.

Bliss, G., 1916, Jacobi's condition for problems of the calculus of variations in parametric form: Transactions of the American Mathematical Society, **17**, 195-206.

Bona, A., and M. Slawinski, 2003, Fermat's principle for seismic rays in elastic media: Journal of Applied Geophysics, **54**, no. 3-4, 445-451.

Cao, J., J. Hu, and H. Wang, 2017, Traveltime computation in TI media using Fermat's principle fast marching: EAGE 79th Conference and Technical Exhibition, Expanded Abstract, **DOI:** 10.3997/2214-4609.201700670.

Casasanta, L., G. Drufuca, C. Andreoletti, and J. Panizzardi, 2008, 3D anisotropic ray tracing by raypath optimization: SEG International Exposition and 78th Annual Meeting, Expanded Abstract, 2161-2165.

Červený, V., 2000, Seismic ray theory: Cambridge University Press, ISBN 978-0521366717.

Christoffel, E., 1877. Über die Fortpflanzung von Stössen durch elastische feste Körper: Annali di Matematica, 8, 193–243, doi: 10.1007/ BF02420789.

Farra, V., and R. Madariaga, 1988, Non-linear reflection tomography: Geophysical Journal, **95**, 135-147.

Farra, V., J. Virieux, and R. Madariaga, 1989, Ray perturbation for interfaces: Geophysical Journal International, **99**, no. 2, 377-390.

Farra, V., and I. Pšenčík, 2016, Weak-anisotropy approximations of P-wave phase and ray velocities for anisotropy of arbitrary symmetry: Studia Geophysica et Geodaetica, **60**, 403-418.
Page 84 of 90

# LIST OF TABLES

# LIST OF FIGURES

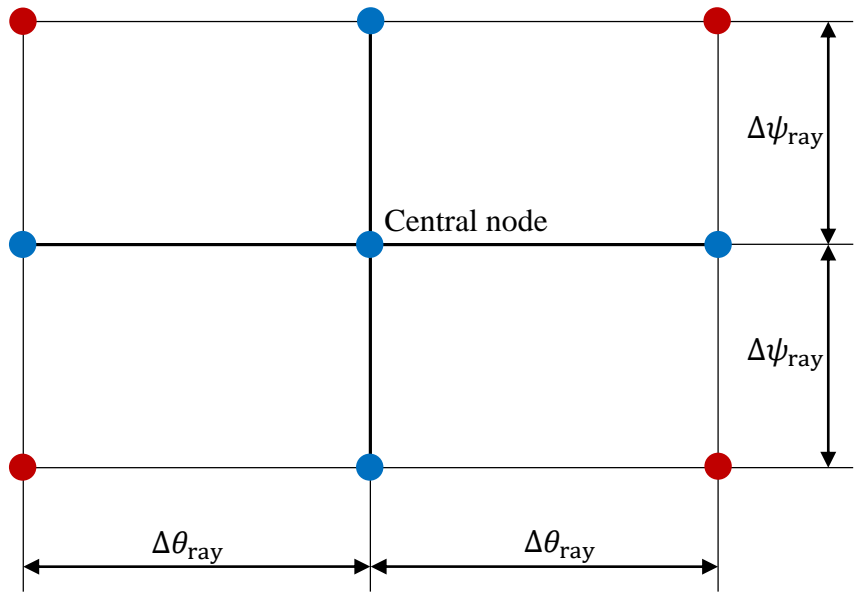

Figure 1. Stencil for directional finite differences of the ray velocity